\documentclass{article}

\usepackage{arxiv}

\usepackage[utf8]{inputenc}  
\usepackage[T1]{fontenc}     
\usepackage{hyperref}        
\usepackage{url}             
\usepackage{booktabs}        
\usepackage{amsmath}
\usepackage{amsfonts}        
\usepackage{nicefrac}        
\usepackage{microtype}       
\usepackage{cleveref}        
\usepackage{lipsum}          
\usepackage{graphicx}
\usepackage{natbib}
\usepackage{doi}
\usepackage{listings}
\usepackage{xcolor}
\usepackage{multirow}

\title{Order-lifted data inversion/retrieval method of neighbor cells to implement general high-order schemes in unstructured-mesh-based finite-volume solution framework}

\renewcommand{\headeright}{Research Paper}
\renewcommand{\undertitle}{Research Paper}

\renewcommand{\shorttitle}{%
Data Order-Lifted Inversion of Neighbor Cells}

\author
{%
  \href{https://orcid.org/0000-0001-6232-690}{\includegraphics[scale=0.06]{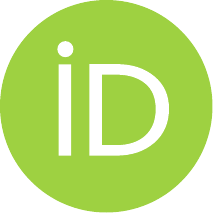}\hspace{1mm}Hao~Guo} \\
	Institute of Engineering Thermophysics \\
	Tsinghua University \\
	Beijing, China \\
	\texttt{guoh19@proton.me} \\
  \AND
  Peixue~Jiang \\
  Institute of Engineering Thermophysics \\
  Tsinghua University \\
	Beijing, China \\
  \texttt{jiangpx@tsinghua.edu.cn} \\
	\And
	Xiaofeng~Ma \\
  Institute of Engineering Thermophysics\\
	Tsinghua University \\
	Beijing, China \\
	\texttt{mxf20@mails.tsinghua.edu.cn} \\
  \And
	Boxing~Hu \\
  Institute of Engineering Thermophysics\\
	Tsinghua University \\
	Beijing, China \\
	\texttt{hbx23@mails.tsinghua.edu.cn} \\
	\And
  Yinhai~Zhu\thanks{Corresponding author.} \\
	Institute of Engineering Thermophysics\\
	Tsinghua University \\
	Beijing, China \\
	\texttt{yinhai.zhu@tsinghua.edu.cn} \\
}

\graphicspath{ {figures/} {./} }

\hypersetup
{%
  colorlinks = true,
  urlcolor   = blue,
  citecolor  = blue,
  pdftitle={Order-lifted data inversion/retrieval method of neighbor cells to implement general high-order schemes in unstructured-mesh-based finite-volume solution framework},
  pdfauthor={Hao~Guo, Peixue~Jiang, Xiaofeng~Ma, Boxing~Hu, Yinhai~Zhu},
  pdfkeywords={Unstructured gird, High-order method, Finite-volume method},
}

\begin{document}

\maketitle

\begin{abstract}
  This study introduces an order-lifted inversion/retrieval method for implementing high-order schemes within the framework of an unstructured-mesh-based finite-volume method. This method defines a special representation called the data order-lifted inversion of neighbor cells (DOLINC) differential, which transforms the degrees of freedom of wide templates into differentials of various orders stored in local grid cells. Furthermore, to retrieve the original far-field information without bias during the reconstruction/interpolation of face values, the corresponding accurate inversion formulas are derived based on the defined DOLINC differentials. The order-lifted inversion method can be applied to multi-dimensional polyhedral-mesh solvers by considering the influence of grid non-uniformity on high-order schemes. It seamlessly accommodates multi-process parallel computing for high-order methods without requiring special consideration for the boundary interface. This method not only enhances the numerical accuracy of second-order finite-volume methods, but also demonstrates a significant computational-speed advantage over similar methods. A series of benchmark cases, including the linear advection, Burgers, and Euler equations, are comprehensively validated to assess the practical performance of the method. The results indicate that the unstructured-mesh high-order schemes implemented based on this method achieve theoretical accuracy in practical computations and substantially reduce computational costs compared with methods that increase grid resolution.
\end{abstract}


\section{Introduction}

Recently, high-order computational fluid dynamics (CFD) methods have gained widespread recognition in the academic community. These methods, which are known for capturing more physical details with fewer grid points, can significantly reduce the computational costs of complex problems. Various fields that involve fluid-simulation problems, such as aerospace, energy, environmental sciences, and biomedical sciences, urgently need more precise simulation methods. Among these, rapidly evolving scale-resolving turbulence-simulation methods such as the detached eddy simulation, large eddy simulation (LES), and direct numerical simulation (DNS) have been increasingly applied to various complex flow problems \citep{Chai:2015, Guo:2023:HMT, Sun:2019}. Increasing the order of the numerical schemes is highly efficient for obtaining higher accuracy in capturing the flow details. For example, when the desired characteristic length reaches 0.001, a second-order scheme requires 100 grid points, whereas a sixth-order scheme only needs 5.71 grid points \citep{Wagner:2007}. This computational-efficiency improvement is particularly attractive for scale-resolving turbulence simulations, which are still constrained by computational costs, making them a focal point in related research.

Research on high-order schemes has been ongoing for nearly 40 years. Since the introduction of the concept of total variation diminishing (TVD) schemes by \citet{Harten:1997}, most high-order scheme research has focused on structured grids. However, CFD has been increasingly applied to complex flow problems; thus, unstructured grids, with their ability to handle intricate geometries and higher computational efficiency, are now widely adopted by most industrial software. The advantage of using an unstructured mesh is the substantial reduction in the required grid information storage. An unstructured mesh can be applied not only to polyhedron meshes of arbitrary shapes, but also to large-problem high-performance computing (HPC) for solving complex problems. However, unstructured grids encounter greater challenges in indirectly obtaining far-field-cell degree-of-freedom (DOF) information compared to structured grids, thereby presenting significant difficulties in implementing high-order schemes.

Barth and Frederickson initially proposed a $k$-exact reconstruction method to implement high-order schemes on unstructured grids \citep{Barth:1990}. The fundamental idea was to utilize a local template comprising multiple mesh cells and Gaussian integration points to construct a $k$-degree interpolation polynomial in the local space, thereby obtaining the corresponding $k+1$ order method. This method later became the foundation for many high-order methods for unstructured meshes \citep{Ekaterinaris:2005, Wang:2007}, including the essentially non-oscillatory (ENO) and weighted essentially non-oscillatory (WENO) methods. The original ENO method was developed based on structured grids, and was first introduced by \citet{Harten:1987}. The basic concept involved selecting the smoothest reconstruction from several alternative templates when constructing the scheme. \citet{Durlofsky:1992} and \citet{Abgrall:1994} subsequently extended the ENO method to unstructured grids. Similarly, based on an unstructured-mesh $k$-exact framework, \citet{Ollivier:1997} established a quasi-ENO scheme using least-squares reconstruction. The WENO method was initially proposed for structured grids by \citet{Liu:1994} and then by \citet{Jiang:1996}. \citet{Friedrich:1998} extended the WENO method to unstructured grids. \citet{Hu:1999} obtained the WENO method for unstructured grids based on a two-dimensional triangular mesh. Dumbser et al. established WENO methods of arbitrary order based on an unstructured finite-volume method (FVM) framework \citep{Dumbser:2007a} and subsequent Gauss-integration-free WENO methods \citep{Dumbser:2007b}. \citet{Li:2011} proposed an improved WENO method for unstructured grids under a $k$-exact framework using quadratic reconstruction. Tsoutsanis et al. implemented WENO methods for unsteady \citep{Tsoutsanis:2011} and viscous flows \citep{Tsoutsanis:2014} based on unstructured hybrid grids. Following Dumbser and Tsoutsanis' methods, \citet{Martin:2018} implemented WENO methods on the open-source C++ library OpenFOAM \citep{FOAM} using the unstructured-mesh-based FVM framework, and subsequent improvements in computational efficiency were made by \citet{Gartner:2020}. Although these high-order methods, which were implemented based on $k$-exact methods, achieved theoretical accuracy on unstructured grids, they did not outperform or persist with high-order schemes on structured grids in terms of computational efficiency. This is because even without increasing the Gaussian integration points, $k$-exact high-order schemes still require local searches for multiple-cell templates and the calculation and storage of a large number of interpolation polynomial coefficients. For example, for a three-dimensional problem, as many as 19 template coefficients of $k=3$ need to be stored. Most CFD industrial software, whether commercial codes, such as ANSYS Fluent \citep{Fluent} and STAR-CCM+ \citep{STAR}, or open-source codes, such as OpenFOAM and code\_saturne \citep{Saturne}, are based almost entirely on second-order FVM frameworks. Implementing $k$-exact high-order schemes on such low-order FVMs involves the challenge of increasing Gaussian integration points as well as the special handling of template-information communication at interfaces during parallel computing, which is unfavorable for both computational cost and stability.

Currently, one of the most prominent and highly promising high-order methods is the correction procedure via reconstruction (CPR) series. In 2007, \citet{Huynh:2007} first introduced the flux reconstruction (FR) approach, unifying the discontinuous Galerkin, spectral volume, and spectral difference methods within a single framework. In 2009, \citet{Huynh:2009} extended the FR method to two-dimensional quadrilateral grids. Wang proposed a lifting collocation penalty formulation in 2009, further generalizing the FR approach to two-dimensional triangular grids and triangular/quadrilateral hybrid grids \citep{Wang:2009}. Subsequently, \citet{Huynh:2011} and \citet{Wang:2011} collectively termed these methods as CPR. Although CPR methods exhibit several excellent characteristics, the differences between CPR and the classical second-order FVM framework are even more pronounced compared to k-exact methods. Implementing the CPR method using most of the existing unstructured-grid FVM industrial software is almost impossible with top-level modifications. To utilize CPR methods, a bottom-up compatibility between the high-order mesh and corresponding solution framework must be rebuilt. Therefore, current applications of CPR methods are primarily implemented in in-house codes developed by researchers specializing in high-order scheme development rather than in widely used industrial software for general applications.

Therefore, although academic research has long developed high-order methods, and various high-order methods have been applied by scholars in different fields to achieve high-precision numerical simulations of complex problems, low-order FVM software still dominates current industrial applications. The widespread adoption of unstructured high-order schemes in industrial settings depends on two factors. First, existing second-order FVM frameworks may not be able to easily implement corresponding high-order methods; rewriting a mature industrial software architecture is a tedious and costly process. Second, the computational efficiency of high-order unstructured grid schemes for improving accuracy must surpass that of a simple grid-refinement strategy. In other words, they must significantly reduce computational costs while maintaining comparable accuracy. Otherwise, industrial software lacks the necessity and motivation to adopt high-order schemes. One method to satisfy these two factors is to use gradient approximations to obtain the required far-field DOFs for high-order schemes. The use of gradient estimation for far-field cell information in unstructured grids originated with \citet{Darwish:2003}, who implemented a TVD limiter. They used the gradient from neighboring cells to compute information for upstream cells. This has become the standard method for the majority of industrial CFD software based on unstructured collocated mesh frameworks to implement TVD schemes. Building on this, \citet{Sheng:2019} utilized a uniform-grid-based least-squares gradient-approximation method to construct a WENO scheme on unstructured grids. However, these gradient-approximation methods present two major challenges. First, the computational formulas were derived based on uniform orthogonal Cartesian grids, and errors may occur even in marginally more complex grid types, thereby limiting the applicability of the method. This error is less pronounced in lower-order TVD, but becomes more prominent in higher-order schemes with wider templates. Second, researchers have attempted to use more accurate methods such as the least-squares approximation to improve the accuracy of gradient calculations. However, changes in the gradient values have caused the estimated cell-averaged values (FVM) of far-field cells to deviate from their true values. Therefore, these uniform-grid gradient-approximation methods perform well only when solving on uniform hexahedral/quadrilateral grids and may experience unpredictable accuracy-degradation issues when applied to other grid types, particularly boundary-layer meshes with significant local size variations \citep{Zhong:2020}. Consequently, such methods have not been widely adopted by low-order FVM software to improve accuracy.

To enhance the solution accuracy of low-order FVM frameworks, this study proposes a method called data order-lifted inversion of neighbor cells (DOLINC). This method can implement various high-order spatial discretization schemes, including the fixed-template reconstruction and ENO/WENO methods, within the framework of an unstructured mesh-based finite-volume method. The fundamental concept of the proposed method is to convert the DOFs required for high-order schemes from multiple grid cells into differential DOFs stored in face-neighboring cells. During the reconstruction/interpolation of face values, the accurate inversion or retrieval of the original far-field information is achieved using strictly derived formulas. This method can be applied to any polyhedral mesh, achieving a practical accuracy no less than those of second-order high-order schemes for solving multi-dimensional flow problems. On non-uniform hexahedral meshes, particularly boundary-layer grids with significant local size variations, this method can accurately reach the theoretical convergence order of high-order schemes. The proposed DOLINC method can be seamlessly integrated into existing low-order FVM frameworks without requiring a special boundary treatment, making it adaptable to large-problem parallel computations. As an illustration, we implemented the DOLINC method based on the open-source C++ library OpenFOAM at a low level and conducted a series of computational tests for various schemes.

The remainder of this paper is organized as follows. Section \ref{sec:method} provides a detailed description of the core concepts and implementation principles of the DOLINC method. It introduces the scope of application of the method and its various characteristics. Section \ref{sec:results} demonstrates the practical performance of high-order schemes implemented based on the DOLINC method in various test cases. Section \ref{ssec:BL} investigates the accuracy of the scheme on a non-uniform boundary layer mesh, and Section \ref{ssec:LAE} studies linear advection problems. Section \ref{ssec:BE} explores the performance of high-order methods in handling discontinuities in solving the Burgers equation. In Section \ref{ssec:Euler}, high-order schemes are applied to solve the Euler equations, encompassing four categories of one-dimensional/two-dimensional problems to comprehensively assess the computational accuracy and cost-effectiveness of the DOLINC method. Finally, Section \ref{sec:conclusions} summarizes this study.

\section{Description of the DOLINC Method}
\label{sec:method}

This section provides a detailed explanation of how the DOLINC method implements high-order finite-volume schemes based on an unstructured collocated mesh solver. The applicability and computational efficiency of this method are briefly discussed.

\subsection{Unstructured-Mesh-Based FVM Framework}

\begin{figure}
	\centering
  \fbox{
    \parbox[b]{11cm}
    {
      \centering
      \includegraphics[width=4.5cm]{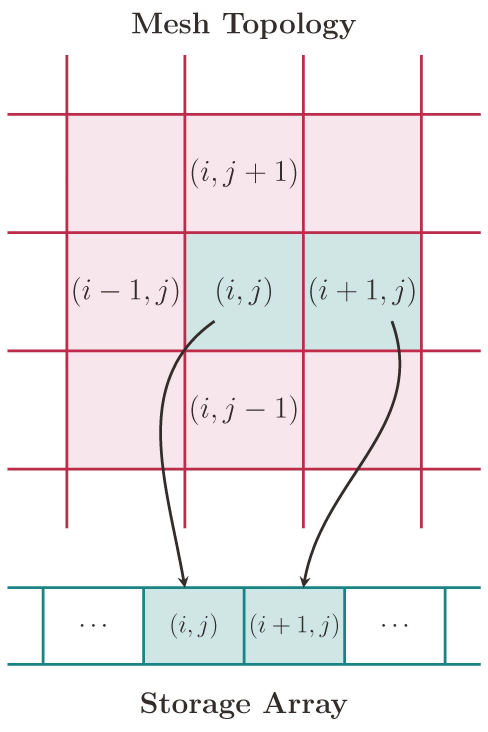}
      \qquad
      \includegraphics[width=4.5cm]{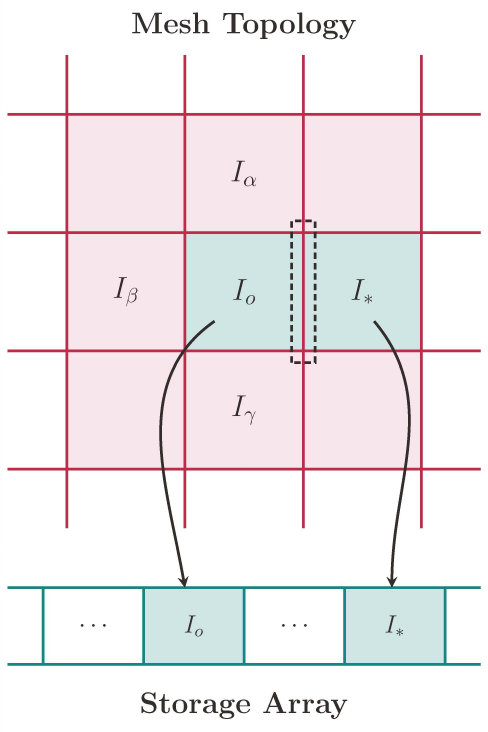}
      \par
      \centering \; (\textit{a}) \hspace{4.7cm} (\textit{b})
    }
  }
	\caption{Differences between structured and unstructured mesh solvers. (\textit{a}) Information indexing based on structured mesh storage, and (\textit{b}) information indexing based on unstructured grid storage.}
	\label{fig:schmtc:meshTopo}
\end{figure}

To date, many studies have demonstrated the differences in structured and unstructured grids. This study delineates unstructured mesh solvers based on the solver's behaviors in creating a grid topology and acquiring data at a low level, rather than relying on the shape of the grid elements, as illustrated in Figure \ref{fig:schmtc:meshTopo}. When using an unstructured mesh architecture for partial differential equation (PDE) solvers to obtain grid-topology information and cell data, even a hexahedral Cartesian grid must be transformed into an unstructured storage format. In Figure \ref{fig:schmtc:meshTopo}, the two types of grids have identical shapes (three-dimensional hexahedral grid or two-dimensional quadrilateral grid); however, different topology-storage methods determine the distinctions between the structured and unstructured grid solvers. For structured grid solvers, the global grid topology is known, enabling the solver to directly access neighboring cell information and determine the relative positions of any two cells in global coordinates. In data storage, this characteristic is typically reflected in adjacent memory-storage units corresponding to adjacent grid cells. Although the storage structure of structured grids is easy to implement for high-order schemes, it is limited by inefficiency in massive grid storage, additional considerations for process/thread topology information communications during parallel computing, and the inability to handle more complex geometric/grid forms. Therefore, primary industrial software, such as the commercial software ANSYS Fluent and STAR-CCM+, as well as open-source codes, such as OpenFOAM and code\_saturne, adopt the unstructured mesh information-storage method. Unstructured meshes treat each grid cell as an independent unit, and the solver cannot directly acquire information about the neighboring cells. It relies on topological information to reconstruct relationships between cells. For example, in Figure \ref{fig:schmtc:meshTopo}(\textit{b}), $I_o$ does not inherently know its neighboring cells; it needs to confirm that $I_*$ is a neighboring cell through the connection information on the grid faces. At the data-storage level, this is reflected in adjacent storage units that do not necessarily contain information about adjacent grid cells. The DOLINC method was developed based on the unstructured grid solvers depicted in Figure \ref{fig:schmtc:meshTopo}(\textit{b}) and serves as an efficient approach for implementing high-order FVM schemes.

\begin{figure}
	\centering
  \fbox{
    \includegraphics[width=7cm]{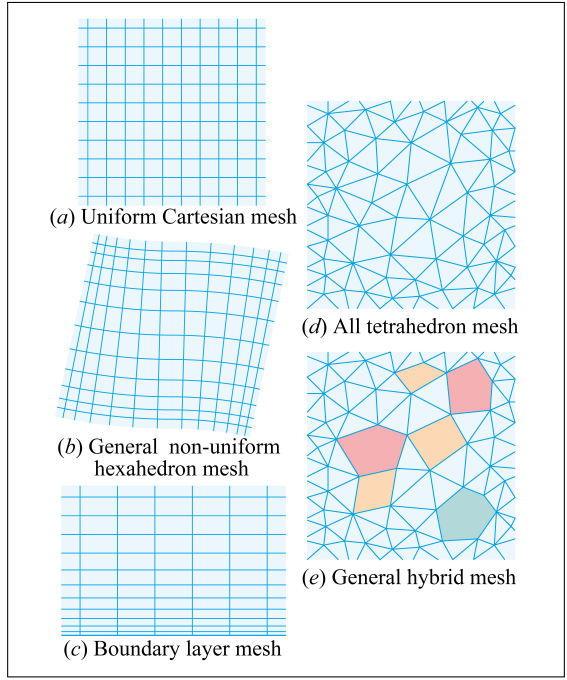}
  }
	\caption{Different shapes of unstructured grids. (\textit{a}) Uniform Cartesian mesh, (\textit{b}) non-uniform hexahedral (three-dimensional)/quadrilateral (two-dimensional) mesh, (\textit{c}) boundary layer mesh, (\textit{d}) pure tetrahedral (three-dimensional)/triangular (two-dimensional) mesh, (\textit{e}) hybrid polyhedral mesh.}
	\label{fig:schmtc:meshShape}
\end{figure}

Based on the same data-storage method, unstructured meshes can be classified into several types according to the differences in their grid cell shapes, as illustrated in Figure \ref{fig:schmtc:meshShape}. The simplest form of an unstructured mesh is the uniform hexahedral (three-dimensional)/quadrilateral (two-dimensional) mesh shown in Figure \ref{fig:schmtc:meshShape}(\textit{a}), which is geometrically identical to a classical structured mesh. The gradient-approximation methods used in previous studies \citep{Sheng:2019, Zhong:2020} apply only to uniform unstructured grids without order or accuracy degeneration. Clearly, the applicability of this grid form is limited because the assumption of uniformity cannot be guaranteed in practical applications. The more common form of the non-uniform hexahedral grid used in various CFD applications is depicted in Figure \ref{fig:schmtc:meshShape}(\textit{b}), exhibiting significant non-uniformity. Gradient-approximation methods neglect this non-uniformity, leading to substantial errors in high-order reconstruction, particularly in the computation of boundary-layer meshes, as shown in Figure \ref{fig:schmtc:meshShape}(\textit{c}). The accurate computation of boundary-layer flows is a fundamental and crucial aspect in most CFD applications. Even in simulations involving complex geometries, such as racing cars or airplanes with polyhedral meshes, as shown in Figure \ref{fig:schmtc:meshShape}(\textit{d}) or Figure \ref{fig:schmtc:meshShape}(\textit{e}), boundary-layer grids are still required near the wall region to correctly resolve near-wall boundary layer flows, ensuring the reliability of the computed drag forces and other results.

Therefore, ensuring the computational accuracy of the boundary-layer meshes is a key aspect in addressing high-order schemes in unstructured-mesh-based finite volume methods. The core issue is the consideration of the local non-uniformity of the mesh. Existing gradient-approximation methods do not consider the size variation of the mesh, leading to uncontrollable accuracy degradation in unstructured meshes other than that depicted in Figure \ref{fig:schmtc:meshShape}(\textit{a}) \citep{Darwish:2003, Sheng:2019}. To guarantee the theoretical accuracy of high-order methods on non-uniform meshes, the proposed DOLINC method employs a strictly mathematically derived result that considers the local mesh growth rate when retrieving far-field cell information. This approach provides accurate results for meshes resembling those in Figures \ref{fig:schmtc:meshShape}(\textit{a}) and \ref{fig:schmtc:meshShape}(\textit{b}). Similar to the gradient methods, DOLINC can implement high-order schemes on all types of unstructured meshes, as depicted in Figure 2, that is, it is applicable to any polyhedral mesh.

\subsection{Spatial Discretization of Finite-Volume Method}

\begin{figure}
	\centering
  \fbox{
    \includegraphics[width=8cm]{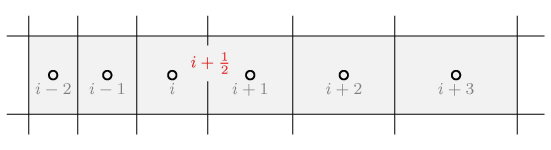}
  }
	\caption{Schematic of non-uniform split multi-dimensional mesh (three-dimensional hexahedral/two-dimensional quadrilateral).}
	\label{fig:schmtc:stencil}
\end{figure}

The derivation of the mathematical model of the DOLINC method is based on the spatial discretization of the finite-volume method, as shown in Figure \ref{fig:schmtc:stencil}, where the mesh size exhibits significant local variations. In the figure, $i+1/2$ marks the current face requiring interpolation/reconstruction, whereas $i$ marks the neighboring mesh cell corresponding to its upwind direction. Assume that the growth rate of the local grid cells remains constant; thus,

\begin{equation}
  g = \frac{\Delta x_{j+1}}{\Delta x_{j}} = \mathrm{const.}, \quad j = \dots, i-1, i, i+1, \dots
\end{equation}

where $g$ represents the cell growth rate, and its value is in the range of $(0,\infty)$. For three-dimensional grids, the center-to-center distance between two adjacent cells on a mesh face is

\begin{equation}
  \begin{aligned}
    \mathbf{d} &= \mathbf{x}_{i+1} - \mathbf{x}_{i}, \\
      d &= \|\mathbf{d}\|.
  \end{aligned}
\end{equation}

In the low-order FVM framework of industrial software, the incorporation of spatial discretization schemes involves the use of weights $\omega$ to introduce the influence of distance variation between the cell center and mesh-face center:

\begin{align}
  \omega =
      \frac{
          ||\mathbf{x}_{i+1} - \mathbf{x}_{i+\frac12}||
      }{
          ||\mathbf{x}_{i} - \mathbf{x}_{i+\frac12}||
        + ||\mathbf{x}_{i+1} - \mathbf{x}_{i+\frac12}||
      }.
\end{align}

The above equation shows that the relationship between the grid growth rate $g$ and $\omega$ can be obtained as follows:

\begin{equation}
  g
      = \frac{\omega}{1 - \omega}
      = \frac{||\mathbf{x}_{i+1} - \mathbf{x}_{i+\frac12}||}{||\mathbf{x}_{i+\frac12} - \mathbf{x}_{i}||}.
\end{equation}

Assuming the exact solution of the PDE in the studied region is the function $\phi(x)$ defined on the domain, the FVM solver obtains the average value $\overline{\phi}_i$ within the grid cell. For the x-split three-dimensional problem, we have

\begin{equation}
  \overline{\phi}_{i} = \int_{x_{i-\frac12}}^{x_{i+\frac12}} \phi(\xi) \; \mathrm{d}\xi.
\end{equation}

In the derivation of the reconstruction formula, the primitive function is used:

\begin{equation}
  \varPhi(x) = \int_{x_0}^x \phi(\xi) \; \mathrm{d}\xi,
\end{equation}

where $x_0$ is an arbitrarily selected reference point. After obtaining the interpolation polynomial $P(x)$ of $\varPhi(x)$, the derivative polynomial $p(x)$ is obtained. We can prove that

\begin{equation}
  \overline{\phi}_{i} = \int_{x_{i-\frac12}}^{x_{i+\frac12}} p(\xi) \; \mathrm{d}\xi.
\end{equation}

Therefore, the interpolation polynomial $p(x)$ can be used to construct function values or various order derivatives on the mesh faces.

\subsection{Basic Principles of DOLINC Method for Inverting Far-Field Data}

Based on the previous discussion, the main challenge in achieving universal high-order schemes on unstructured grids is efficiently obtaining relevant data from far-field elements or their corresponding multi-DOF information. The computational cost of high-order reconstruction methods is undoubtedly higher than that of low-order FVM methods because of the necessity of additional DOFs when constructing interpolation polynomials. Therefore, a feasible high-order unstructured-grid finite-volume method must meet three efficiency requirements: a) easy implementation based on the current low-order FVM; b) minimal additional runtime storage; and c) a short time needed to acquire additional DOFs.

The fundamental idea behind the DOLINC method is to store DOF-transformed order-lifted differentiations and precisely invert the original data based on this storage. This corresponds to two core steps. The first step involves storing multi-grid information from the high-order reconstruction template as localized high-order differential data, and the second step utilizes the stored differential data to accurately calculate and recover the transformed DOFs. Throughout the computation process, the grid topology or solver matrix assembly within the low-order FVM framework do not required modification. All computational steps are completed within the interpolation/reconstruction phase of the equation discretization.

In the second-order FVM, linear interpolation combined with Gauss's theorem is commonly employed to calculate the gradient of grid cells (referred to as the central differencing gradient or central differencing derivative). Considering the local non-uniform variation of the mesh cell, the following exact expression can be derived:

\begin{align}
\label{eq:recursion}
  \overline{\phi}_{i+n+2} &= g (\overline{\phi}_{i+n} + 2 g^{n} d_{j}\, \overline{\phi}_{i+n+1,\,j}) + (1 - g)\overline{\phi}_{i+n+1},
\end{align}

where $n = 0, 1, 2, \dots$, and $\overline{\phi}_{i+n+1,\,j}$ represents the $j$-th component of the first-order central differencing derivative of $\overline{\phi}_{i+n+1}$. In the above equation, the right-hand side only includes information from grid cells $i+n$ and $i+n+1$, successfully obtaining the average value of the far-field cells on the left-hand side of the equation using only first-order derivatives. Note that the central differencing gradient in Equation \eqref{eq:recursion} appears in pairs with the vector of the cell-center distances and can be expressed in a differential form. We define a representation called the DOLINC differential as follows:

\begin{equation}
  \begin{aligned}
    \varphi_{n}^{(0)} &= \varphi_{n} = \overline{\phi}_{i+n}, \\
    \varphi_{n}^{(m)} &= 2^{m} \overline{\phi}_{i+n,\,j_1 j_2 \cdots j_m} d_{j_1} d_{j_2} \cdots d_{j_m}, \quad m = 1, 2, 3, \dots
  \end{aligned}
\end{equation}

Equation \eqref{eq:recursion} can be further extended to include a recursive relationship involving high-order differential information, as follows:

\begin{align}
  \varphi_{n+2}^{(m)} = g(\varphi_{n}^{(m)} + g^{n} \varphi_{n+1}^{(m+1)}) + (1-g)\varphi_{n+1}^{(m)}.
\end{align}

Similarly, for the mesh on the side of $n = 0, -1, -2, \dots$, we obtain

\begin{align}
  \varphi_{n-1}^{(m)} = g^{-1}(\varphi_{n+1}^{(m)} - g^{n} \varphi_{n}^{(m+1)}) + (1-g^{-1})\varphi_{n}^{(m)}.
\end{align}

These two recursive expressions can ultimately be unified into a universal DOLINC differentiation relationship:

\begin{align}
\label{eq:diffDLC}
  \varphi_{n+1}^{(m)} - g \varphi_{n-1}^{(m)} = g^{n} \varphi_{n}^{(m+1)} + (1-g)\varphi_{n}^{(m)}, \quad m = \dots, -2, -1, 0, 1, 2, 3, \dots
\end{align}

Through multiple recursive applications of Equation \eqref{eq:diffDLC}, any far-field cell information can be transformed into the differential information of various-order $\varphi_0^{(m)}$ and $\varphi_1^{(m)}$ on the mesh face of two adjacent grid cells. This transformation can be used for constructing high-order schemes. In terms of DOFs, the originally required 0th-order wide-range DOFs are transformed into local high-order ones, and the number of DOFs remains unchanged. In terms of accuracy, by neglecting floating-point errors in the calculation, the calculated $\varphi_{n-1}^{(m)}$ and $\varphi_{n+1}^{(m)}$ on the left-hand side of Equation \eqref{eq:diffDLC} are consistent with the original values $\varphi_{n-1}^{(m)}$ and $\varphi_{n+1}^{(m)}$ stored on the far-field mesh cells involved in the computations of $\varphi_{n}^{(m+1)}$ on the right-hand side. Notably, the significance of DOLINC differentials is in providing a medium or an intermediate variable for retrieving far-field cell information, and they do not require the approximation of the exact values of the actual function differentials. In contrast, using more accurate gradient-calculation methods, such as the least-squares approximation, introduces errors to the right-hand side of Equation \eqref{eq:diffDLC}, leading to a deviation of the recalculated far-field data from the initial values. Additionally, using the exact average values of the function gradient in the calculation formula still results in a decrease in the accuracy of high-order schemes. This is the reason we refer to this process as an inversion rather than a gradient approximation: the accuracy of the result depends on whether the inversion formula is strictly valid, not on how accurate the estimated gradient used is. In terms of storage, DOLINC differentials and the physical quantity $\phi$ to be solved belong to the same-order tensors, requiring the same data length, whereas the tensor order increases continuously when calculating gradients of $\phi$. For a $D$-dimensional problem, the data storage required for calculating the Nth-order DOLINC differentials is N+1 times that of the original, and the data-storage required for calculating Nth-order gradients is $\tfrac{D^{N+1} - 1}{D - 1}$ times that of the original.

When the required data for the corresponding high-order scheme do not exceed the split four-grid-cell template, only the storage of first-order DOLINC differentials is necessary. The inversion formula is as follows:

\begin{align}
  \varphi_{-1} &= g^{-1}(\varphi_{1} - \varphi_{0}^{(1)}) + (1-g^{-1})\varphi_{0} = g^{-1} \varphi_{1}^{(0)} + (1-g^{-1})\varphi_{0}^{(0)} - g^{-1} \varphi_{0}^{(1)},  \\
  \varphi_{2} &= g(\varphi_{0} + \varphi_{1}^{(1)}) + (1-g)\varphi_{1} = g \varphi_{0}^{(0)} + (1-g)\varphi_{1}^{(0)} + g \varphi_{1}^{(1)}.
\end{align}

At this point, the split-mesh template $\{\varphi_{-1},\varphi_{0},\varphi_{1},\varphi_{2}\}$ is transformed into a DOLINC template $\{\varphi_{0}^{(0)},\varphi_{0}^{(1)},\varphi_{1}^{(0)},\varphi_{1}^{(1)}\}$ with the same number of DOFs. This type of high-order scheme, which only requires the use of first-order DOLINC differentials, is referred to as a first-order DOLINC scheme (the order of the DOLINC scheme is different from the convergence order of the high-order scheme; for example, a fourth-order accurate fixed-template reconstruction scheme belongs to the first-order DOLINC scheme). To achieve a form similar to conventional unstructured-mesh finite-volume schemes, the inversion formula for the far-field cells in the DOLINC scheme is organized into the following general form:

\begin{align}
\label{eq:formDLC}
  \overline{\phi}_{j} = a_{0}^{j-i} \overline{\phi}_{(i)} + a_{1}^{j-i} \overline{\phi}_{(i+1)} + A_{\delta}^{j-i}, \quad j = \dots, i-2, i-1, i+2, i+3, \dots
\end{align}

Thus, for first-order DOLINC schemes, we have

\begin{equation}
\label{eq:lCoeffDLC1}
  \begin{aligned}
    a_{0}^{-1} &= 1 - g^{-1},  \\
    a_{1}^{-1} &= 1 - a_{0}^{-1},  \\
    A_{\delta}^{-1} &= - g^{-1} \varphi_{0}^{(1)},
  \end{aligned}
\end{equation}

and

\begin{equation}
\label{eq:rCoeffDLC1}
  \begin{aligned}
    a_{0}^{2} &= g,  \\
    a_{1}^{2} &= 1 - a_{0}^{2},  \\
    A_{\delta}^{2} &= g \varphi_{1}^{(1)}.
  \end{aligned}
\end{equation}

When high-order reconstruction requires the use of second-order DOLINC differentials (second-order DOLINC scheme), such as in the case of a fixed central-template reconstruction with sixth-order accuracy, the additional inversion formula for the far-field cells is as follows:

\begin{align}
  \label{eq:UW2}
  \varphi_{-2} &= g^{-1} \varphi_{0}^{(0)} + (1-g^{-1})\varphi_{-1}^{(0)} - g^{-2} \varphi_{-1}^{(1)},  \\
  \label{eq:DW2}
  \varphi_{3} &= g \varphi_{1}^{(0)} + (1-g)\varphi_{2}^{(0)} + g^{2} \varphi_{2}^{(1)},
\end{align}

where

\begin{align}
  \label{eq:UW10}
  \varphi_{-1}^{(0)} &= g^{-1} \varphi_{1}^{(0)} + (1-g^{-1})\varphi_{0}^{(0)} - g^{-1} \varphi_{0}^{(1)},  \\
  \label{eq:UW11}
  \varphi_{-1}^{(1)} &= g^{-1} \varphi_{1}^{(1)} + (1-g^{-1})\varphi_{0}^{(1)} - g^{-1} \varphi_{0}^{(2)},  \\
  \label{eq:DW10}
  \varphi_{2}^{(0)} &= g \varphi_{0}^{(0)} + (1-g)\varphi_{1}^{(0)} + g \varphi_{1}^{(1)},  \\
  \label{eq:DW11}
  \varphi_{2}^{(1)} &= g \varphi_{0}^{(1)} + (1-g)\varphi_{1}^{(1)} + g \varphi_{1}^{(2)}.
\end{align}

Substituting Equations \eqref{eq:UW10}--\eqref{eq:DW11} into Equations \eqref{eq:UW2} and \eqref{eq:DW2}, the inversion formulas become

\begin{equation}
  \begin{aligned}
    \varphi_{-2} &= g^{-1}(1-g^{-1}) \varphi_{1}^{(0)} - g^{-3} \varphi_{1}^{(1)}  \\
    &+ (1 - g^{-1} + g^{-2}) \varphi_{0}^{(0)} - g^{-1}(1 - g^{-2}) \varphi_{0}^{(1)} + g^{-3} \varphi_{0}^{(2)},
  \end{aligned}
\end{equation}

and

\begin{equation}
    \begin{aligned}
        \varphi_{3} &= g(1-g) \varphi_{0}^{(0)} + g^3 \varphi_{0}^{(1)}  \\
        &+ (1 - g + g^2) \varphi_{1}^{(0)} + g(1 - g^2) \varphi_{1}^{(1)} + g^3 \varphi_{1}^{(2)}.
    \end{aligned}
\end{equation}

At this point, the split six-grid template $\{\varphi_{-2},\varphi_{-1},\varphi_{0},\varphi_{1},\varphi_{2},\varphi_{3}\}$ is transformed into a DOLINC template $\{\varphi_{0}^{(0)},\varphi_{0}^{(1)},\varphi_{0}^{(2)},\varphi_{1}^{(0)},\varphi_{1}^{(1)},\varphi_{1}^{(2)}\}$ with the same number of DOFs. By introducing the general form of Equation \eqref{eq:formDLC}, we obtain

\begin{equation}
\label{eq:lCoeffDLC2}
  \begin{aligned}
    a_{0}^{-2} &= (1 - g^{-1} + g^{-2}), \\
    a_{1}^{-2} &= 1 - a_{0}^{-2},  \\
    A_{\delta}^{-2} &= - g^{-3} \varphi_{1}^{(1)} - g^{-1}(1 - g^{-2}) \varphi_{0}^{(1)} + g^{-3} \varphi_{0}^{(2)},
  \end{aligned}
\end{equation}

and

\begin{equation}
\label{eq:rCoeffDLC2}
  \begin{aligned}
    a_{0}^{3} &= g - g^2,  \\
    a_{1}^{3} &= 1 - a_{0}^{3},  \\
    A_{\delta}^{3} &= g^3 \varphi_{0}^{(1)} + g(1 - g^2) \varphi_{1}^{(1)} + g^3 \varphi_{1}^{(2)}.
  \end{aligned}
\end{equation}

Following the inversion procedure for far-field elements using the aforementioned DOLINC method, the corresponding DOLINC version of high-order schemes can be constructed based on expressions for high-order reconstruction.

\subsection{Implementation of High-Order Schemes Based on DOLINC Method (DOLINC Schemes)}

Generally, if the index of the leftmost cell adopted by a reconstruction template is denoted as $i-r$, then the $k$th-order reconstruction method can be represented as

\begin{equation}
  \phi_{i+\frac{1}{2}}
    = p(x_{i+\frac12})
    = \sum_{j = 0}^{k-1} \; c_{rj}^{k} \; \overline{\phi}_{i-r+j},
\end{equation}

where

\begin{equation}
\label{eq:coeff}
  c_{rj}^k =
    \left[
      \sum_{m=j+1}^{k}
        \frac{
          \displaystyle\sum_{
            \substack{l=0 \\ l \neq m}
          }^{k}
          \displaystyle\prod_{
            \substack{q=0 \\ q \neq m,l}
          }^{k}
          \Big( x_{i+\frac12} - x_{i-r+q-\frac12} \Big)
        }{
          \displaystyle\prod_{
            \substack{l=0 \\ l \neq m}
          }^{k}
          \Big(
            x_{i-r+m-\frac12} - x_{i-r+l-\frac12}
          \Big)
        }
    \right]
    \Delta x_{i-r+j}.
\end{equation}

Detailed derivations are available in the literature \citep{Shu:1998}. For simplicity, the superscripts can be omitted when the order of reconstruction is clear, and the coefficient is represented by $c_{rj}$.

Using the third-order reconstruction of $k=3,r=1$ as an example, a corresponding DOLINC scheme was constructed. In this case, considering the non-uniformity of the grid, we can derive through Equation \eqref{eq:coeff} that

\begin{align}
  c_{10} &= \frac{-g^3}{G_2 G_3},  \\
  c_{11} &= \frac{g (1 + 2g + 2g^2)}{G_2 G_3},  \\
  c_{12} &= \frac{1}{G_3}.
\end{align}

In the above expression, $G_n$ is employed to simplify the coefficient representation, and its specific definition is given by

\begin{equation}
    \begin{aligned}
        G_n &= 1 + g + g^2 + g^3 + \cdots + g^{n-1}  \\
            &= \frac{g^n - 1}{g - 1}, \quad n = 1, 2, 3, \dots
    \end{aligned}
\end{equation}

Noticing that when $g = 1$, i.e., the grid size remains unchanged, $G_n = n$, the coefficients degenerate to the constants -1/6, 5/6, and 1/3. This non-uniform mesh third-order reconstruction employs a fixed template one-cell biased towards the upwind direction. Hereafter, we denote this reconstruction scheme as the third-order upwind stencil reconstruction (USR3). The basic representation of USR3 can be organized into the weight-representation method commonly used in lower-order FVM using Equation \eqref{eq:lCoeffDLC1}:

\begin{equation}
\label{eq:USR3}
  \begin{aligned}
    \phi_{i+\frac12}
      &= c_{10} \overline{\phi}_{i-1} + c_{11} \overline{\phi}_{i} + c_{12} \overline{\phi}_{i+1}  \\
      &= (c_{10} a_{0}^{-1} + c_{11}) \overline{\phi}_{i} + (c_{10} a_{1}^{-1} + c_{12}) \overline{\phi}_{i+1} + c_{10} A_{\delta}^{-1}  \\
      &= w_s \overline{\phi}_{i} + (1-w_s) \overline{\phi}_{i+1} + c_{r},
  \end{aligned}
\end{equation}

where

\begin{align}
  w_s &= c_{10} a_{0}^{-1} + c_{11},  \\
  c_r &= c_{10} A_{\delta}^{-1}.
\end{align}

This representation style, derived for high-order schemes through the DOLINC approach, is referred to as the DOLINC representation. In Equation \eqref{eq:USR3}, the DOLINC representation of USR3 is consistent with the logical form commonly used in lower-order FVM codes for face-interpolation schemes, requiring no additional adjustments to its architecture. In practical implementation, the USR3 scheme can be realized through an embedding approach similar to the second-order upwind or TVD schemes. If time advancement is explicitly conducted in solving the PDEs \citep{Guo:2023:arXiv}, the DOLINC-implemented USR3 is entirely equivalent to the classical third-order accurate reconstruction. When employing an implicit time-integration scheme for PDEs, as deduced above, the DOLINC method essentially provides a modified semi-implicit third-order scheme that contributes to the stability of the matrix to be solved.

Owing to the lack of advanced high-order schemes, industrial software typically recommends the use of a second-order central differencing scheme (CD2) to mitigate the impact of numerical dissipation on the results of turbulence LESs or DNSs \citep{FOAM}. Therefore, we consider the example of a fourth-order central stencil reconstruction (CSR4) scheme to provide the corresponding DOLINC representation; it shares similarities with CD2, but offers higher accuracy. In this case, $k=4,r=1$, and through derivation, we obtain

\begin{equation}
\label{eq:CSR4}
  \begin{aligned}
    \phi_{i+\frac12}
      &= c_{10} \overline{\phi}_{i-1} + c_{11} \overline{\phi}_{i} + c_{12} \overline{\phi}_{i+1} + c_{13} \overline{\phi}_{i+2}  \\
      &= w_s \overline{\phi}_{i} + (1-w_s) \overline{\phi}_{i+1} + c_{r},
  \end{aligned}
\end{equation}

where

\begin{align}
  w_s &= c_{10} a_{0}^{-1} + c_{11} + c_{13} a_{0}^{2},  \\
  c_r &= c_{10} A_{\delta}^{-1} + c_{13} A_{\delta}^{2}.
\end{align}

In the above equation, the basic representation of CSR4 also considers the non-uniformity of the mesh for each reconstruction coefficient, and the calculation formula is as follows:

\begin{align}
  c_{10} &= \frac{-g^5}{G_3 G_4},  \\
  c_{11} &= \frac{g^2 (1 + 2g + 2g^2 + 2g^3)}{G_3 G_4},  \\
  c_{12} &= \frac{2 + 2g + 2g^2 + g^3}{G_3 G_4},  \\
  c_{13} &= \frac{-1}{G_3 G_4}.
\end{align}

Similarly, when $g=1$, the above coefficients degenerate to the constants -1/12, 7/12, 7/12, and -1/12.

USR3 and CSR4 in the above derivation process only involve the DOLINC inversion coefficients from Equations \eqref{eq:lCoeffDLC1} and \eqref{eq:rCoeffDLC1}, making them first-order DOLINC schemes. Second-order DOLINC schemes such as USR5 or CSR6 would require the use of DOLINC inversion coefficients from Equations \eqref{eq:lCoeffDLC2} and \eqref{eq:rCoeffDLC2}. Based on the DOLINC fixed-stencil schemes, achieving DOLINC versions of the ENO and WENO methods only requires supplementing stencil-smoothness indicators or a stencil-weight-calculation approach. Similarly, the calculation of indicators or weights must consider local variations in grid sizes. For example, in the third-order accurate ENO method (ENO3) using the classical stencil-selection approach \citep{Harten:1987}, the smoothness indicator for the required split two-grid stencil is

\begin{align}
  &
  \begin{aligned}
    \Delta_0^2 \varPhi[x_{i-\frac32}, x_{i-\frac12}, x_{i+\frac12}]
      &= g (\overline{\phi}_{i} - \overline{\phi}_{i-1})  \\
      &= \overline{\phi}_i - \overline{\phi}_{i+1} + \varphi_{0}^{(1)},
  \end{aligned}  \\
  &\Delta_0^2 \varPhi[x_{i-\frac12}, x_{i+\frac12}, x_{i+\frac32}] = \overline{\phi}_{i+1} - \overline{\phi}_{i}.
\end{align}

Similarly, the smoothness indicator for the split three-grid stencil is obtained as follows:

\begin{align}
  \Delta_{-1}^3 \Delta_{0}^2 \varPhi[x_{i-\frac52}, x_{i-\frac32}, x_{i-\frac12}, x_{i+\frac12}] &=
    g^2 [\overline{\phi}_{i} - (1+g) \overline{\phi}_{i-1} + g \overline{\phi}_{i-2}],  \\
  \Delta_{-1}^3 \Delta_{0}^2 \varPhi[x_{i-\frac32}, x_{i-\frac12}, x_{i+\frac12}, x_{i+\frac32}] &=
    \overline{\phi}_{i+1} - (1+g) \overline{\phi}_{i} + g \overline{\phi}_{i-1},  \\
  \Delta_{-1}^3 \Delta_{0}^2 \varPhi[x_{i-\frac12}, x_{i+\frac12}, x_{i+\frac32}, x_{i+\frac52}] &=
    g^{-2}[\overline{\phi}_{i+2} - (1+g) \overline{\phi}_{i+1} + g \overline{\phi}_{i}].
\end{align}

The indicator calculation formulas above require the use of the corresponding DOLINC inversion formulas. Similarly, by incorporating stencil-weight calculations, WENO reconstruction methods of various orders of accuracy can be obtained. ENO3-DOLINC and WENO5-DOLINC (hereafter distinguished by this representation to denote the same high-order scheme implemented using different methods) involve multiple alternative stencils; thus, they are classified as second-order DOLINC schemes.

\subsection{Error Analysis of Uniform-Mesh-Based Least-Squares Gradient-Approximation Method}
\label{ssec:error}

Based on the previous derivation, the DOLINC scheme can achieve a theoretical convergence accuracy for the corresponding reconstruction. Two key points are notable in this regard: first, both the inversion formula and scheme coefficients in the method must consider local mesh growth, and second, the DOLINC differentials must be rigorously calculated. Owing to its inability to ensure these two points, the uniform-mesh least-squares gradient-approximation method (hereinafter referred to as the UG method) may exhibit uncontrollable accuracy degradation in the finite-volume scheme obtained. The next step involves an error analysis of the UG method through mathematical derivation.
First, we consider the influence of non-uniformity. Let the reconstructed values obtained using the DOLINC method be denoted as $\phi_{i+\frac{1}{2}}^{DLC}$. For a $k$th-order reconstruction, we have

\begin{equation}
\label{eq:foam1DLC}
  \phi_{i+\frac{1}{2}}^{DLC}
    = \sum_{j = 0}^{k-1} \; c_{rj}^{DLC} \; \overline{\phi}_{i-r+j}.
\end{equation}

For the non-uniform grid exhibited in Figure \ref{fig:schmtc:stencil}, we have

\begin{equation}
  \phi_{i+\frac12} = \phi_{i+\frac12}^{DLC} + O(\Delta x^k),
\end{equation}

where $\phi_{i+\frac12}$ represents the exact function values. Similarly, if we follow the UG method and assume a uniform mesh to calculate various coefficients, a reconstruction method can be obtained (temporarily assuming no errors in the estimation of the phantom points in the UG method). The reconstruction results obtained using this method are denoted as $\phi_{i+\frac{1}{2}}^{UG}$. At this point, the UG scheme can also be expressed as

\begin{equation}
\label{eq:form1UG}
  \phi_{i+\frac{1}{2}}^{UG}
    = \sum_{j = 0}^{k-1} \; c_{rj}^{UG} \; \overline{\phi}_{i-r+j}.
\end{equation}

This non-DOLINC method induces the error

\begin{equation}
  \phi_{i+\frac{1}{2}}^{UG} - \phi_{i+\frac{1}{2}}^{DLC}
    = \sum_{j = 0}^{k-1} \; \Big(c_{rj}^{UG} - c_{rj}^{DLC}\Big) \; \overline{\phi}_{i-r+j}.
\end{equation}

Let $\overline{\phi}_{i-r+j} = \phi_{0} + \delta\phi_{i-r+j}$, where $\phi_0$ is an arbitrary reference value that can be considered the exact value $\phi_{i+\frac12}$. Notice that

\begin{equation}
  \sum_{j = 0}^{k-1} \Big( c_{rj}^{UG} - c_{rj}^{DLC} \Big) = 0,
\end{equation}

then, we have

\begin{equation}
\label{eq:error1}
  \begin{aligned}
    \phi_{i+\frac{1}{2}}^{UG} - \phi_{i+\frac{1}{2}}^{DLC}
      &= \sum_{j = 0}^{k-1} \; \Big(c_{rj}^{UG} - c_{rj}^{DLC}\Big) \; \Big( \phi_{0} + \delta\phi_{i-r+j} \Big),  \\
      &= \sum_{j = 0}^{k-1} \; \Big(c_{rj}^{UG} - c_{rj}^{DLC}\Big) \; \delta\phi_{i-r+j}.
  \end{aligned}
\end{equation}

Further examination of the above errors shows that the first component of the error is

\begin{equation}
\label{eq:errC}
  \epsilon_{c} = c_{rj}^{UG} - c_{rj}^{DLC} = f(g) \sim O(1),
\end{equation}

which is solely a function of the mesh growth rate $g$. Therefore, even as the grid size decreases continuously, $\epsilon_c$ will remain constant as long as non-uniformity exists in the grid. This implies that the magnitude of $\epsilon_c$ in the UG method is entirely determined by the growth rate. When the UG method is applied to a uniform grid, this error term disappears; however, when local grid sizes significantly vary, $\epsilon_c$ cannot be reduced in another manner. For the second component of the error,

\begin{equation}
\label{eq:error12}
  \epsilon_{\phi} = \delta\phi_{i-r+j} \sim O(\Delta x),
\end{equation}

is why mesh refinement can improve non-DOLINC schemes in practical applications. As the mesh gradually refines, the physical range of the local template decreases, leading to a reduction in the local variation of the solution within the template. However, as indicated by Equation \eqref{eq:error12}, the convergence accuracy of this error term is only first-order. Introducing a first-order error term directly into a $k$th-order scheme results in uncontrollable accuracy degradation of the reconstruction \citep{Zhong:2020}.

In the above analysis, Equation \eqref{eq:form1UG} directly utilizes $\overline{\phi}_{i-r+j}$; however, in the UG method, $\overline{\phi}_{i-r+j}$ cannot be correctly inverted because of the fundamentally incorrect inversion caused by the use of non-DOLINC differential calculations. By employing the representation of the DOLINC scheme in Equation \eqref{eq:CSR4}, an expression equivalent to Equation \eqref{eq:foam1DLC} can be obtained:

\begin{equation}
  \phi_{i+\frac{1}{2}}^{DLC} = w_s^{DLC} \overline{\phi}_{i} + (1-w_s^{DLC}) \overline{\phi}_{i+1} + c_{r}^{DLC},
\end{equation}

Similarly, the reconstructed value through the UG method can be transformed into

\begin{equation}
  \phi_{i+\frac{1}{2}}^{UG} = w_s^{UG} \overline{\phi}_{i} + (1-w_s^{UG}) \overline{\phi}_{i+1} + c_{r}^{UG}.
\end{equation}

Then, we have

\begin{align}
  \phi_{i+\frac{1}{2}}^{UG} - \phi_{i+\frac{1}{2}}^{DLC}
    &= \Big( w_s^{UG} - w_s^{DLC} \Big) \Big( \overline{\phi}_{i} - \overline{\phi}_{i+1} \Big) + \Big( c_{r}^{UG} - c_r^{DLC} \Big).
\end{align}

Similarly,

\begin{equation}
  \epsilon_{w} = w_s^{UG} - w_s^{DLC} = f(g) \sim O(1),
\end{equation}

Apart from the non-uniform errors in the reconstruction coefficients, as described in Equation \eqref{eq:errC}, an additional contribution of $\epsilon_{w}$ exists from neglecting the non-uniformity in the calculation of phantom points in the UG method. Similarly, at this point,

\begin{equation}
  \epsilon_{\phi} = \overline{\phi}_{i} - \overline{\phi}_{i+1} \sim O(\Delta x).
\end{equation}

However, unlike the previous derivation, an additional term is introduced into the error at this point:

\begin{equation}
  \epsilon_{\varphi} = c_{r}^{UG} - c_r^{DLC} = f(g, \varphi_0^{(1)}, \varphi_0^{(2)}, \cdots, \varphi_1^{(1)}, \varphi_1^{(2)}, \cdots).
\end{equation}

From the above equation, regardless of the accuracy of the gradient calculation in $c_r^{UG}$, introducing a new error term is unavoidable as long as it does not adopt DOLINC differentials. When the UG method employs the least-squares approximation to compute the gradient, it may seem to obtain more accurate gradient data on the surface. However, in reality, this leads to a larger discrepancy between the calculated phantom point data and original field data, causing further uncontrollable accuracy degradation in the reconstruction scheme. Therefore, we emphasize once again that, as long as all information of the far-field element can be accurately inverted, the accuracy of the derivatives/gradients/differentials involved in the inversion formula is irrelevant. This is the core idea that distinguishes the DOLINC method from general gradient-approximation methods.

\subsection{Boundary Handling of DOLINC Method}

A good algorithm should be efficient, accurate, and stable, and the boundary handling required for the DOLINC scheme should exhibit these characteristics. For a low-order FVM framework to implement DOLINC schemes, no special treatment of the boundary mesh cells or modifications are required for the existing boundary-condition codes. The DOLINC scheme can be applied directly without any adjustments.

To handle coupled boundaries, such as processor boundaries or periodic boundaries, the key to implementation lies in ensuring accurate and efficient data exchange between the two sides of the interface. In the high-order schemes implemented by the DOLINC method, coupled boundaries still only require the communication of data between adjacent grids on both sides of the boundary, without considering far-field elements. Compared to the low-order finite-volume scheme, the only difference in the DOLINC scheme is the addition of DOLINC differentials to the exchanged data.

For non-coupled boundaries, such as the Dirichlet or Neumann boundaries, the handling approach of the low-order FVM still ensures the correct implementation of the DOLINC scheme at the boundary. In this case, the DOLINC differentials use the boundary values directly, and the inversion calculation yields auxiliary cells located on the other side of the boundary. The relationship between the auxiliary and boundary-adjacent cells strictly adheres to the boundary conditions. For example, for a given outward normal gradient q on a Neumann boundary, assuming the distance from the body center of the boundary cell to the center of the boundary face is $d_{fC}$, and the value of the boundary cell is $T_C$, the interpolated data on the boundary face is obtained as $T_f = T_C + q\,d_{fC}$. With known values of $T_f$ and $T_C$, the inversion formula yields the value of the auxiliary element on the other side as $T_O = T_C + 2(T_f - T_C)$. At this point, the outward normal gradient on the boundary can be recalculated using $T_O$ and $T_C$ as follows:

\begin{equation}
  \frac{\partial T}{\partial n_f}
= \frac{T_O - T_C}{2 d_{fC}}
= \frac{T_f - T_C}{d_{fC}}
= q.
\end{equation}

Therefore, in the DOLINC method, the Neumann boundary remains unchanged. Similarly, at the Dirichlet boundary, $T_f$ obtained through the linear interpolation of $T_O$ and $T_C$ remains constant. Overall, the DOLINC method does not require the treatment of non-coupled boundaries to ensure its computational accuracy and efficiency.

\section{Numerical Results and Discussion}
\label{sec:results}

This section demonstrates the performance of the DOLINC method when applied to numerical solutions of different problems. Using the open-source C++ library, OpenFOAM, we successfully integrated the DOLINC method into a low-order FVM framework. We compared the computational results of the DOLINC scheme with those of classical finite-volume schemes such as second-order central differencing, second-order upwind, and TVD limiters. The various solvers used in this section employed either a fourth-order classical Runge--Kutta scheme or third-order TVD Runge--Kutta scheme to minimize the impact of temporal discretization on the final results. Details of the implementation are available in the literature \citep{Guo:2023:arXiv}. The HPC clusters used for the validation cases consists of multiple computational nodes. Each node comprised two Intel$^{\circledR}$ Xeon$^{\circledR}$ Platinum 9242 processors, totaling 96 physical cores. The maximum memory capacity of each node was 384 GB. Additional HPC settings for the numerical examples are available in the literature \citep{Guo:2023:JFM}.

\subsection{Boundary Layer Flows on Non-Uniform Grids}
\label{ssec:BL}

\begin{figure}
	\centering
  \fbox{
    \includegraphics[width=10cm]{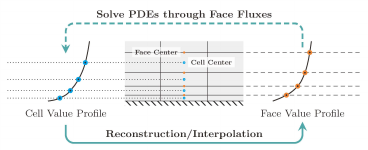}
  }
	\caption{Typical boundary layer mesh reconstruction/interpolation and PDE-solution iterations.}
	\label{fig:schmtc:bLayerMesh}
\end{figure}

As mentioned in Section \ref{ssec:error}, the DOLINC reconstruction scheme exhibits higher accuracy than non-DOLINC reconstruction schemes, especially in non-uniform grids with local growth. To verify this, we considered the reconstruction process in the context of a typical boundary-layer-flow problem on a non-uniform mesh. Figure \ref{fig:schmtc:bLayerMesh} illustrates the solution iterations on a typical boundary-layer mesh. In the reconstruction phase, the solver calculated face values based on the cell averages obtained from the previous iteration or time step. The reconstructed face values in the form of the scheme influenced the update of the cell values at the current iteration or time step in the PDE-solving algorithm. The updated cell values were then used for reconstruction in the next iteration or timestep.

Two points must be noted regarding this process. First, the errors introduced by the reconstruction scheme are repeatedly incorporated into the solution process and do not gradually disappear during the computation process. In fact, when solving problems that are sensitive to initial values, single-step errors may accumulate and increase over time, eventually causing the computation to diverge and preventing it from continuing. Second, the computed results of the PDE are affected not only by the reconstruction scheme, but also by various other factors in the solving algorithm. Therefore, reconstruction errors in the results may be evident or concealed in different test cases. Consequently, if a strict analysis of the difference in accuracy of the reconstruction scheme is required, specific solving algorithms should not be involved.

\begin{figure}
	\centering
  \fbox{
    \parbox[b]{16cm}{
      \includegraphics[width=7.9cm]{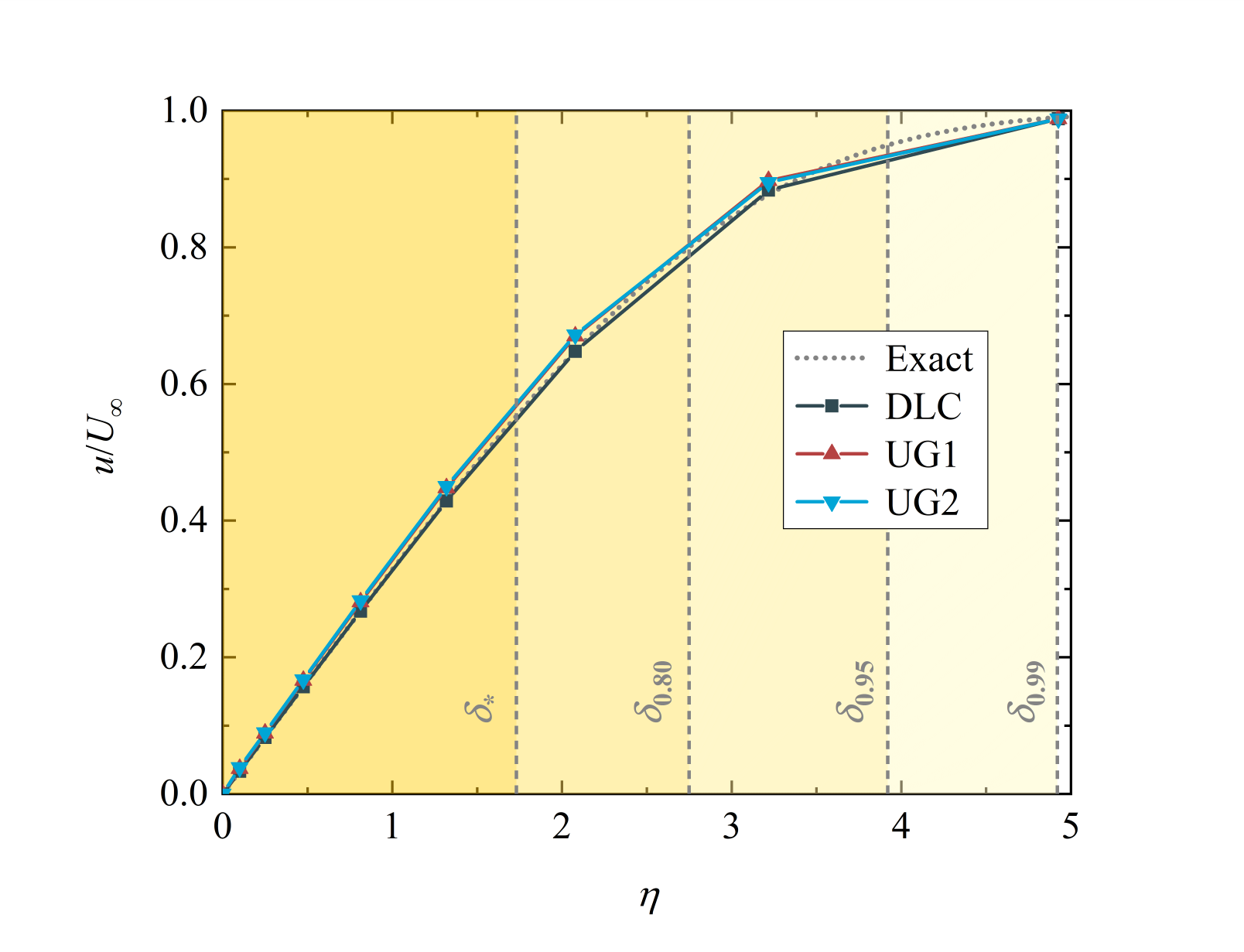}
      \includegraphics[width=7.9cm]{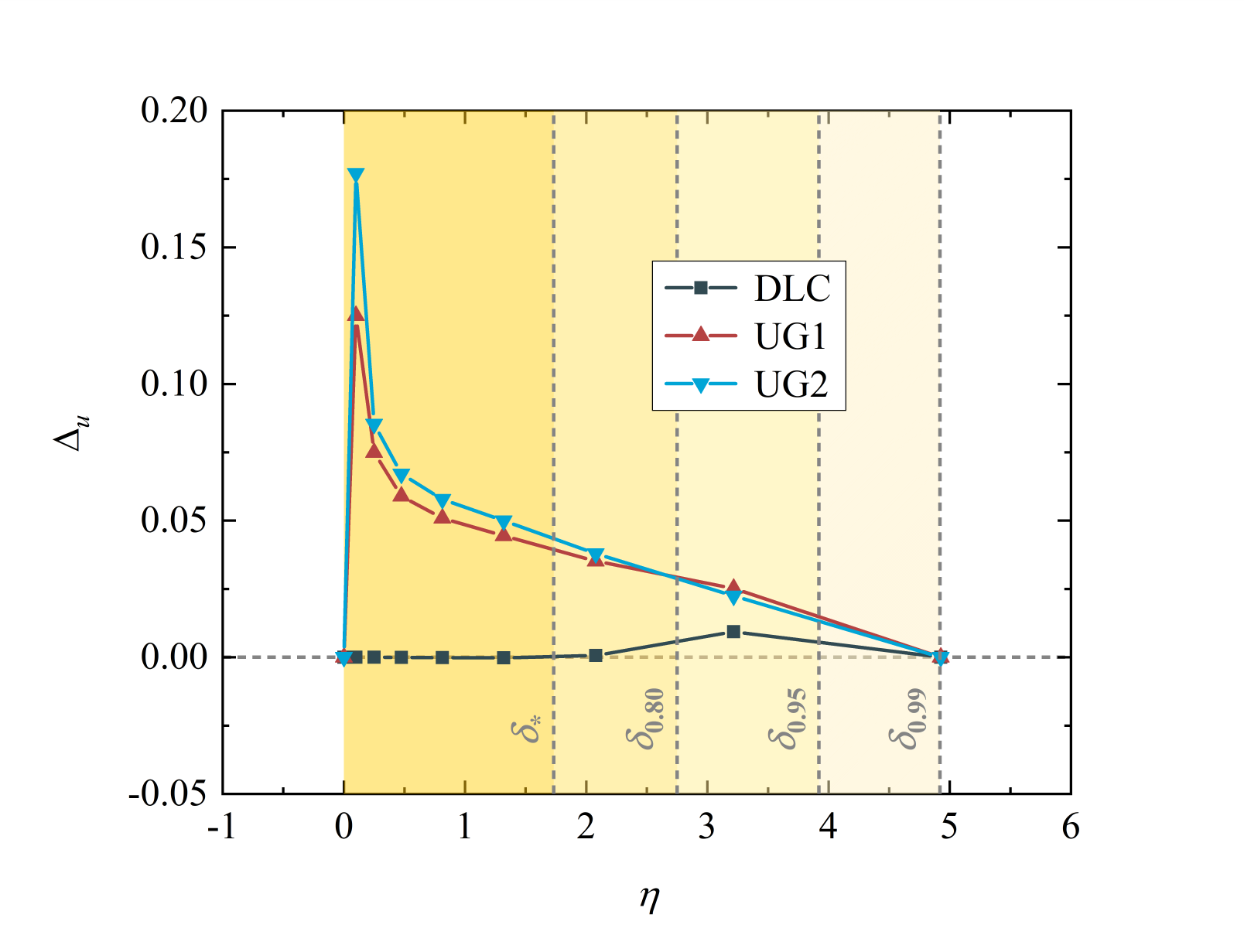}
      \par
      \centering \; (\textit{a}) \hspace{7.4cm} (\textit{b})
    }
  }
	\caption{Streamwise velocity component of mesh face obtained through single-step reconstruction and the exact solution. (\textit{a}) Absolute values of dimensionless velocity, (\textit{b}) relative error of dimensionless velocity compared to the exact solution.}
	\label{fig:BL:U}
\end{figure}

\begin{figure}
	\centering
  \fbox{
    \parbox[b]{16cm}{
      \includegraphics[width=7.9cm]{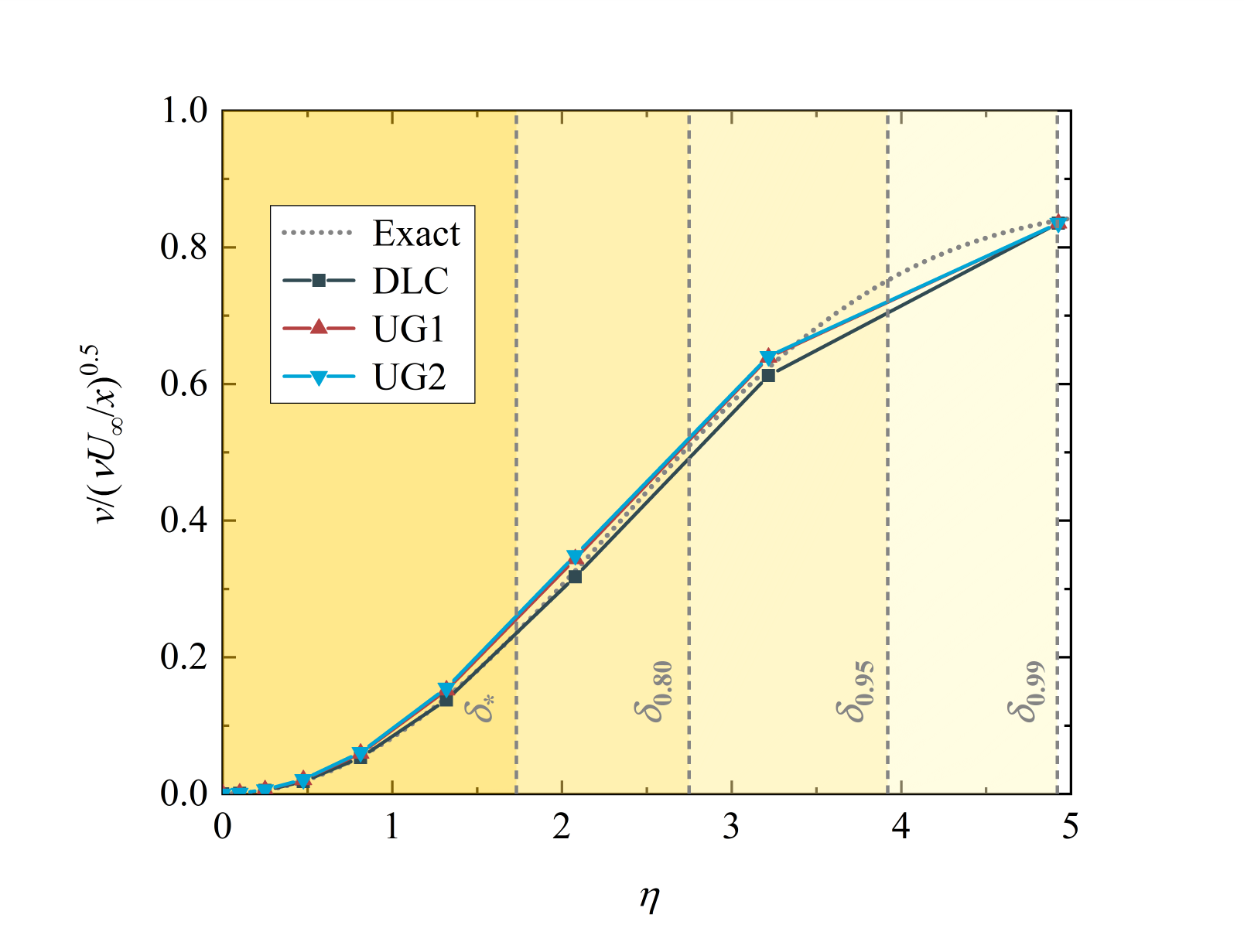}
      \includegraphics[width=7.9cm]{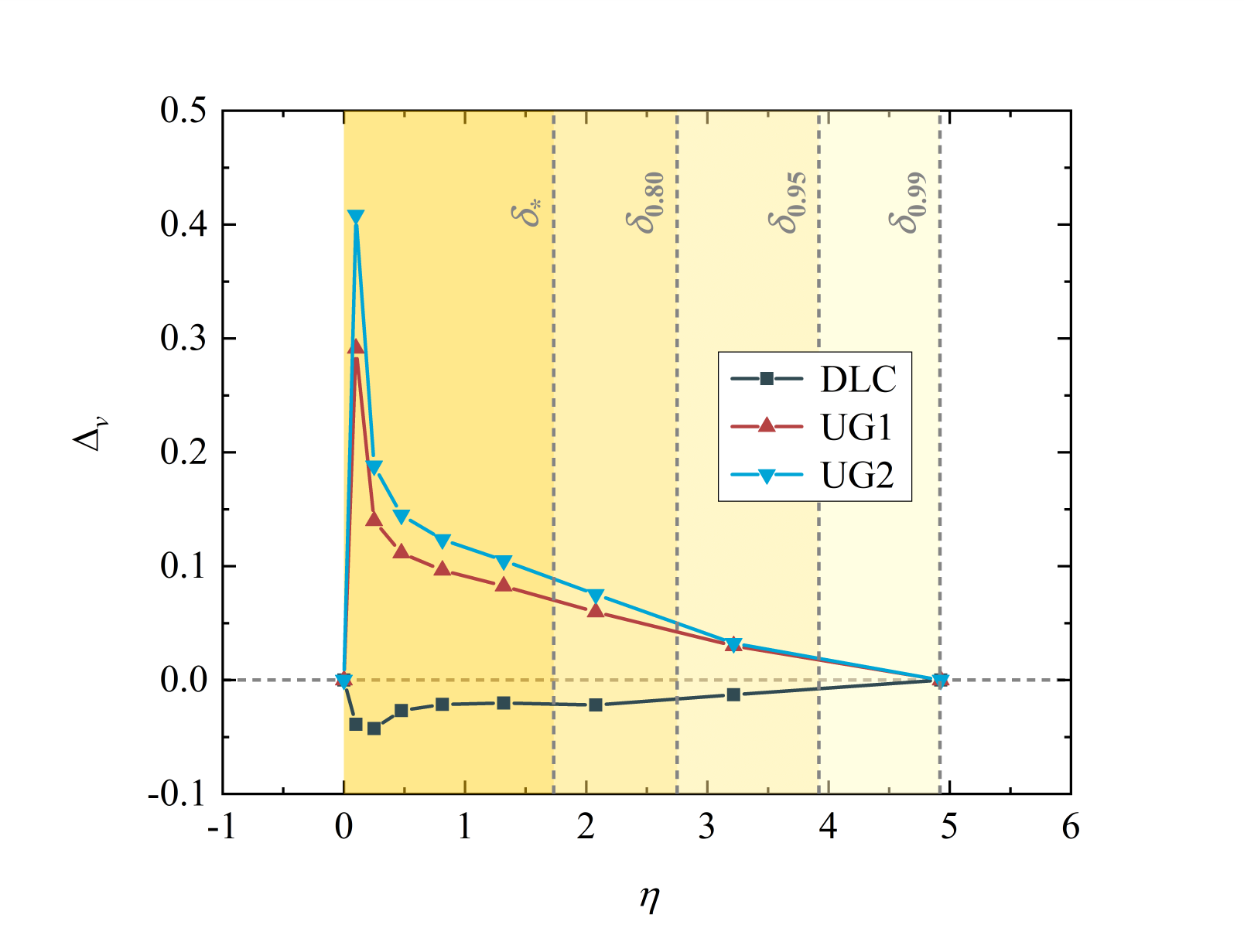}
      \par
      \centering \; (\textit{a}) \hspace{7.4cm} (\textit{b})
    }
  }
	\caption{Wall-normal velocity component of mesh face obtained through single-step reconstruction and the exact solution. (\textit{a}) Absolute values of dimensionless velocity, (\textit{b}) relative error of dimensionless velocity compared to the exact solution.}
	\label{fig:BL:V}
\end{figure}

Considering the aforementioned factors, DOLINC and non-DOLINC schemes were used only for single-step reconstruction. The data of the mesh cells on which the reconstruction was based directly adopted the exact solution of the laminar Blasius equation (obtained using a high-resolution ODE solver). Different reconstruction schemes were applied to obtain face values, which were then compared with the Blasius solution at the face-center positions. Figures \ref{fig:BL:U} and \ref{fig:BL:V} illustrate the reconstruction results on a boundary layer mesh with a vertical wall growth rate of 1.5 based on the USR3 scheme. Different methods were employed to implement the same high-order scheme: DLC represents USR3 implemented based on the DOLINC method, whereas UG1 and UG2 represent USR3 implemented based on the uniform-mesh-based gradient-approximation method. The difference between UG1 and UG2 is the specific gradient-calculation method, where the former considers the non-uniform characteristics of the mesh in the gradient calculation, resulting in more accurate gradient data than the latter.

Figures \ref{fig:BL:U} and \ref{fig:BL:V} show significant differences between the DLC and UG results, regardless of whether the velocity component is reconstructed parallel or perpendicular to the wall. In Figures \ref{fig:BL:U}(\textit{b}) and \ref{fig:BL:V}(\textit{b}), the relative errors of the velocity reconstructed by the DOLINC scheme are consistently low throughout the entire boundary layer, whereas both UG methods exhibit large reconstruction/interpolation errors in the strong shear region within the boundary layer. This accuracy difference in the single-step reconstruction accumulates as the computation progresses and ultimately has a significant impact on the computed results. Although the absolute values presented in the plots in Figures \ref{fig:BL:U}(\textit{a}) and \ref{fig:BL:V}(\textit{a}) may seem similar, the distribution pattern of the velocities within the boundary layer fundamentally changed because of the errors introduced by the UG method. Comparing the results of the two velocity components, the differences between the DOLINC and UG methods in the wall-normal direction were more pronounced, as shown in Figure \ref{fig:BL:V}. Therefore, if the computed boundary layer flow is still in a developing state, the uniform-mesh gradient-approximation method leads to larger errors, resulting in a less accurate distribution of the boundary layer velocities. Notably, UG1 and UG2, which used two different gradient-calculation methods, exhibited very small differences. The use of a more accurate gradient in UG1 did not effectively reduce the impact of the results produced by the UG method. In fact, at some locations, the relative error of UG1 surpassed that of UG2. Therefore, non-DOLINC methods cannot eliminate errors in non-uniform grids solely by improving gradient estimates; they can only marginally alter the reconstruction results and cannot guarantee that the altered results will always be more accurate.

\begin{figure}
	\centering
  \fbox{
    \parbox[b]{16cm}{
      \includegraphics[width=7.9cm]{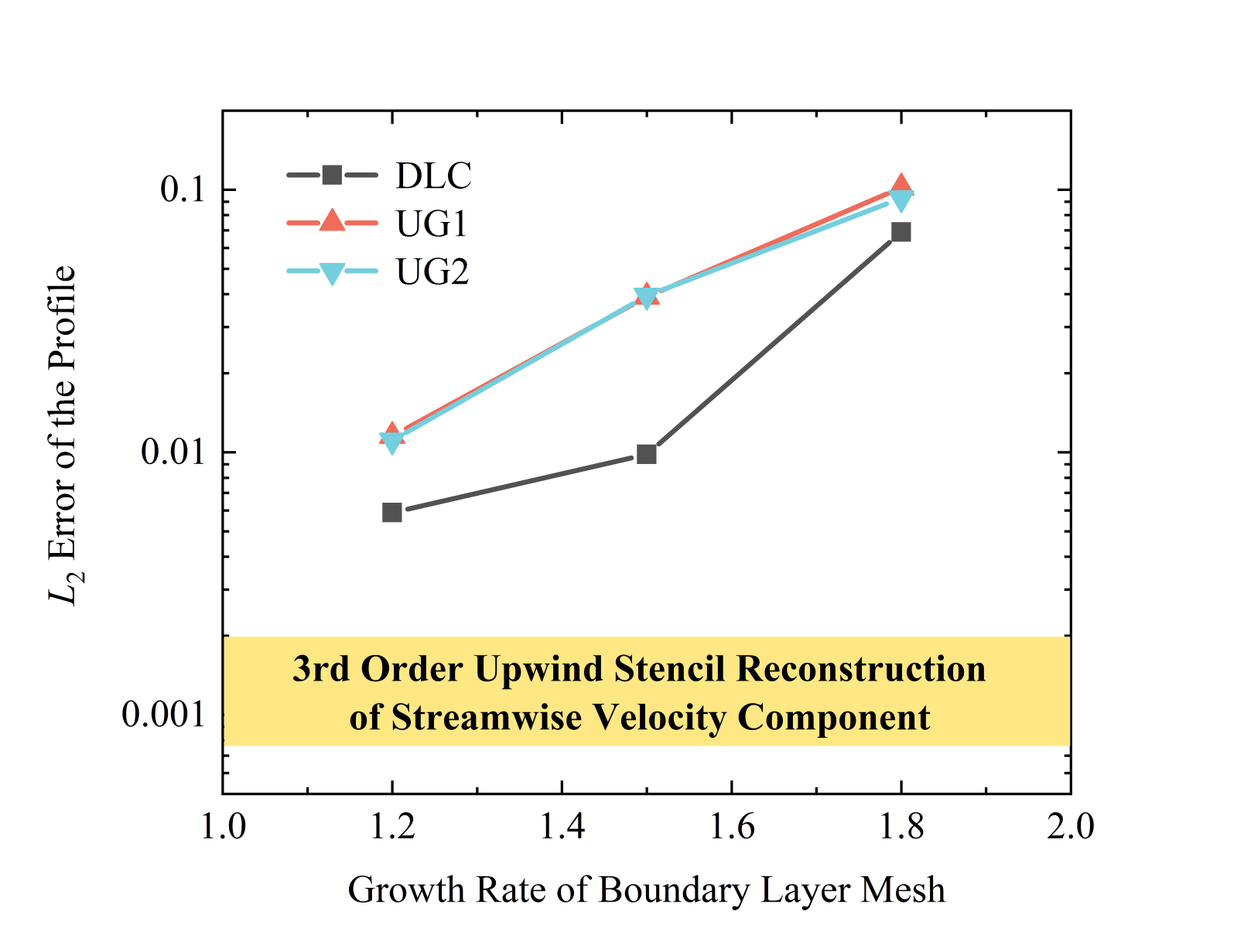}
      \includegraphics[width=7.9cm]{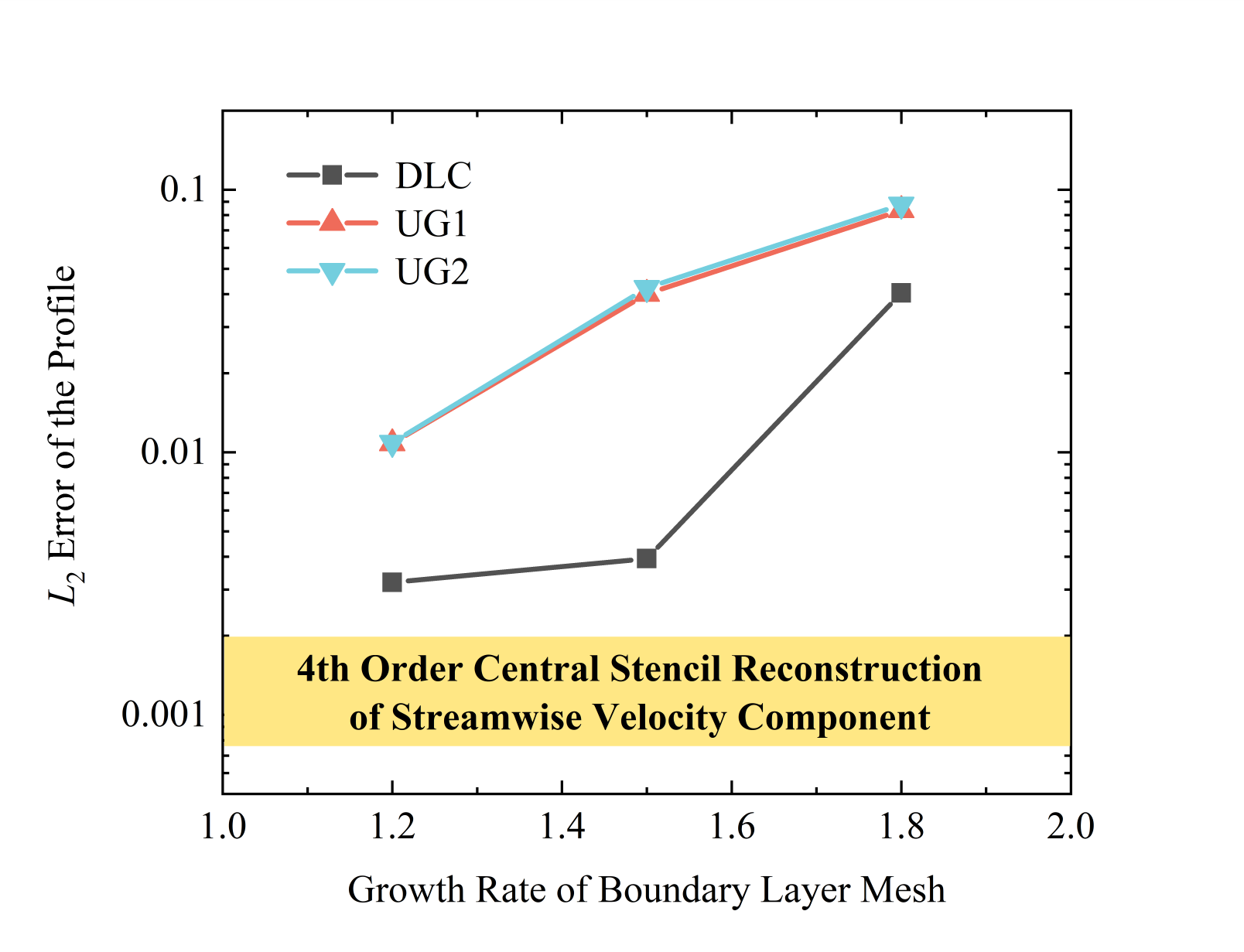}
      \par
      \centering \; (\textit{a}) \hspace{7.4cm} (\textit{b})
    }
  }
	\caption{Global errors of the streamwise velocity component with respect to mesh growth rate. (\textit{a}) Third-order fixed-stencil reconstruction, (\textit{b}) fourth-order fixed-stencil reconstruction.}
	\label{fig:BL:gErrU}
\end{figure}

\begin{figure}
	\centering
  \fbox{
    \parbox[b]{16cm}{
      \includegraphics[width=8cm]{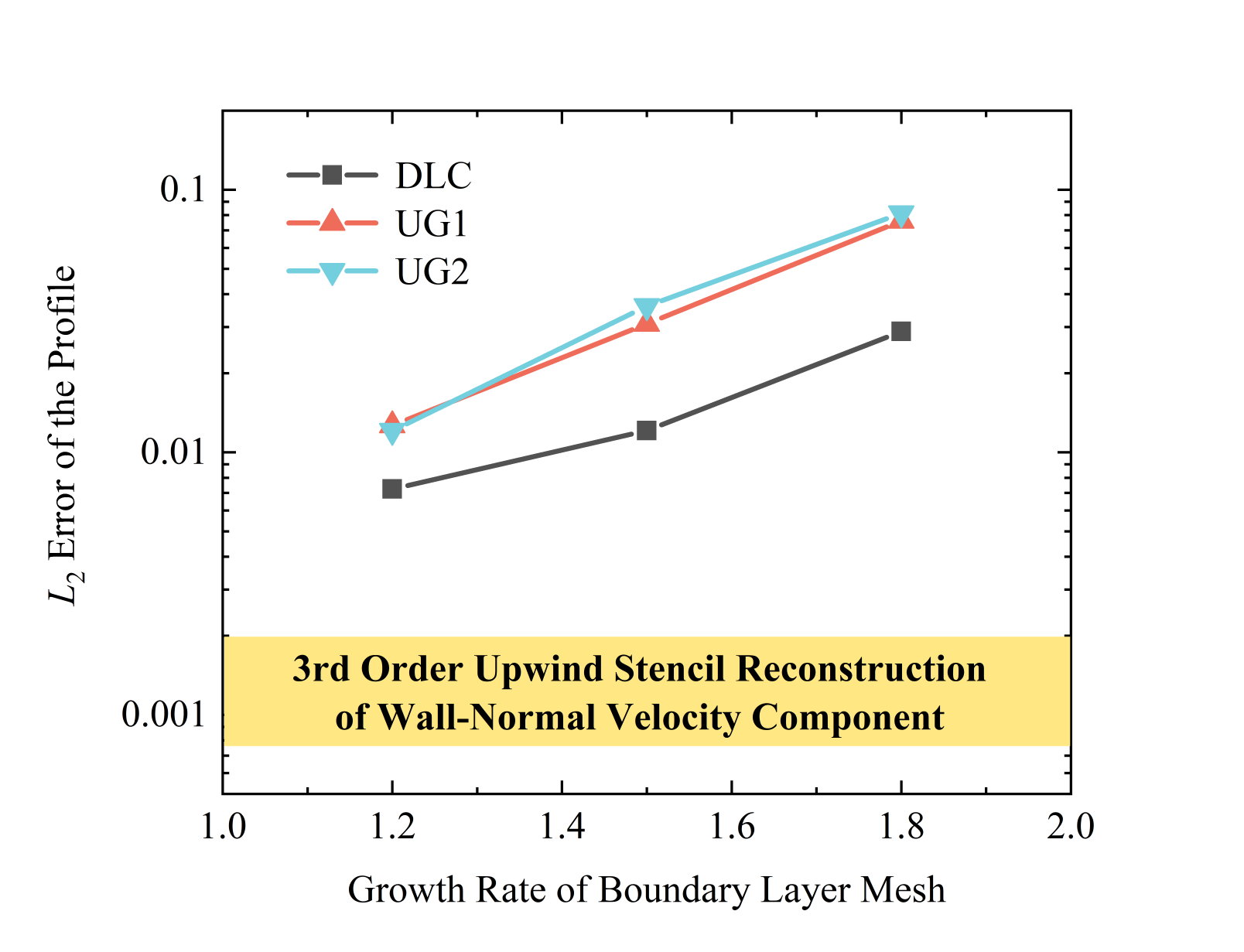}
      \includegraphics[width=8cm]{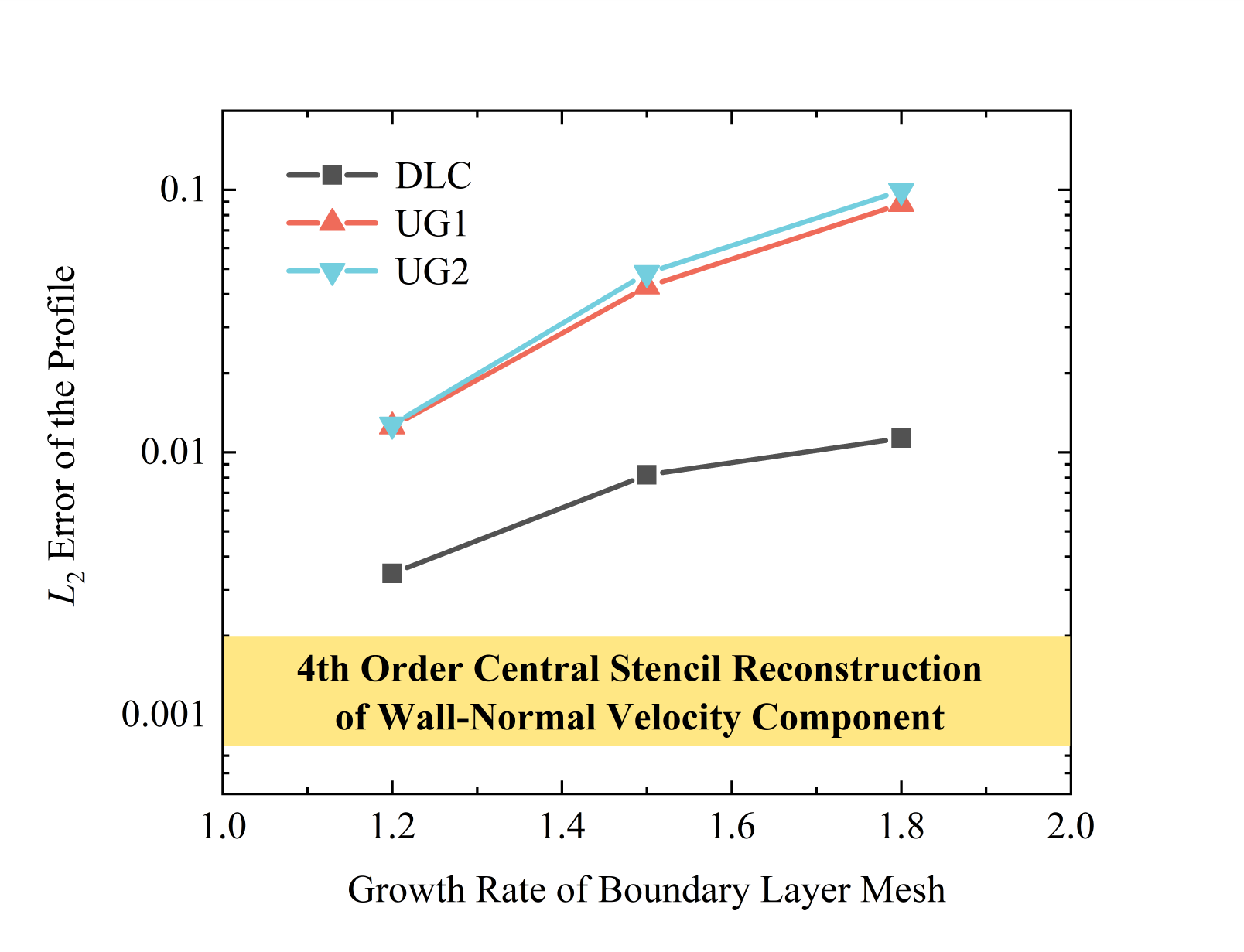}
      \par
      \centering \; (\textit{a}) \hspace{7.4cm} (\textit{b})
    }
  }
	\caption{Global errors of the wall-normal velocity component with respect to mesh growth rate. (\textit{a}) Third-order fixed-stencil reconstruction, (\textit{b}) fourth-order fixed-stencil reconstruction.}
	\label{fig:BL:gErrV}
\end{figure}

The superiority of the DOLINC method, as shown in Figures \ref{fig:BL:U} and \ref{fig:BL:V} is not specific to a particular grid growth rate or high-order reconstruction. We further investigated the global errors for different grid growth rates and different-order reconstruction schemes, as shown in Figures \ref{fig:BL:gErrU} and \ref{fig:BL:gErrV}. The errors in the figures were calculated using the L2 norm. Overall, the global errors in the computed results increased with the grid growth rate. This is because, when the range of the boundary layer is fixed and the height of the first layer of the grid remains constant, a large grid-growth rate represents fewer grid cells. For different growth rates, the DOLINC scheme exhibited significantly lower global errors in reconstructing the wall-parallel and wall-normal velocity components compared with the non-DOLINC schemes. Notably, the global errors of UG1 and UG2 were almost identical, again indicating that the difference in the gradient-calculation approaches had minimal impact on the errors. However, the DOLINC method, which was based on increasing the reconstruction accuracy from third-order to fourth-order precision, showed a considerable reduction in global errors. This implies that the DOLINC method accurately achieved high-order precision in non-uniform grids. In contrast, this precision improvement was not observed when higher-order reconstruction schemes were implemented using the UG method; the error level of USR3-UG was consistent with that of CSR4-UG. This indicates that the main error of the UG method far surpassed the error of the interpolation polynomial itself, and the error in the UG scheme was primarily due to the error in the UG method rather than the error in USR3. Therefore, even if USR3 is upgraded to CSR4, the precision of the UG scheme cannot be enhanced because the error in the UG method itself does not improve. Considering that the inherent precision of USR3 is only third-order, UG-USR3 may have decayed to second-order or even lower precision.

\subsection{Linear Advection Problem}
\label{ssec:LAE}

The linear advection problem is one of the simplest hyperbolic conservation systems, and its multi-dimensional equation is formulated as follows:

\begin{equation}
  \partial_{t} u + a_j u_{,j} = 0.
\end{equation}

where $a_j$ denotes the constant propagation velocity. In the advection transport of this problem, the original waveform/distribution maintains its shape during propagation with a uniform and steady propagation velocity across the entire domain. In the computation of an incompressible flow, pure heat convection and mass-advection transport, neglecting diffusion, generally satisfy the linear advection transport equation. Therefore, the performance of numerical schemes in solving the linear advection equation reflects their practical effectiveness in characterizing the generalized passive scalar transport results in the computations of incompressible flow problems. For the one-dimensional case, the linear advection equation is simplified to

\begin{equation}
  \frac{\partial u}{\partial t} + a \frac{\partial u}{\partial x} = 0.
\end{equation}

\begin{figure}
	\centering
  \fbox{
    \parbox[b]{16cm}
    {
      \includegraphics[width=7.9cm]{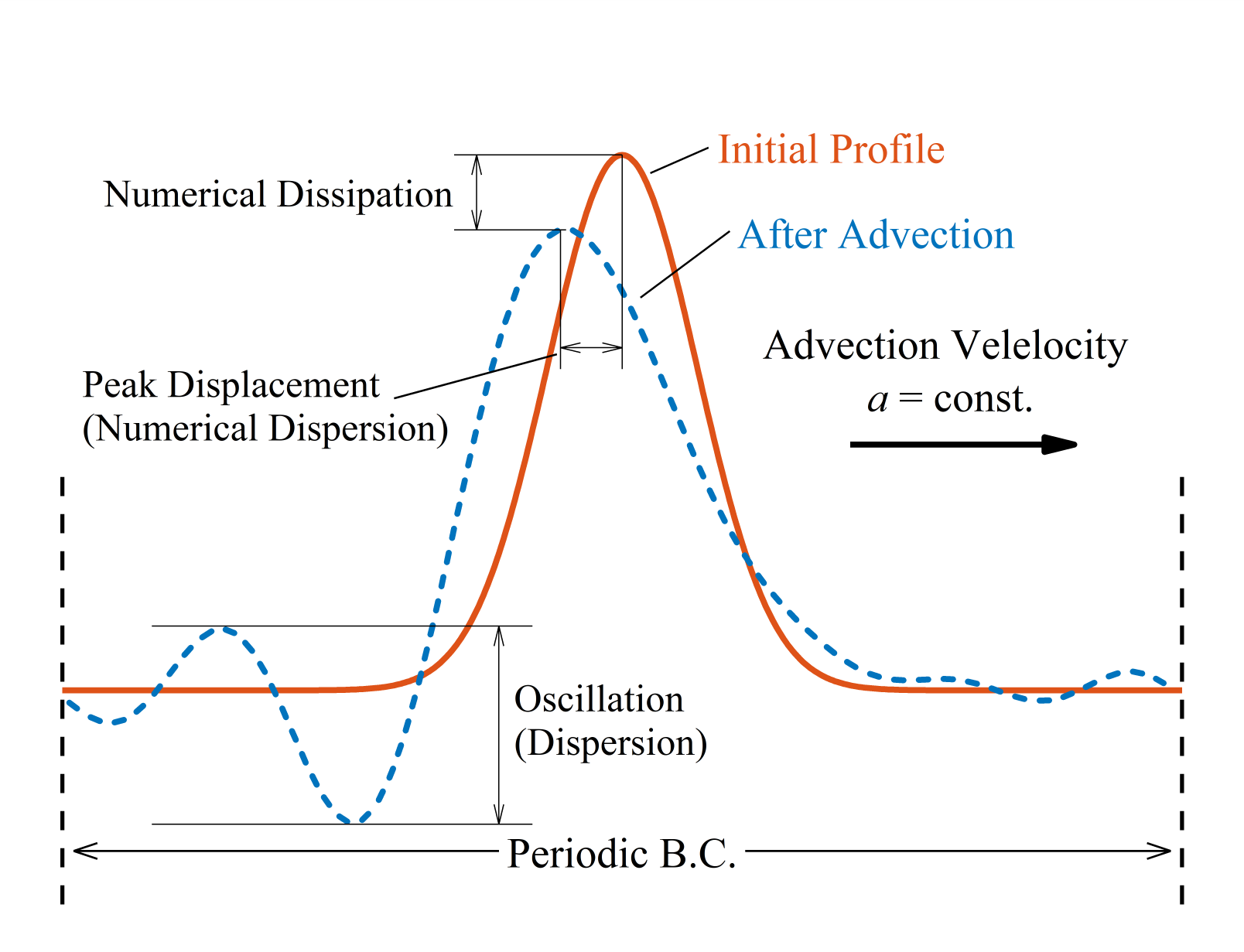}
      \includegraphics[width=7.9cm]{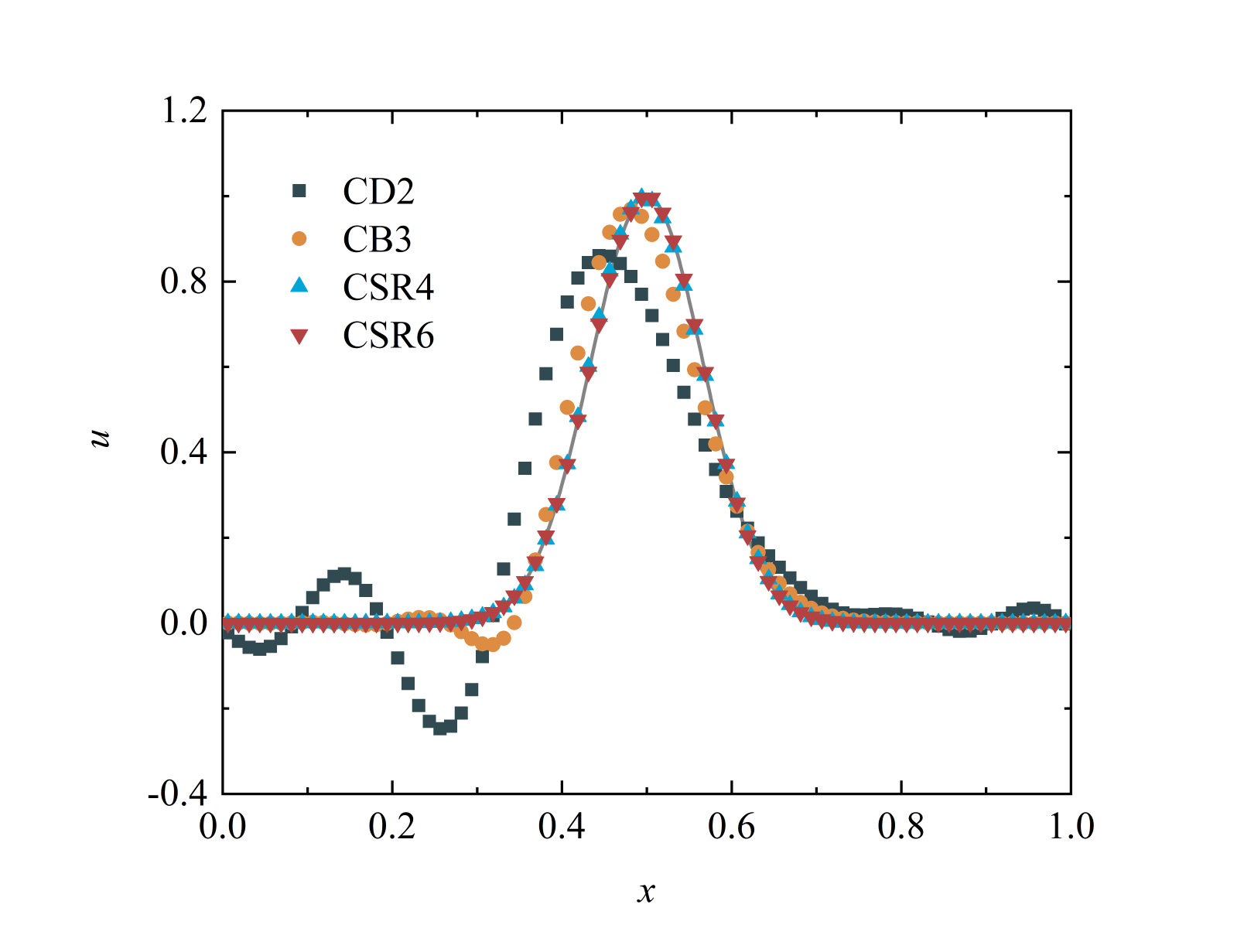}
      \par
      \centering \; \hspace{8cm} (\textit{a})
      \par
      \includegraphics[width=7.9cm]{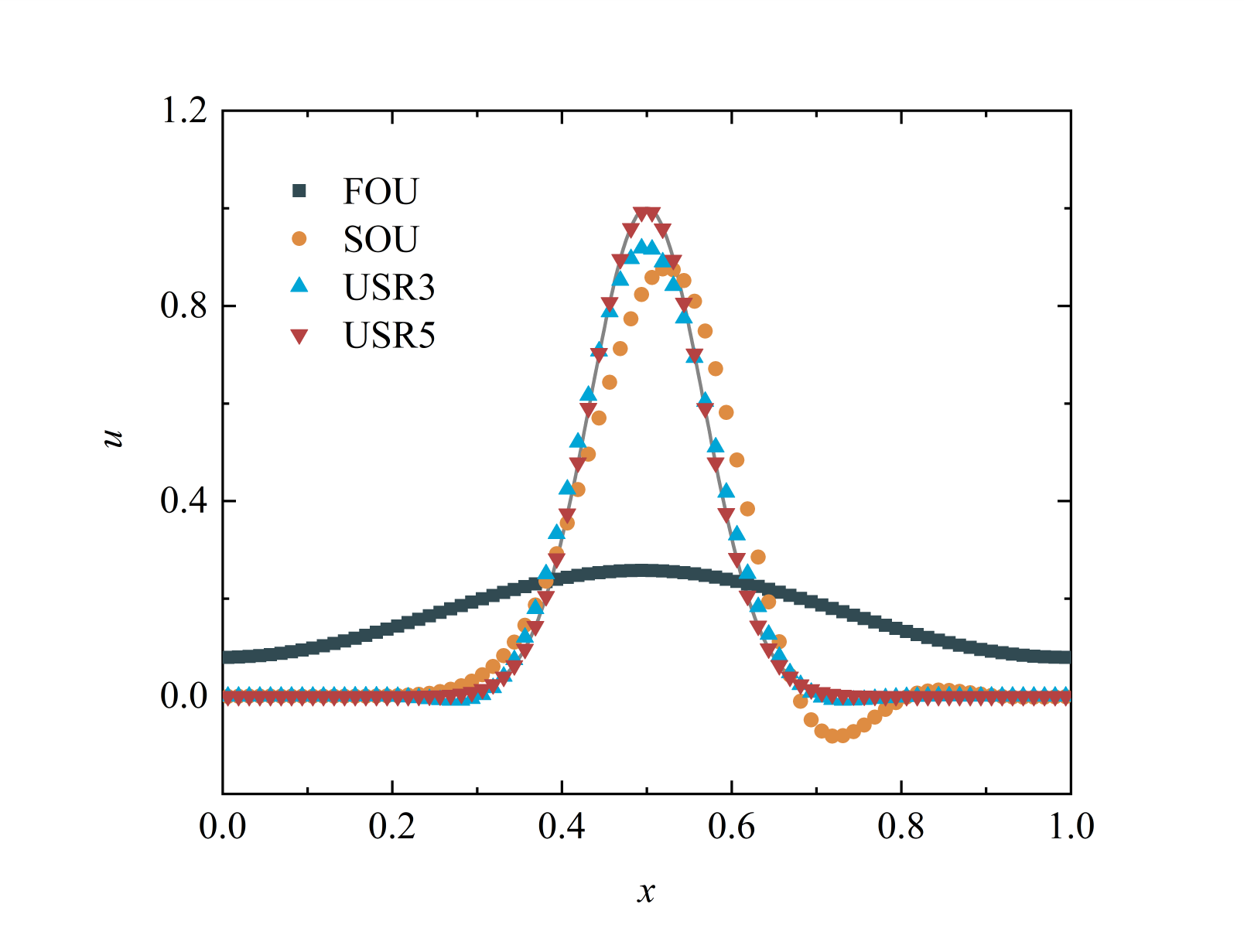}
      \includegraphics[width=7.9cm]{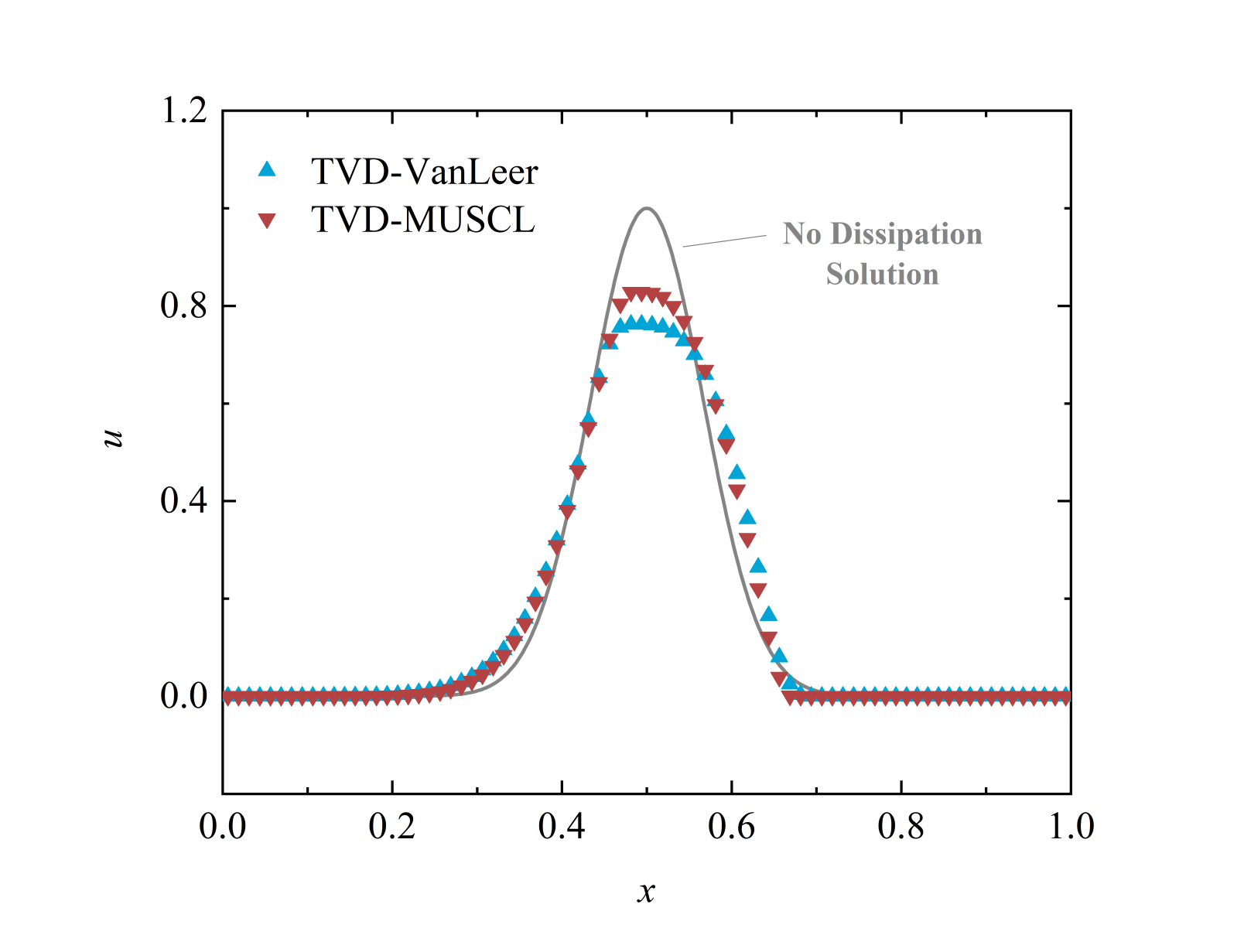}
      \par
      \centering \; (\textit{b}) \hspace{7.4cm} (\textit{c})
    }
  }
	\caption{Linear advection transport of a smooth Gaussian distribution. (\textit{a}) Central schemes, (\textit{b}) upwind schemes, (\textit{c}) non-oscillatory schemes.}
	\label{fig:LAE:GaussW2}
\end{figure}

An unstructured mesh solver was employed to solve the initial-value problem of the one-dimensional linear advection equation. Periodic boundary conditions were applied on both sides, allowing the initial waveform/distribution to re-enter from one endpoint after leaving the other, which is referred to as one transport period. To explore the performance of the high-order reconstruction schemes on smooth functions, Figure \ref{fig:LAE:GaussW2} presents the computed results for the initial waveform corresponding to a Gaussian function of $\sigma = 0.0667$ after five transport periods. The top-left schematic provides a reference for the impact of numerical dissipation and dispersion on the resulting waveform. Figure \ref{fig:LAE:GaussW2}(\textit{a}) illustrates the differences among various central schemes during the smooth function transport computation. CSR4 and CSR6 represent the fourth- and sixth-order reconstruction schemes implemented using the DOLINC method, respectively. CD2 is a common second-order accuracy central-differencing scheme for second-order FVM. CB3 is a scheme in the official release of OpenFOAM that utilizes cubic polynomial interpolation. In Figure \ref{fig:LAE:GaussW2}(\textit{a}), CD2 exhibits typical numerical dissipation and dispersion phenomena associated with the central schemes. After five computation cycles, not only were the extrema significantly reduced, but numerical oscillations were observed on both the upwind and downwind sides. When using the CB3 scheme based on higher-order interpolation polynomials, both the numerical dispersion and dissipation were noticeably improved. However, as CB3 itself does not achieve fourth-order accuracy in practice because of its non-DOLINC implementation, differences in accuracy were still observed compared with the fourth-order CSR4 belonging to the DOLINC schemes. In this problem, the high-order reconstructions CSR4 and CSR6 exhibit almost no apparent numerical dissipation or oscillations. Figure \ref{fig:LAE:GaussW2}(\textit{b}) shows the results for several upwind schemes, including the first-order upwind scheme (FOU), second-order upwind scheme (SOU), third-order scheme USR3, and fifth-order scheme USR5 implemented through the DOLINC approach. Excessively strong numerical dissipation were observed in the FOU results in a complete deviation of the transport outcome from the initial waveform. Compared with the conventional SOU scheme, the DOLINC USR3 scheme demonstrates better performance. While it has lower numerical dissipation, it exhibits almost no numerical oscillations on the downwind side. Similar to the central schemes CSR4 and CSR6, USR5 maintained its initial distribution shape well after five periods with minimal dissipation and oscillations. Regardless of whether they were central or upwind schemes, even if the construction principles of the schemes were so simple that they were not inherently non-oscillation types, simply increasing the order of the method could significantly reduce numerical dissipation and dispersion in practice. For the two non-oscillation TVD schemes in Figure \ref{fig:LAE:GaussW2}(\textit{c}), the results for the VanLeer and MUSCL limiters both showed significant peak dissipation, greatly affecting the overall accuracy of the results. This again demonstrated the major drawback of TVD schemes in terms of accuracy degradation at smooth extrema points, leading to an inability to capture some small-scale continuous flow structures.

\begin{figure}
	\centering
  \fbox{
    \parbox[b]{16cm}
    {
      \includegraphics[width=7.9cm]{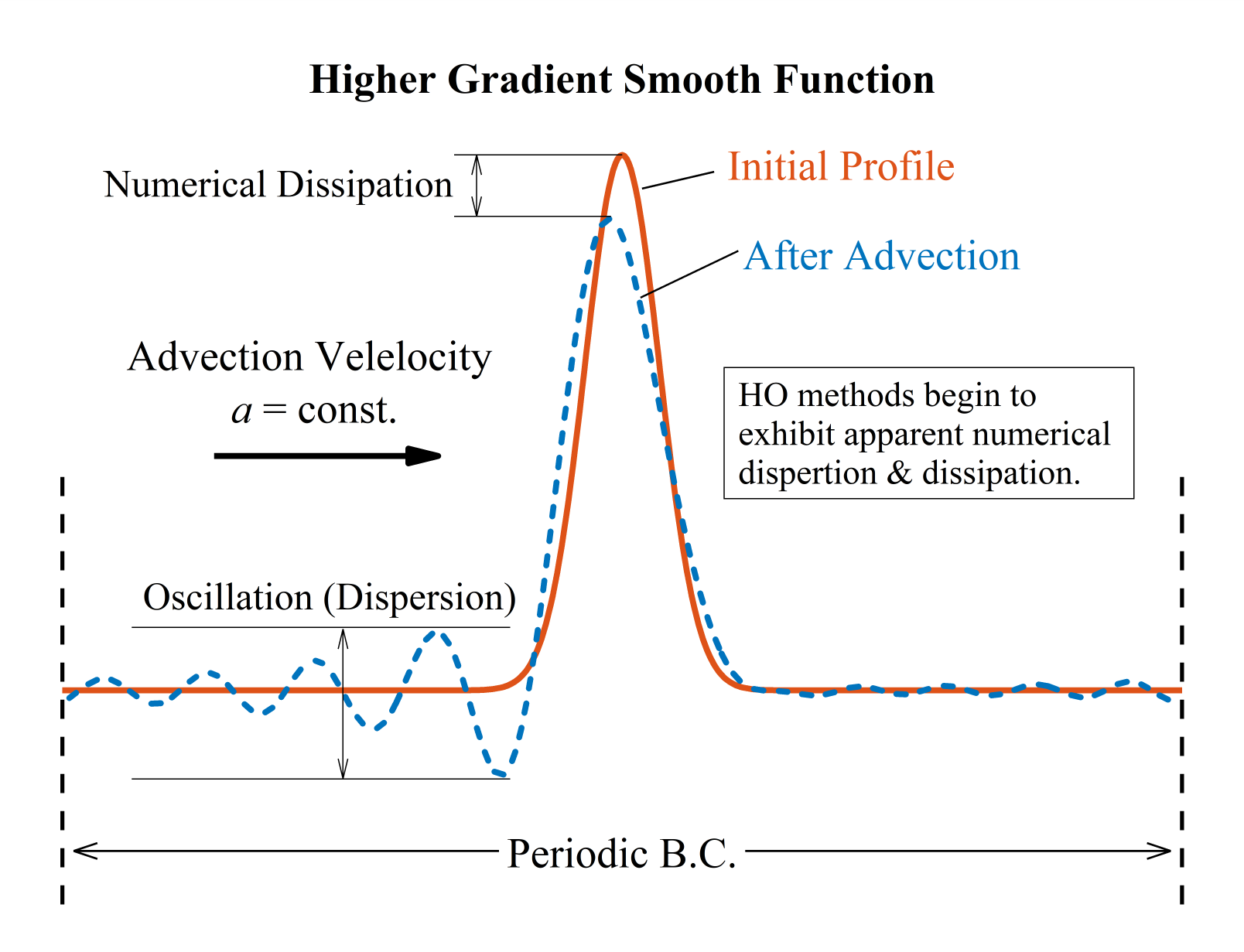}
      \includegraphics[width=7.9cm]{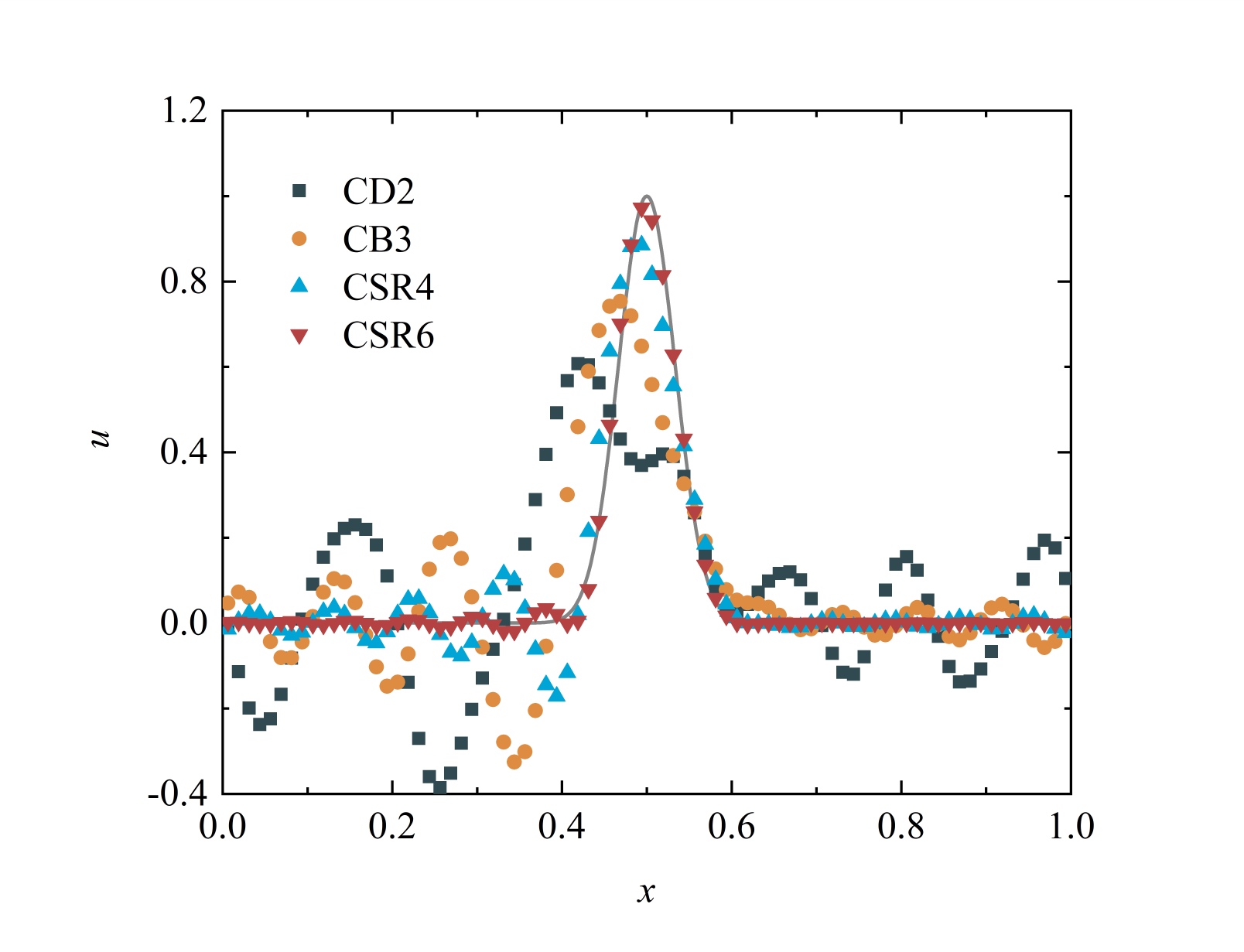}
      \par
      \centering \; \hspace{8cm} (\textit{a})
      \par
      \includegraphics[width=7.9cm]{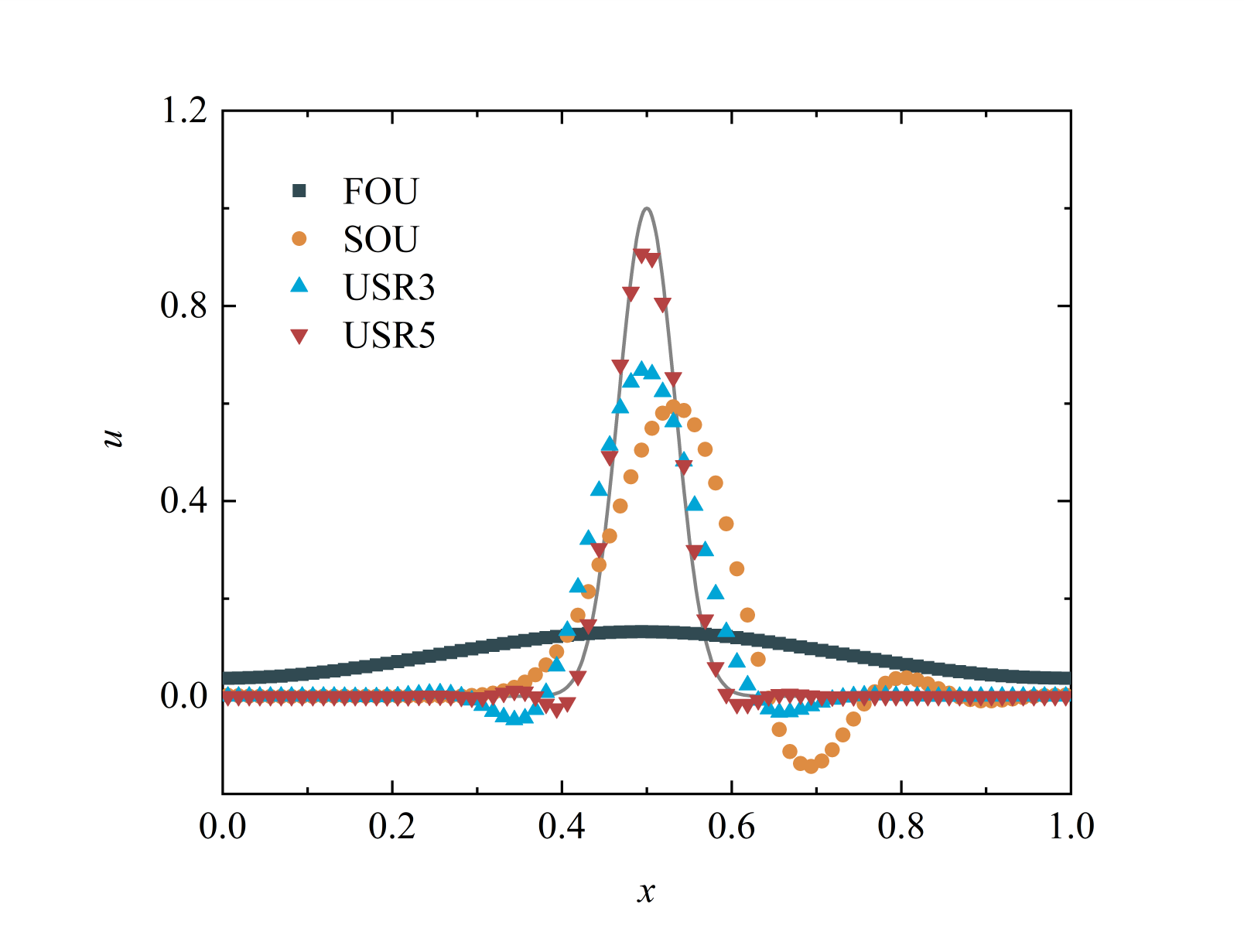}
      \includegraphics[width=7.9cm]{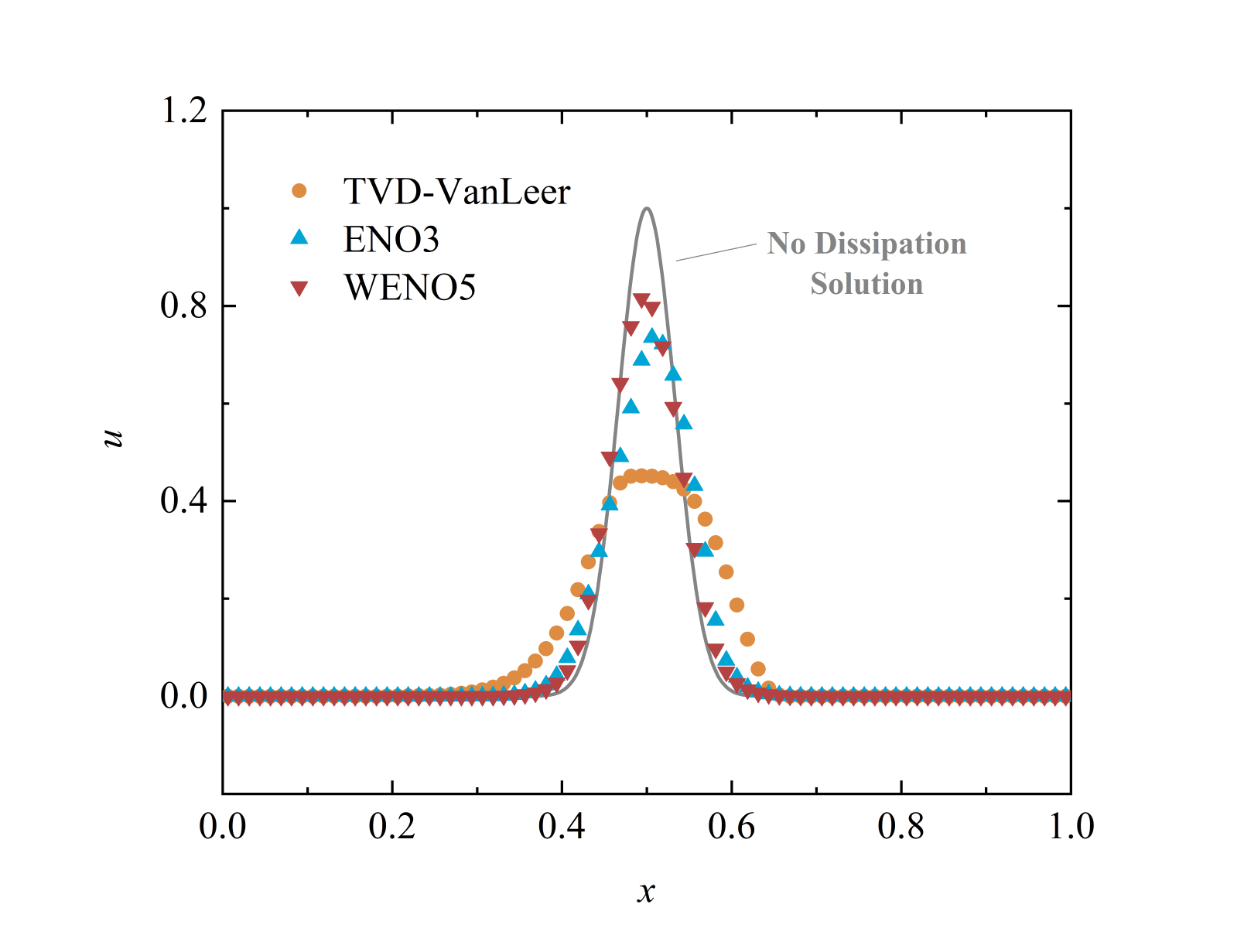}
      \par
      \centering \; (\textit{b}) \hspace{7.4cm} (\textit{c})
    }
  }
	\caption{Linear advection transport of a steep Gaussian distribution. (\textit{a}) Central schemes, (\textit{b}) upwind schemes, (\textit{c}) non-oscillatory schemes.}
	\label{fig:LAE:GaussW1}
\end{figure}

Although the Gaussian distribution is theoretically smooth, the discretized physical field after FVM discretization is effectively a piecewise function. Therefore, a discontinuous step on the grid faces in the discretized space becomes apparent when the width of the Gaussian distribution decreases to a certain extent. This causes the results to exhibit characteristics similar to those of computations of discontinuous functions, namely exacerbated numerical dissipation and numerical dispersion. As shown in Figure \ref{fig:LAE:GaussW1}, when the half-width of the Gaussian function decreases to $\sigma=0.0333$, the high-order schemes that do not exhibit numerical oscillations and dissipation in Figure \ref{fig:LAE:GaussW1} noticeable peak decay and oscillations. As shown in Figure \ref{fig:LAE:GaussW1}(\textit{a}), the oscillations and dissipation of CD2 and CB3 continued to worsen; at this point, CSR4 also showed dissipation and oscillations. Compared with CSR4, the higher-order central reconstruction, CSR6, exhibited marginal oscillations; however, it accurately characterized the initial distribution owing to its higher accuracy. Similarly, in Figure \ref{fig:LAE:GaussW1}(\textit{b}), while USR3 was still superior to SOU, the dissipation and oscillations on both sides were enhanced. At this point, USR5 also began to exhibit slight dissipation and oscillations. However, compared with the lower-order USR3, its performance advantage remained evident. Overall, when the discontinuity characteristics of the physical field were strengthened, simply increasing the order of the schemes could effectively mitigate the dissipation/dispersion phenomena, resulting in significantly improved computational results. The results of the non-oscillation schemes shown in Figure \ref{fig:LAE:GaussW1}(\textit{c}) show the TVD scheme with the VanLeer limiter and demonstrate the performances of the third-order ENO reconstruction (ENO3) and fifth-order WENO reconstruction (WENO5) implemented using the DOLINC method for this problem. Evidently, the accuracy-degradation problem of the TVD schemes was more severe near the discontinuous positions close to the peak. In contrast, ENO3 and WENO5 exhibited significantly smaller numerical dissipation, maintaining the original wave series while strictly ensuring non-oscillation characteristics.

\begin{figure}
	\centering
  \fbox{
    \parbox[b]{16cm}
    {
      \includegraphics[width=7.9cm]{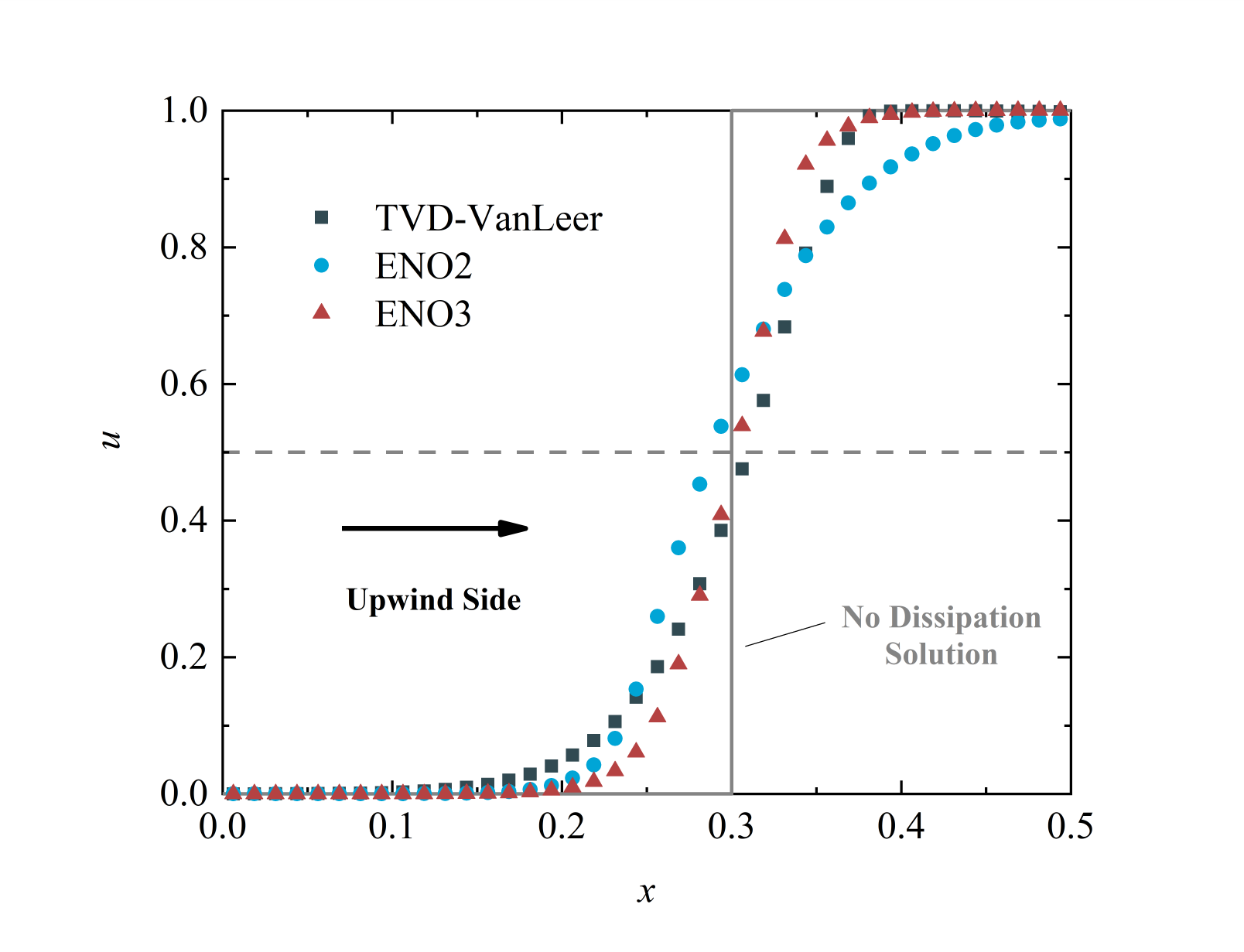}
      \includegraphics[width=7.9cm]{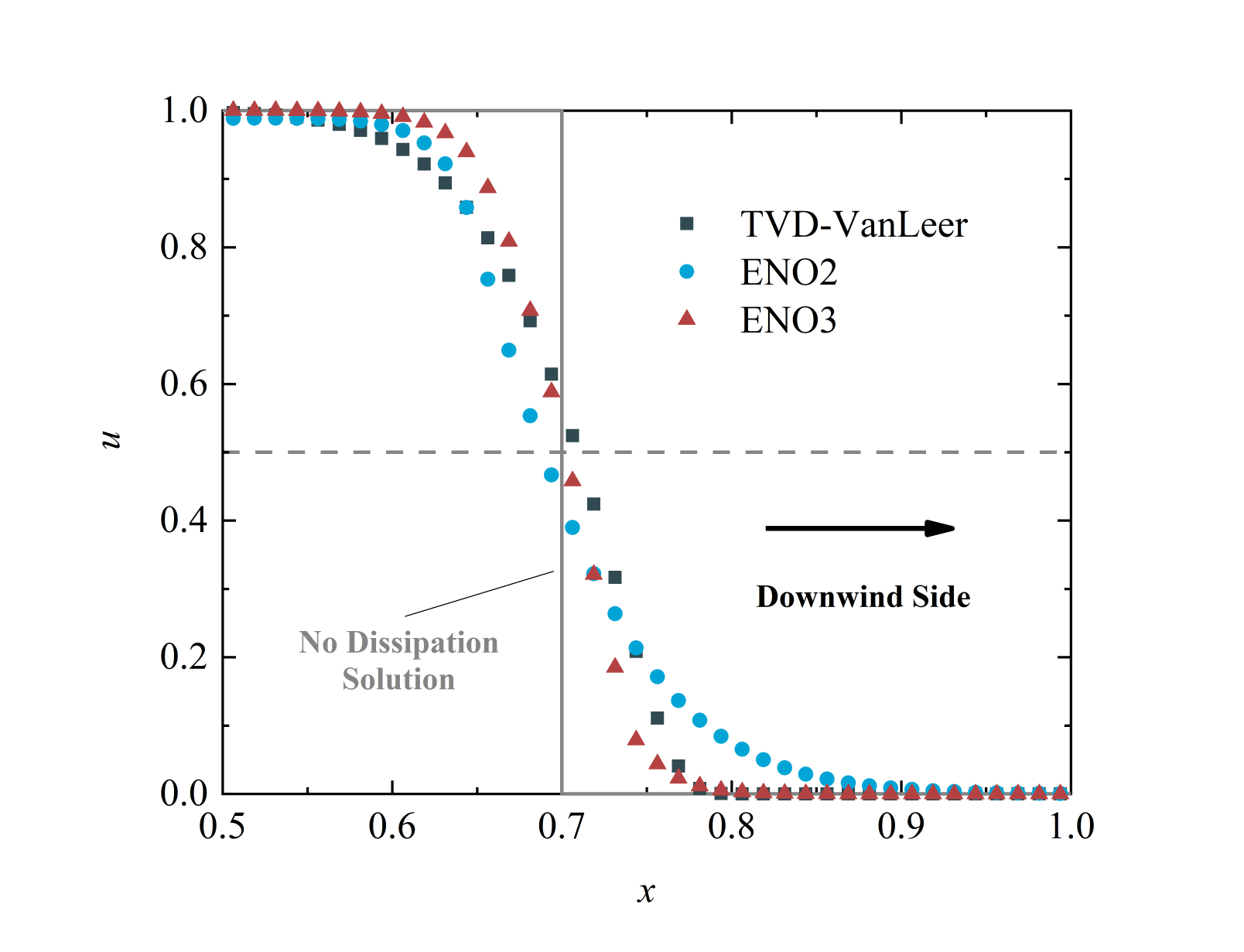}
      \par
      \centering \; (\textit{a}) \hspace{7.4cm} (\textit{b})
    }
  }
	\caption{Results after linear advection of the initial-step distribution calculated using the ENO method. (\textit{a}) Upstream discontinuity on the windward side, (\textit{b}) downstream discontinuity on the leeward side.}
	\label{fig:LAE:stepENO}
\end{figure}

\begin{figure}
	\centering
  \fbox{
    \parbox[b]{16cm}
    {
      \includegraphics[width=7.9cm]{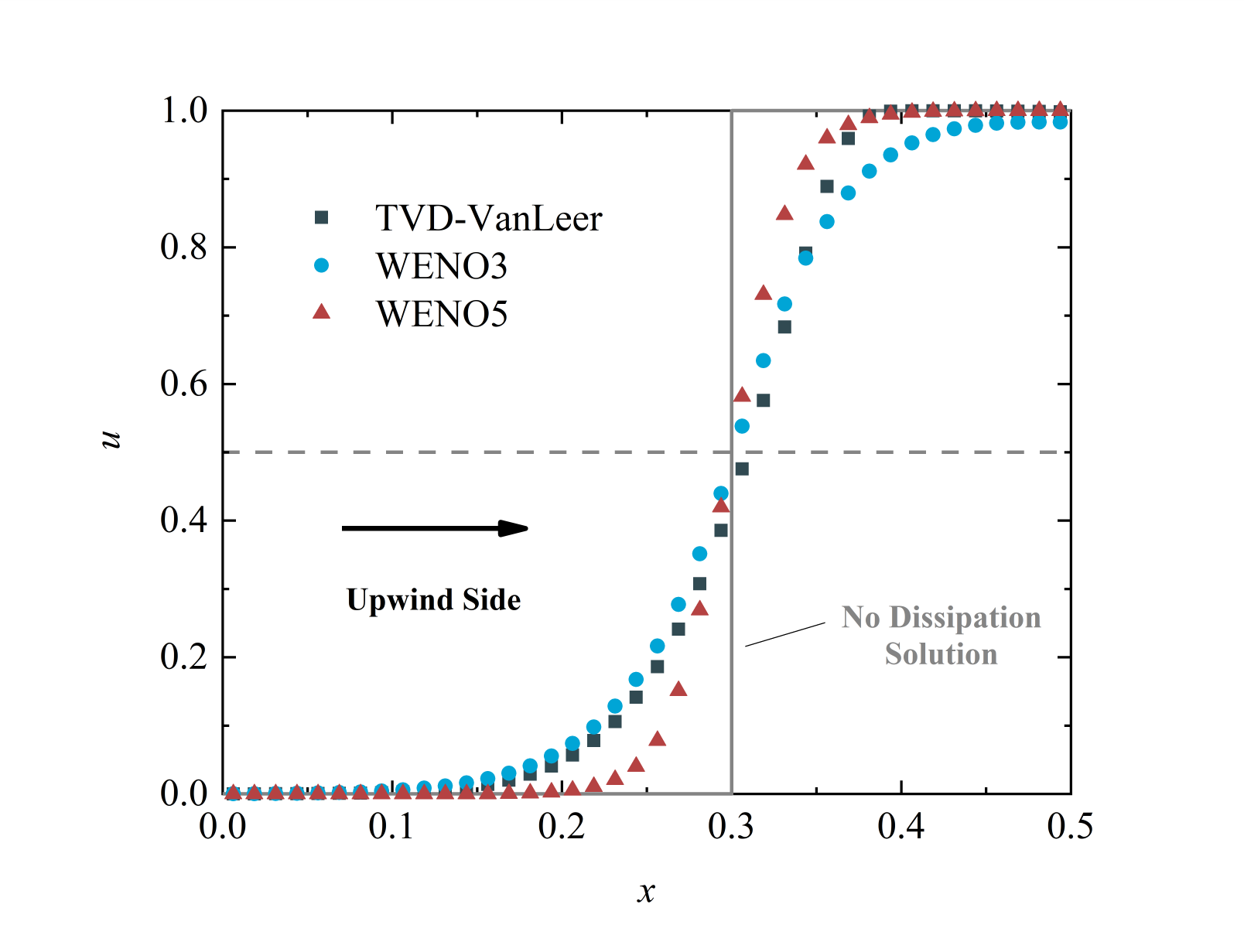}
      \includegraphics[width=7.9cm]{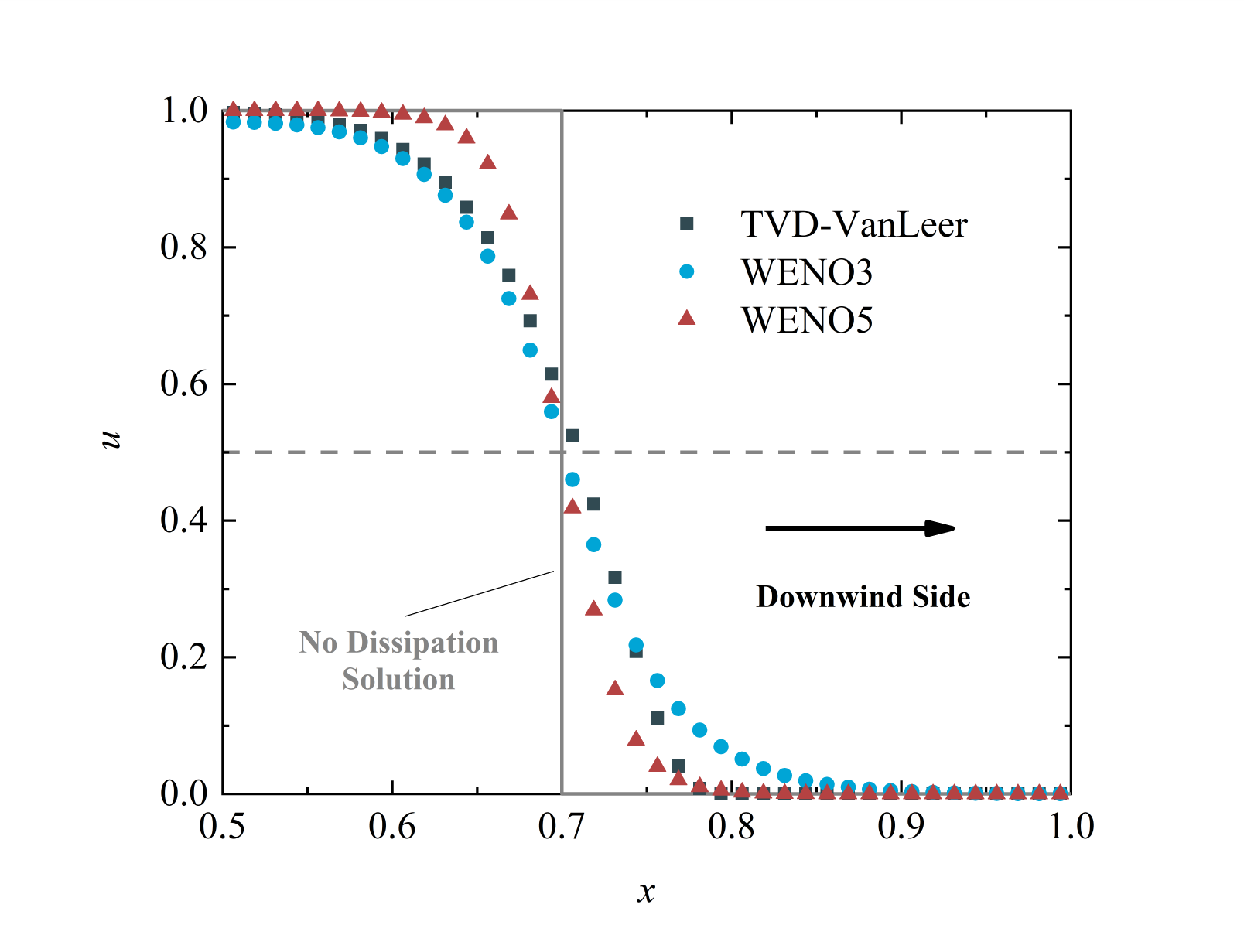}
      \par
      \centering \; (\textit{a}) \hspace{7.4cm} (\textit{b})
    }
  }
	\caption{Results after linear advection of the initial-step distribution calculated using the WENO method. (\textit{a}) Upstream discontinuity on the windward side, (\textit{b}) downstream discontinuity on the leeward side.}
	\label{fig:LAE:stepWENO}
\end{figure}

The results of advective transport with a rectangular-wave initial distribution were compared between the ENO/WENO and TVD schemes to further investigate the performance of the different schemes in handling strong discontinuities, as shown in Figures \ref{fig:LAE:stepENO} and \ref{fig:LAE:stepWENO}. The figures show that ENO3 and WENO5, both of which belong to the second-order DOLINC scheme, exhibited significantly higher numerical accuracy at the discontinuity than the TVD-VanLeer scheme, with WENO5 showing a more pronounced advantage. ENO2 and WENO3, which rely on second-order reconstructions in the template for handling the discontinuity area, have fewer obvious advantages than the TVD scheme. Thus, the ENO/WENO series methods only demonstrated significantly better characteristics than the TVD scheme in handling discontinuities when they reached a particular high order \citep{Shi:2003}. Therefore, on unstructured grids, methods such as the DOLINC approach, which can achieve arbitrarily high orders, are more valuable than implementing a specific third-order scheme. Furthermore, the ENO and WENO schemes exhibited better symmetry in both the upwind and downwind discontinuities compared with the TVD scheme. The VanLeer limiter used in this case satisfied the property of the symmetry $\psi(1/r)=\psi(r)/r$. Other TVD limiters that do not satisfy this relationship may exhibit a poorer symmetry. Therefore, if only the capture of flow-field discontinuities is considered, the TVD scheme performs well, and in terms of cost-effectiveness, industrial software using the TVD scheme has a clear advantage. However, when flow problems involve the capture of shockwaves and fine vortex structures simultaneously (such as typical supersonic turbulent flows), the most fatal issue manifested by the TVD scheme compared with higher-order schemes, such as ENO/WENO, is the previously mentioned accuracy degradation at the extremum points. Thus, the main drawback of the TVD scheme in industrial applications is its excessive dissipation of small-scale flow structures, leading to an inability to correctly handle fine vortex structures that may appear owing to instability in strong-discontinuity problems, rather than its handling of strong discontinuities. This is also a reason for caution when using the TVD scheme in LESs and DNSs for strong compressible flows.

\begin{figure}
	\centering
  \fbox{
    \parbox[b]{16cm}
    {
      \includegraphics[width=7.9cm]{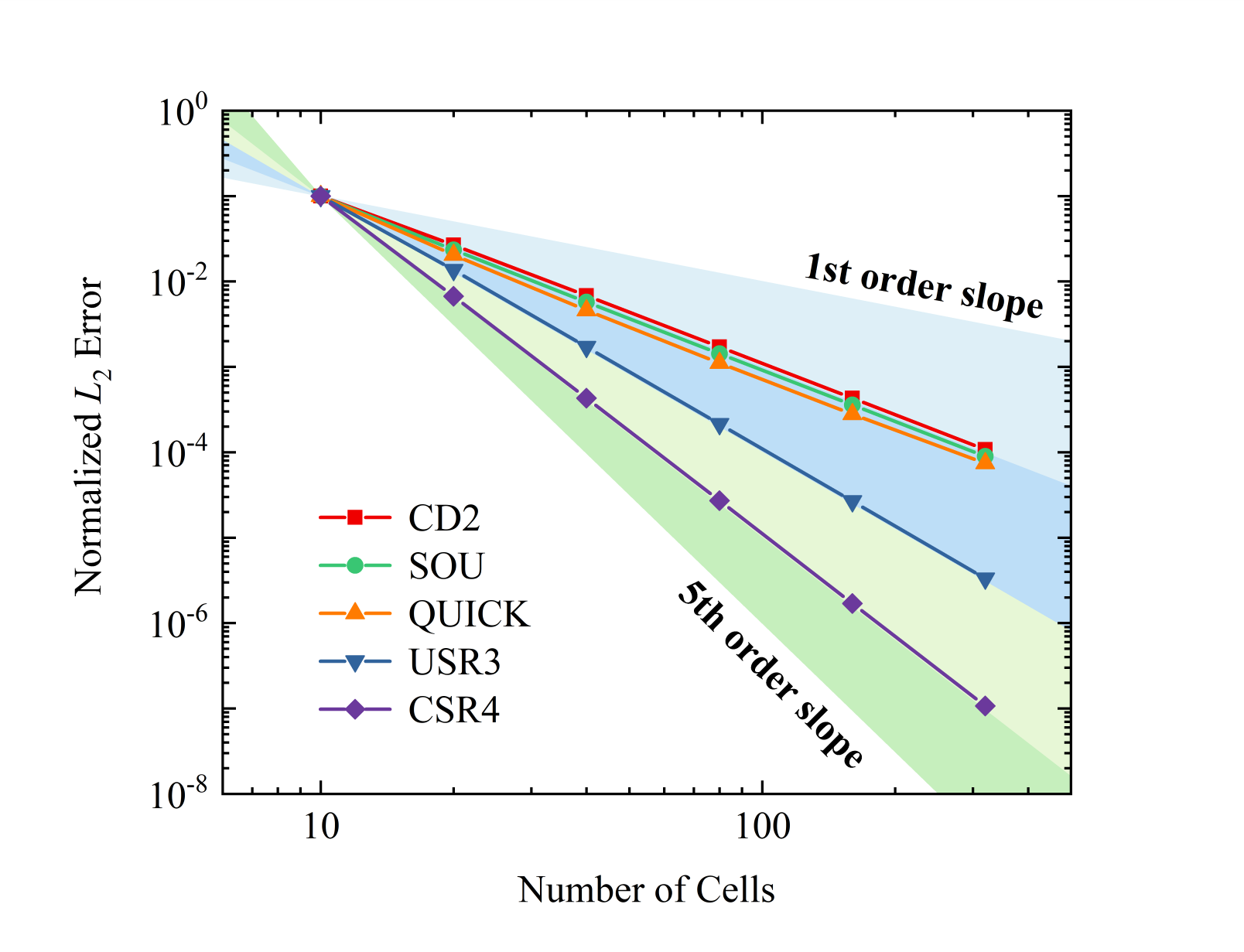}
      \includegraphics[width=7.9cm]{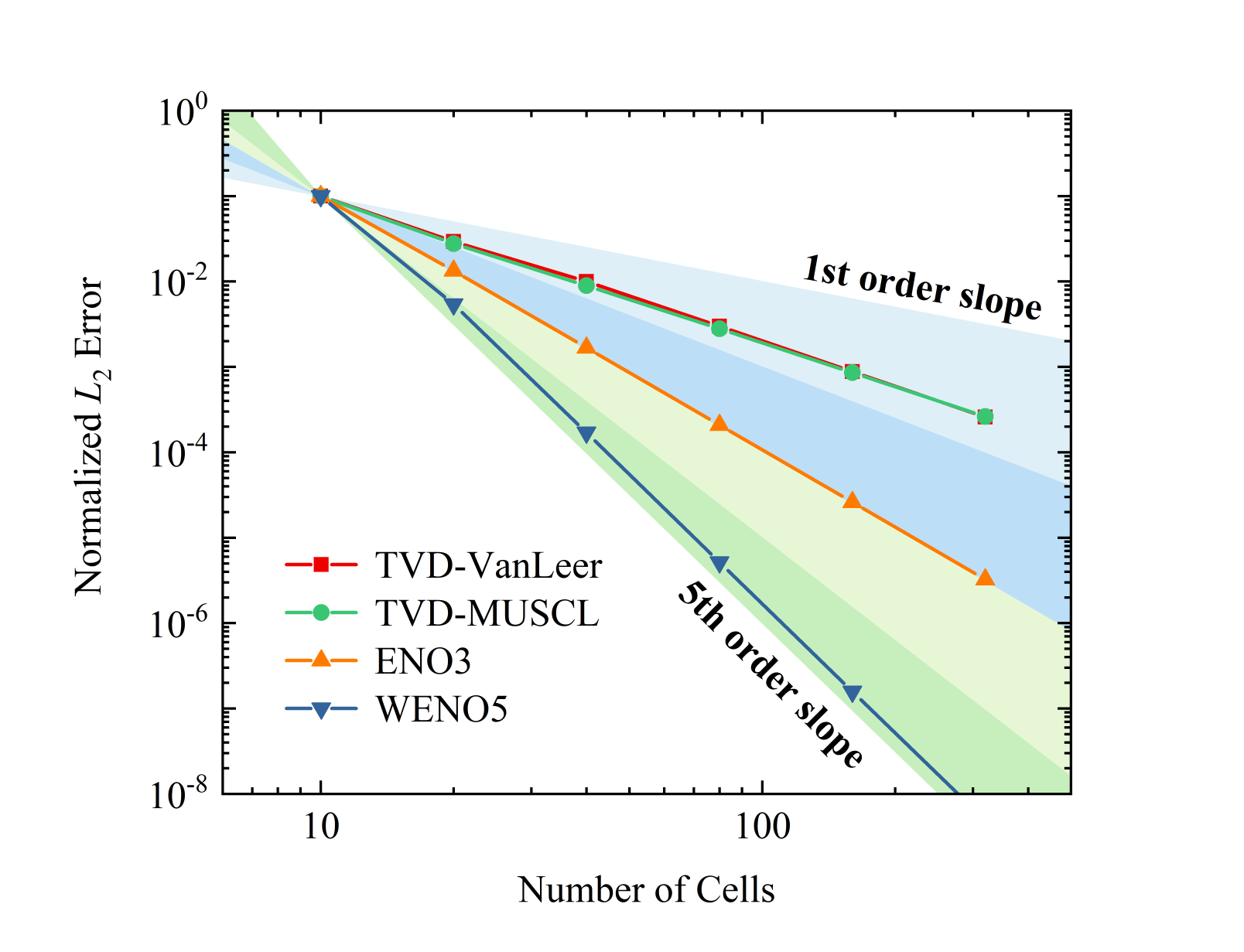}
      \par
      \centering \; (\textit{a}) \hspace{7.4cm} (\textit{b})
    }
  }
	\caption{Error convergence order of accuracy for different schemes. (\textit{a}) Primitive scheme, (\textit{b}) non-oscillatory scheme.}
	\label{fig:LAE:order}
\end{figure}

To validate the actual convergence order of the DOLINC scheme, its global error was calculated based on an initial problem using a sine function. The initial errors of the results for each scheme with 10 grid cells were normalized to the same reference value to facilitate the comparison of the error-reduction rate with an increasing number of grid cells. At this point, the absolute-error size after refining the grid reflects the convergence speed of the scheme, as shown in Figure \ref{fig:LAE:order}. The original error data and corresponding convergence accuracy without normalization are presented in Tables \ref{tab:orderPScheme} and \ref{tab:orderHOScheme}, respectively. In the series of basic schemes shown in Figure \ref{fig:LAE:order}(\textit{a}), commonly used low-order FVM schemes in industrial software, such as CD2 and SOU, and some special schemes, such as QUICK, all exhibit an actual convergence order of approximately two. The QUICK scheme, which is based on a quadratic interpolation polynomial derived from finite differencing, cannot achieve third-order theoretical accuracy when applied to the finite-volume method. In contrast, the fixed template schemes, USR3 and CSR4, implemented using the DOLINC scheme, achieved theoretical convergence accuracies of the third and fourth orders, respectively. In non-oscillation schemes, TVD schemes are limited by the first-order accuracy near the local extrema, resulting in a global error that does not exceed the second order. Therefore, the two TVD-limited schemes shown in Figure \ref{fig:LAE:order}(\textit{b}) demonstrated orders of accuracy lower than two. In contrast, the two non-oscillation schemes, ENO3 and WENO5, implemented using the DOLINC method, achieved third- and fifth-order theoretical convergence accuracies, respectively. Notably, the convergence order of WENO5 quickly converged from approximately 4.2 on coarse grids to approximately 5, as shown in Table \ref{tab:orderHOScheme}.

\begin{table}
  \caption{$L_2$ Error and convergence order for the primitive schemes.}
  \centering
  \begin{tabular}{ccccccccc}
    \toprule
    \multirow{2}{*}{Mesh} & \multicolumn{2}{c}{CD2} & \multicolumn{2}{c}{SOU} & \multicolumn{2}{c}{USR3} & \multicolumn{2}{c}{CSR4}  \\
    \cmidrule{2-9}
                & Error        & Order    & Error        & Order    & Error        & Order     & Error        & Order     \\
    \midrule
    10          & 0.264391     &          & 0.15908      &          & 0.084241     &           & 0.020641     &           \\
    20          & 0.070505     & 1.91     & 0.03729      & 2.09     & 0.011493     & 2.87      & 0.001384     & 3.90      \\
    40          & 0.017996     & 1.97     & 0.009122     & 2.03     & 0.001446     & 2.99      & 8.87E-05     & 3.96      \\
    80          & 0.004536     & 1.99     & 0.002276     & 2.00     & 0.00018      & 3.00      & 5.59E-06     & 3.99      \\
    160         & 0.001138     & 1.99     & 0.00057      & 2.00     & 2.25E-05     & 3.00      & 3.51E-07     & 3.99      \\
    320         & 0.000285     & 2.00     & 0.000143     & 2.00     & 2.81E-06     & 3.00      & 2.21E-08     & 3.99      \\
    \bottomrule
  \end{tabular}
  \label{tab:orderPScheme}
\end{table}

\begin{table}
  \caption{$L_2$ Error and convergence order for the non-ocsillatory schemes.}
  \centering
  \begin{tabular}{ccccccccc}
    \toprule
    \multirow{2}{*}{Mesh} & \multicolumn{2}{c}{TVD-VanLeer} & \multicolumn{2}{c}{TVD-MUSCL} & \multicolumn{2}{c}{ENO3} & \multicolumn{2}{c}{WENO5} \\
    \cmidrule{2-9}
                & Error            & Order        & Error           & Order       & Error        & Order     & Error         & Order     \\
    \midrule
    10          & 0.250299         &              & 0.176635        &             & 0.086452     &           & 0.031202      &           \\
    20          & 0.073412         & 1.77         & 0.049033        & 1.85        & 0.011572     & 2.90      & 0.001682      & 4.21      \\
    40          & 0.025002         & 1.55         & 0.01582         & 1.63        & 0.001449     & 3.00      & 5.30E-05      & 4.99      \\
    80          & 0.007493         & 1.74         & 0.004935        & 1.68        & 0.00018      & 3.01      & 1.61E-06      & 5.04      \\
    160         & 0.002205         & 1.76         & 0.001516        & 1.70        & 2.25E-05     & 3.00      & 4.93E-08      & 5.03      \\
    320         & 0.000647         & 1.77         & 0.000463        & 1.71        & 2.81E-06     & 3.00      & 1.53E-09      & 5.01      \\
    \bottomrule
  \end{tabular}
  \label{tab:orderHOScheme}
\end{table}

\begin{figure}
	\centering
  \fbox{
    \parbox[b]{16cm}
    {
      \includegraphics[width=7.9cm]{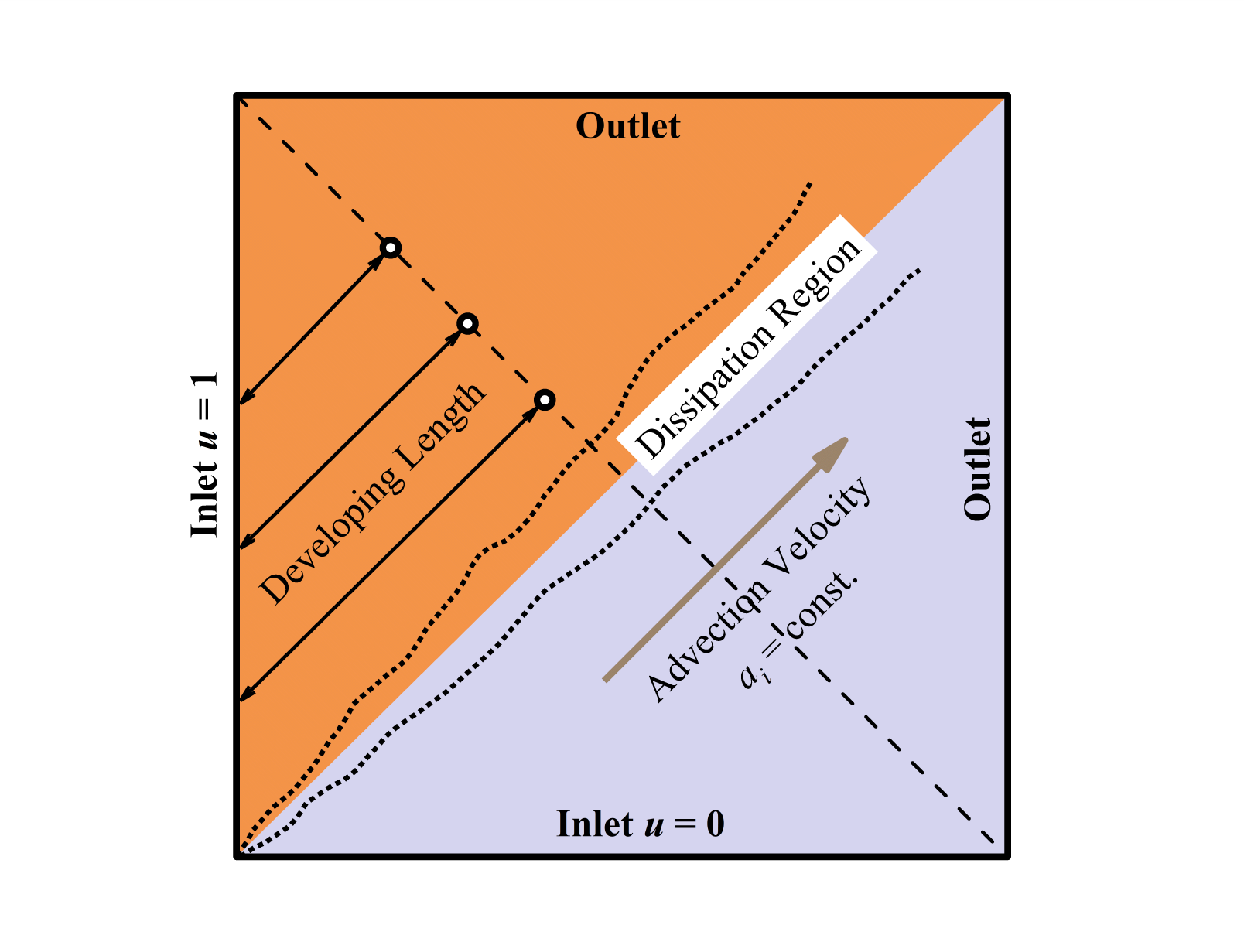}
      \includegraphics[width=7.9cm]{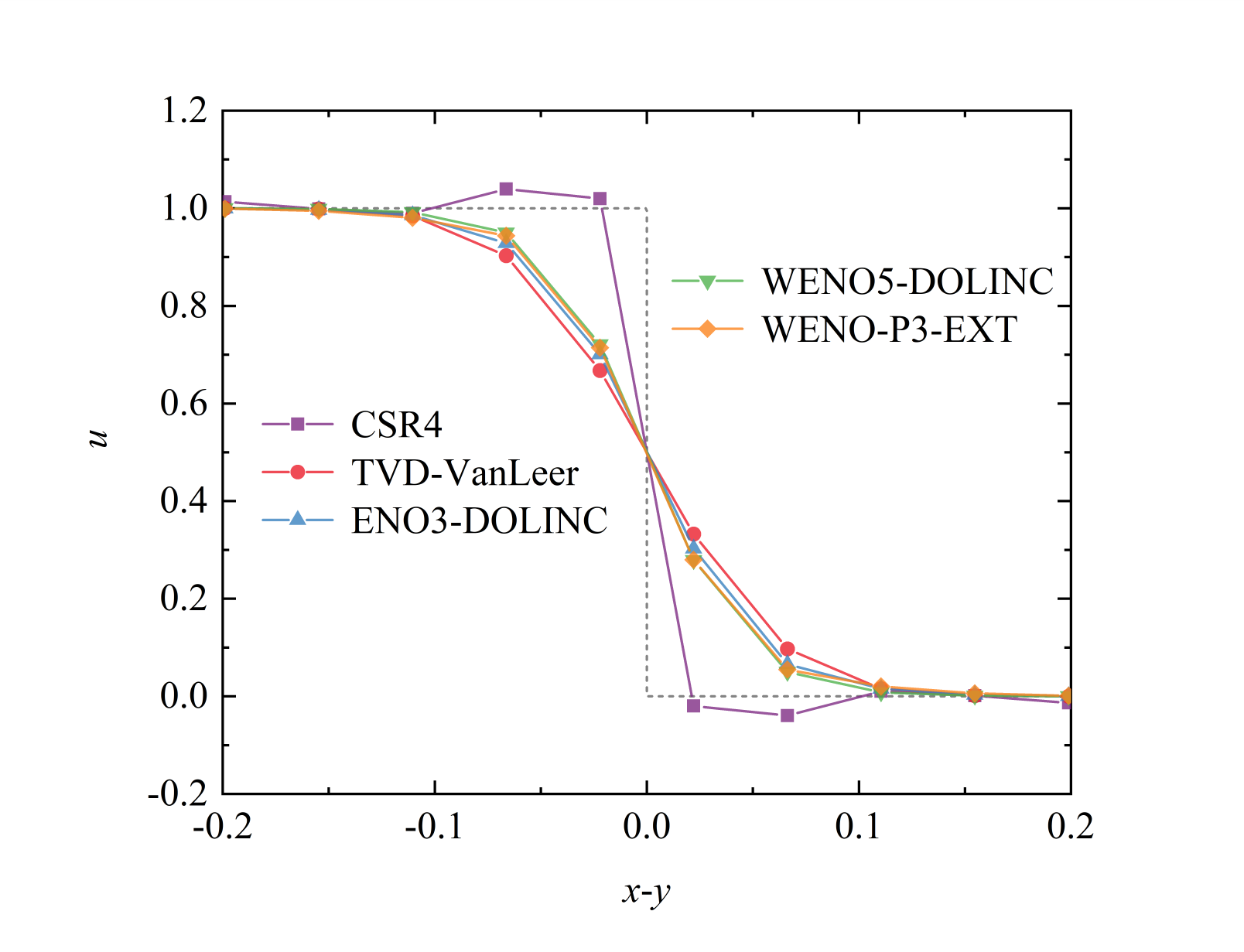}
      \par
      \centering \; (\textit{a}) \hspace{7.4cm} (\textit{b})
    }
  }
	\caption{Two-dimensional linear advection case. (\textit{a}) Computational domain and boundary conditions setup, (\textit{b}) enlarged view of the computed results along the dashed diagonal line.}
	\label{fig:LAE:2D}
\end{figure}

In some validation cases, the differences between the high- and low-order FVM schemes are insignificant. This is not because the performances of the schemes are similar, but rather because the chosen problem is relatively mild. A frequently used two-dimensional convection transport problem for comparing scheme differences in industrial CFD software is shown in Figure \ref{fig:LAE:2D}(\textit{a}). It computes a pure advection problem within a square region. The advantage of this case is its simple geometry and boundary settings, which make it convenient to demonstrate some characteristics of the numerical schemes visually. Quantitative comparisons of the numerical dissipation can also be made by comparing the diagonal distribution perpendicular to the flow direction. Notably, this problem eventually reaches a steady state. Additionally, because the computational domain itself is not large, and the diagonal used for quantitative analysis is relatively close to the inlet boundary, the influence of numerical dissipation in this problem is relatively weak. Unlike the one-dimensional convection problem discussed earlier, where the dissipation becomes more significant as the computation time progresses, the numerical dissipation in this two-dimensional case, after reaching a steady state, is constrained within a certain range. Even at the most dissipative position in the upper-right corner, the experienced dissipation path is limited to a diagonal length that is considerably less than the dissipation path scale achievable in multiple periods of the one-dimensional convection problem. Therefore, in this mild case, the differences in the computed results between the different schemes were not significant. As shown in Figure \ref{fig:LAE:2D}(\textit{b}), the advantages of the high-order schemes are not reflected well in this case. Even low-order FVM schemes yield satisfactory results for this problem. The results shown in Figure \ref{fig:LAE:2D}(\textit{b}) represent an enlarged distribution in the central region, and the actual total length of the diagonal was 2.0. In addition to the ENO3/WENO5 schemes implemented based on the DOLINC method, this case also used the WENO-P3 scheme based on the unstructured-mesh k-exact method for computation. WENO-P3 also uses an interpolation polynomial for weighting and exhibits accuracy similar to that of WENO5 (supported by both Figure \ref{fig:LAE:2D} and Section \ref{sssec:RTI}). The specific implementation of the unstructured-grid WENO methods can be found in the literature \citep{Martin:2018}. To distinguish between the two WENO schemes, they are referred to as WENO5-DOLINC and WENO-P3-EXT. The figure shows that the resulting curves of the WENO5-DOLINC and WENO-P3-EXT basically overlapped, and no significant difference in accuracy was observed. Except for CSR4, which marginally overpredicted owing to numerical dispersion at the discontinuity, the other numerical schemes did not exhibit numerical oscillations.

In addition to verifying the accuracy of the DOLINC scheme, tests were conducted to assess its computational efficiency. Based on the linear advection equation, we compared the computational-time differences between the native schemes provided in the OpenFOAM release and high-order schemes implemented using the DOLINC method in one-dimensional and two-dimensional cases. The results are listed in Tables \ref{tab:speed1D} and \ref{tab:speed2D}. In Table \ref{tab:speed1D}, we show the increase in computational cost when improving the accuracy with different schemes using the second-order central differencing scheme CD2 as the baseline. CD2 was chosen as the baseline because, in low-order FVM industrial software, when conducting incompressible turbulent LES or DNS simulations, CD2 is often recommended to avoid excessive numerical dissipation, even if numerical oscillations may occur. Owing to the lack of higher-accuracy high-order schemes, the only strategy for reducing errors and improving accuracy in the second-order FVM is to the continuous refinement of the grid. When the number of grid cells was increased to twice the baseline value (while keeping the CFL number constant), as shown in Table \ref{tab:speed1D}, the computational time increased by a factor of 1.5 times the original (an increment of 49\%). By further refining the grid to four times the baseline results in a nearly 170\% increase in computational time, the accuracy of CD2 approached that of CSR4. Choosing other high-order schemes while keeping the grid resolution constant yielded a clear advantage in terms of computational efficiency compared to simply refining the grid. Even for the most time-consuming WENO5-DOLINC scheme, the time only increased by 44\%. The basic schemes, USR and CSR, implemented using the DOLINC method, exhibited computational times approximately equivalent to those of the native SOU scheme without a significant increase owing to the use of high-order schemes. Considering that the DOLINC method can further improve the efficiency of the actual code-implementation algorithm, USR3 and CSR4 had shorter computational times than the second-order upwind scheme. This is reflected by the native CB3 scheme, which also uses a third-order polynomial, requiring more computational time than the CSR4 scheme, which has better computational accuracy.

\begin{table}
  \caption{Compuational-time increment using different schemes for a one-dimensional case (with CD2 as the reference).}
  \centering
  \begin{tabular}{ccccccccc}
    \toprule
    Primitive Schemes   & CPU Time Increase & DOLINC Schemes & CPU Time Increase \\
    \midrule
    CD2                 & Baseline          & USR3           & 0.16              \\
    CD2 (2x Refinement) & 0.49              & USR5           & 0.29              \\
    CD2 (4x Refinement) & 1.69              & CSR4           & 0.18              \\
    CB3                 & 0.42              & CSR6           & 0.29              \\
    SOU                 & 0.24              & WENO5          & 0.44              \\
    \bottomrule
  \end{tabular}
  \label{tab:speed1D}
\end{table}

\begin{table}
  \caption{Computation time of various schemes for a two-dimensional case at different grid resolutions.}
  \centering
  \begin{tabular}{ccc}
    \toprule
    \multirow{2}{*}{Schemes} & \multicolumn{2}{c}{CPU Time} \\
    \cmidrule{2-3}
                             & Standard mesh & Refined mesh \\
    \midrule
    TVD-VanLeer              & 8.05          & 50.49        \\
    CSR4-DOLINC              & 11.57         & 77.28        \\
    ENO3-DOLINC              & 31.57         & 261.35       \\
    WENO5-DOLINC             & 48.41         & 324.49       \\
    WENO-P3-EXT              & 429.72        & 4514.03      \\
    \bottomrule
  \end{tabular}
  \label{tab:speed2D}
\end{table}

With the introduction of advanced methods such as ENO/WENO, the computational times of high-order schemes tend to increase compared to basic schemes. Table \ref{tab:speed2D} shows a significant difference in the computational times between the CSR4-DOLINC and ENO/WENO-DOLINC. The additional computational cost was primarily due to the more complex implementation procedure of these non-oscillation methods, rather than being closely related to the DOLINC method. However, compared with the WENO-P3 scheme, the ENO/WENO schemes implemented using the DOLINC method demonstrated a significant speed advantage. The computational cost of WENO-P3 in the table is 50--90 times that of the TVD scheme, whereas that of WENO5 is approximately six times that of the TVD scheme. This gap continues to widen in terms of the computational time after grid refinement. Therefore, high-order schemes with similar accuracy implemented based on the DOLINC method exhibit higher computational efficiency compared to unstructured-grid high-order schemes implemented using multi-grid templates.

\subsection{Discontinuities in the Burgers Equation}
\label{ssec:BE}

The Burgers equation is another typical hyperbolic conservation system that is commonly employed to study the characteristics of shocks in fluid dynamics. The multi-dimensional form of the Burgers equation is given by:

\begin{equation}
  \partial_t u_i + u_j u_{i,j} = 0.
\end{equation}

The Burgers equation is used to assess the performance of the spatial schemes in handling the generation and propagation of discontinuities. In one dimension, the corresponding governing equation is as follows:

\begin{equation}
  \frac{\partial u}{\partial t} + u \frac{\partial u}{\partial x} =
    \frac{\partial u}{\partial t} + \frac{\partial}{\partial x} \bigg( \frac12 u^2 \bigg) = 0
\end{equation}

\begin{figure}
	\centering
  \fbox{
    \includegraphics[width=8.0cm]{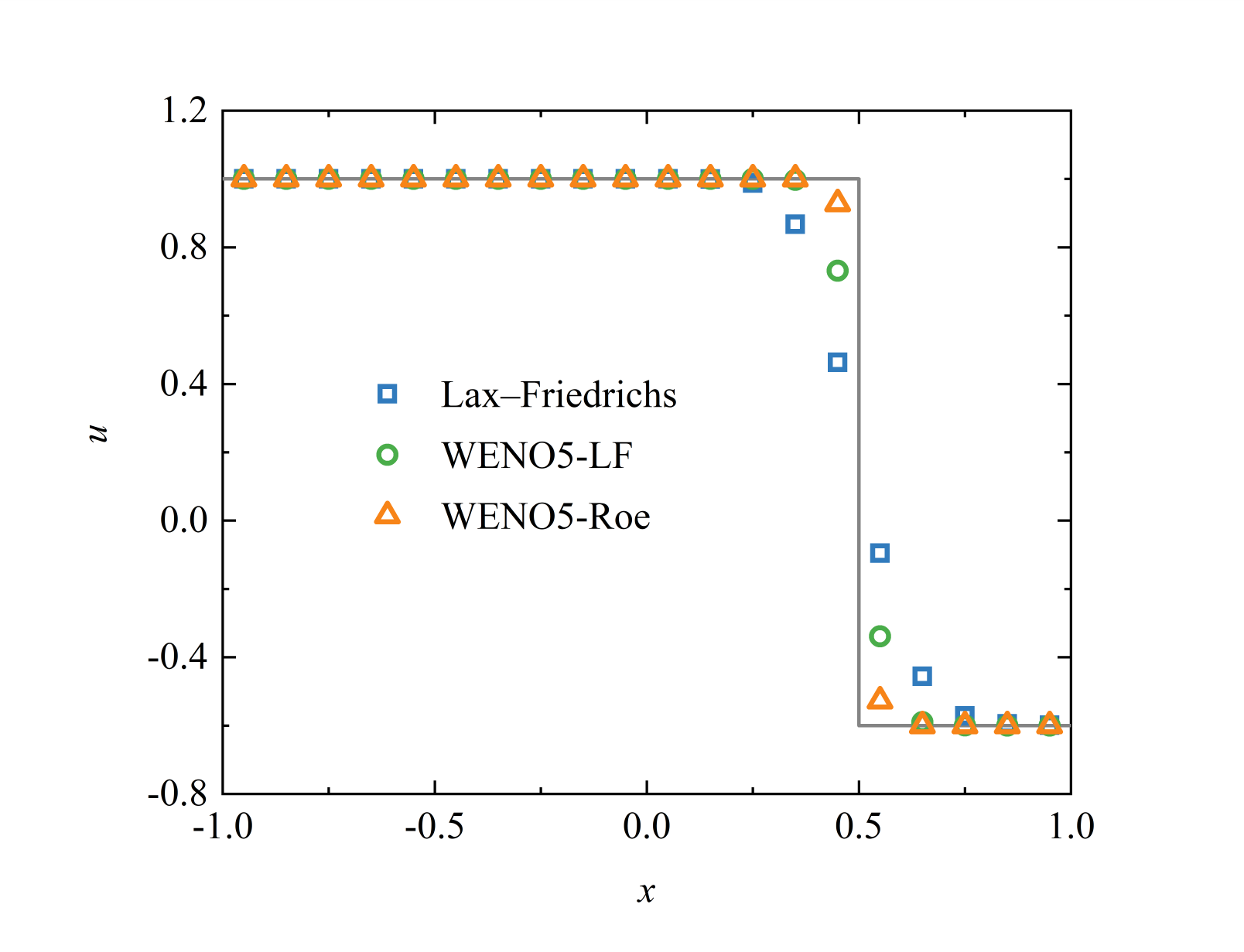}
  }
	\caption{Propagation results of the pre-existing discontinuity problem.}
	\label{fig:BE:shock}
\end{figure}

The first solved problem involves the propagation of a discontinuity initially located at $x = -0.5 \;\mathrm{m}$ with a propagation speed of 0.2 m/s, constituting a Riemann problem for the Burgers equation. Figure \ref{fig:BE:shock} shows the propagation results at $t = 5 \;\mathrm{s}$ obtained using different schemes. The results from the straightforward Lax--Friedrichs (LF) scheme exhibited significant numerical dissipation, leading to a thickness of six grid cells for the discontinuity after 5 s. By incorporating the WENO5-DOLINC reconstruction into the LF scheme, replacing the original method that directly used cell averages for computation, the obtained WENO5-LF scheme significantly improved the numerical dissipation of the LF scheme and enhanced the accuracy of discontinuity capture. Thus, the WENO5-LF scheme achieved a good balance between computational speed and accuracy. Combining the WENO5 reconstruction with a more accurate approximate Riemann solver yielded better performance. For example, as shown in the figure, the results from the WENO5-Roe scheme demonstrate a further improvement in accuracy compared to the WENO5-LF scheme. The waveform obtained by WENO5-Roe closely matched the theoretical discontinuity, and the error was confined to one grid cell on either side of the discontinuity.

\begin{figure}
	\centering
  \fbox{
    \parbox[b]{16cm}
    {
      \includegraphics[width=7.9cm]{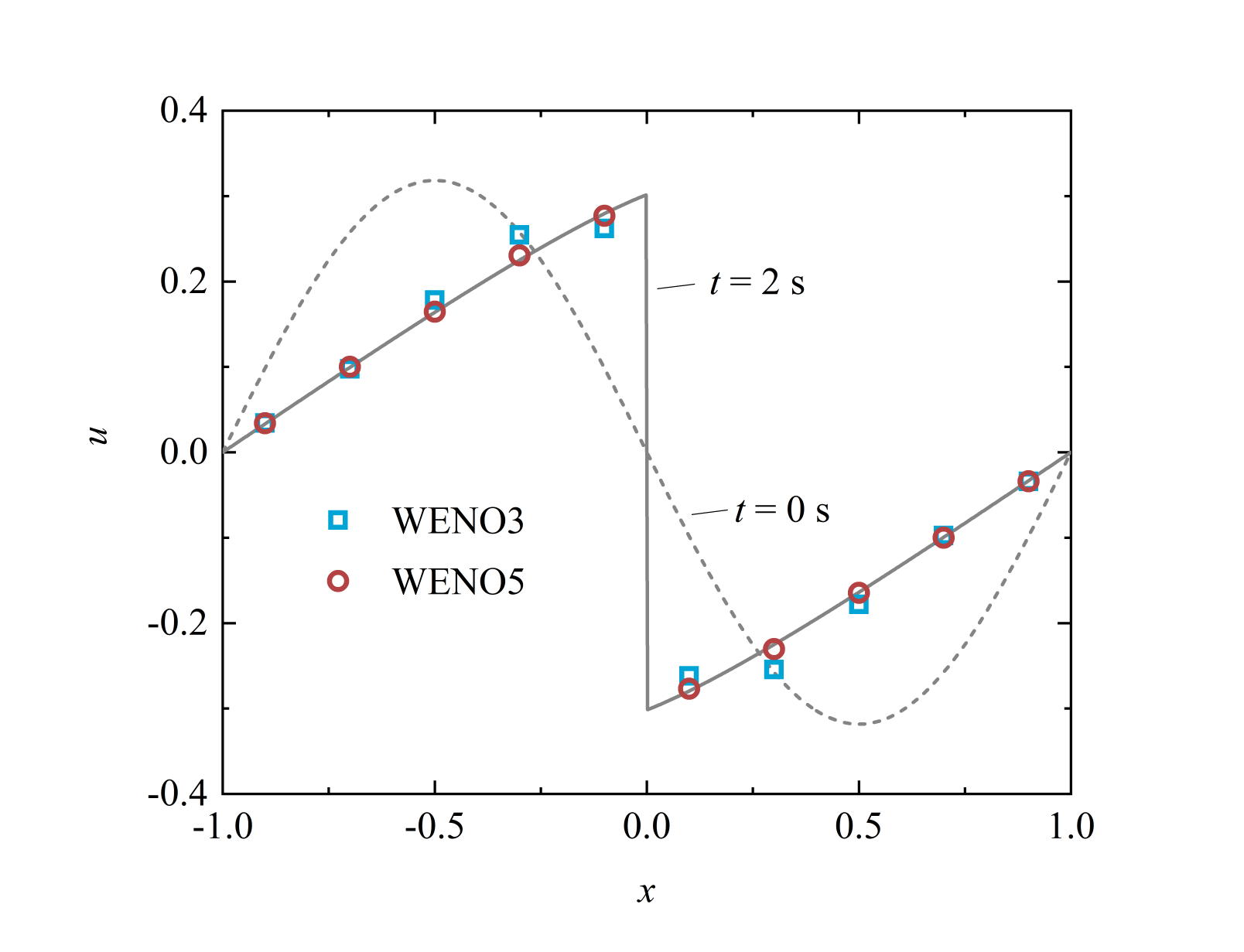}
      \includegraphics[width=7.9cm]{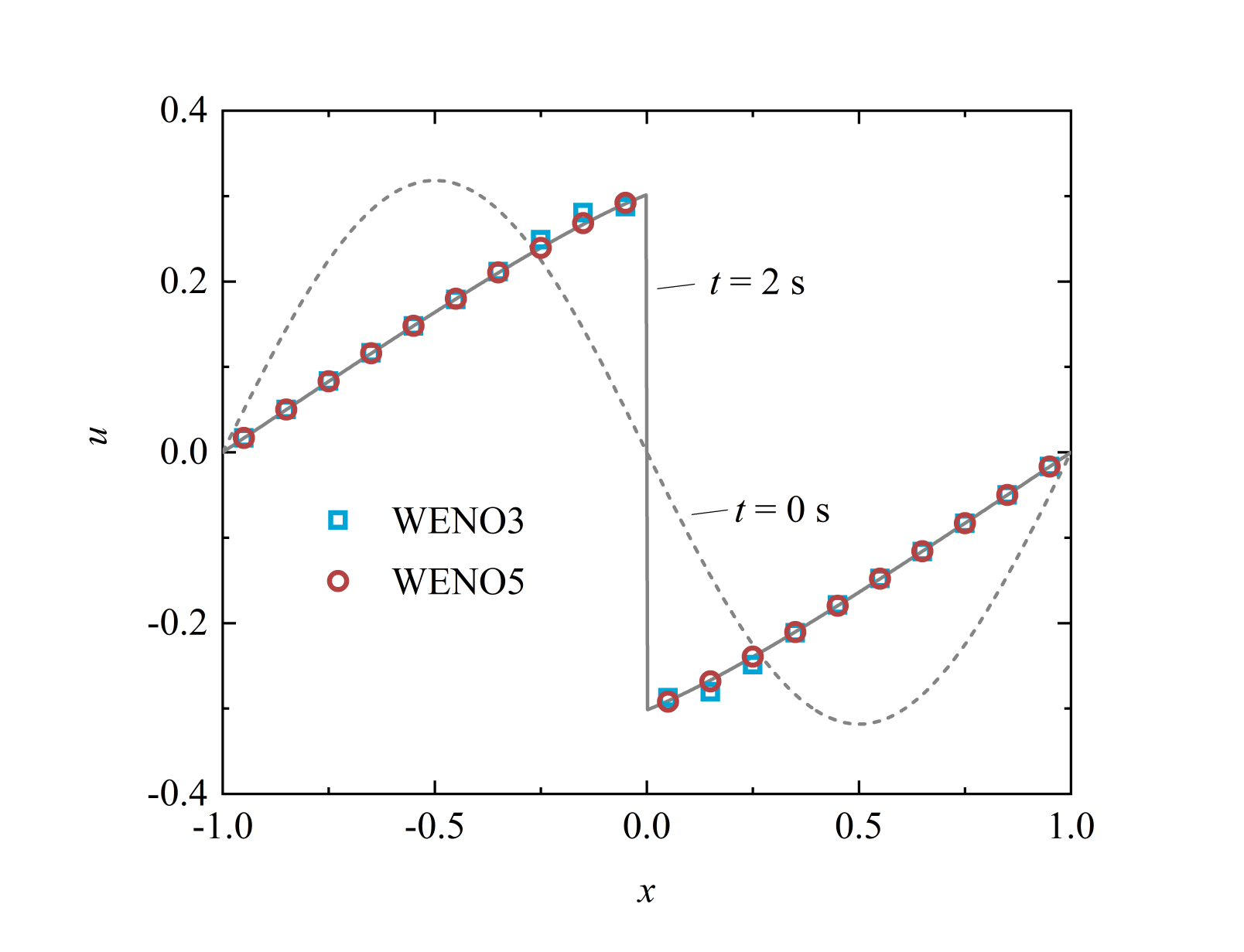}
      \par
      \centering \; (\textit{a}) \hspace{7.4cm} (\textit{b})
    }
  }
	\caption{Results of discontinuity generation in a smooth initial field under different grid resolutions. (\textit{a}) Coarse gird and (\textit{b}) fine grid.}
	\label{fig:BE:sine}
\end{figure}

The second problem solved employed a sine function as the initial field to study the discontinuity-generation process in a smooth field. The computational results are presented in Figure \ref{fig:BE:sine}, where the dashed lines represent the initial field distributions. The results for different grid resolutions and various orders of WENO methods were compared. The WENO5 scheme, implemented based on the DOLINC method, accurately captured the generated discontinuity even on coarse grids and exhibited an accuracy comparable to that of finer grids. The lower-order WENO3 scheme performed well on finer grids when coupled with the solver, but errors near the discontinuity increased marginally on coarser grids compared to the precise results obtained with WENO5. The non-oscillatory finite-volume scheme implemented using the DOLINC method on unstructured grids showed that increasing the order of the scheme can achieve high-precision results with a lower grid density, highlighting the accuracy of the DOLINC method in the inversion calculation.

\subsection{Solving the Euler Equations}
\label{ssec:Euler}

Finally, we apply the DOLINC scheme to solve the inviscid-fluid governing equations, known as the Euler equations. The Euler equations for hyperbolic conservation systems are as follows:

\begin{equation}
  \begin{gathered}
    \partial_t \rho + ( \rho u_j )_{,\,j} = 0,  \\
    \partial_t (\rho u_i) + ( \rho u_i u_j + p \delta_{ij})_{,\,j} = 0,  \\
    \partial_t (\rho E) + ( \rho u_j E + u_j p )_{,\,j} = 0.
  \end{gathered}
\end{equation}

x-split multi-dimensional Euler equations have the following form:

\begin{equation}
  \begin{gathered}
    \frac{\partial \rho}{\partial t} + \frac{\partial \rho u}{\partial x} = 0,  \\
    \frac{\partial \rho u}{\partial t} + \frac{\partial \rho u u}{\partial x} + \frac{\partial p}{\partial x} = 0,  \\
    \frac{\partial \rho v}{\partial t} + \frac{\partial \rho u v}{\partial x} = 0,  \\
    \frac{\partial \rho w}{\partial t} + \frac{\partial \rho u w}{\partial x} = 0,  \\
    \frac{\partial \rho E}{\partial t} + \frac{\partial (\rho E + p)u}{\partial x} = 0.
  \end{gathered}
\end{equation}

Euler equations for one-dimensional problems can be simplified to

\begin{equation}
  \begin{gathered}
    \frac{\partial \rho}{\partial t} + \frac{\partial}{\partial x} (\rho u) = 0,  \\
    \frac{\partial \rho u}{\partial t} + \frac{\partial}{\partial x}(\rho u^2 + p) = 0,  \\
    \frac{\partial \rho E}{\partial t} + \frac{\partial}{\partial x}(\rho uE + pu) = 0.
  \end{gathered}
\end{equation}

To better demonstrate the improvement in numerical accuracy achieved by the DOLINC scheme, the multi-dimensional Euler equations solvers discussed in this section do not employ approximate Riemann solvers, such as the HLLC or Roe schemes. Instead, they are combined with a more generally efficient LF scheme or local LF scheme \citep{Kurganov:2001, Kurganov:2000}. This treatment highlights the enhanced computational accuracy of the DOLINC method for low-order finite-volume methods on unstructured grids.

\subsubsection{One-dimensional Riemann Problem}

\begin{figure}
	\centering
  \fbox{
    \parbox[b]{16cm}
    {
      \includegraphics[width=7.9cm]{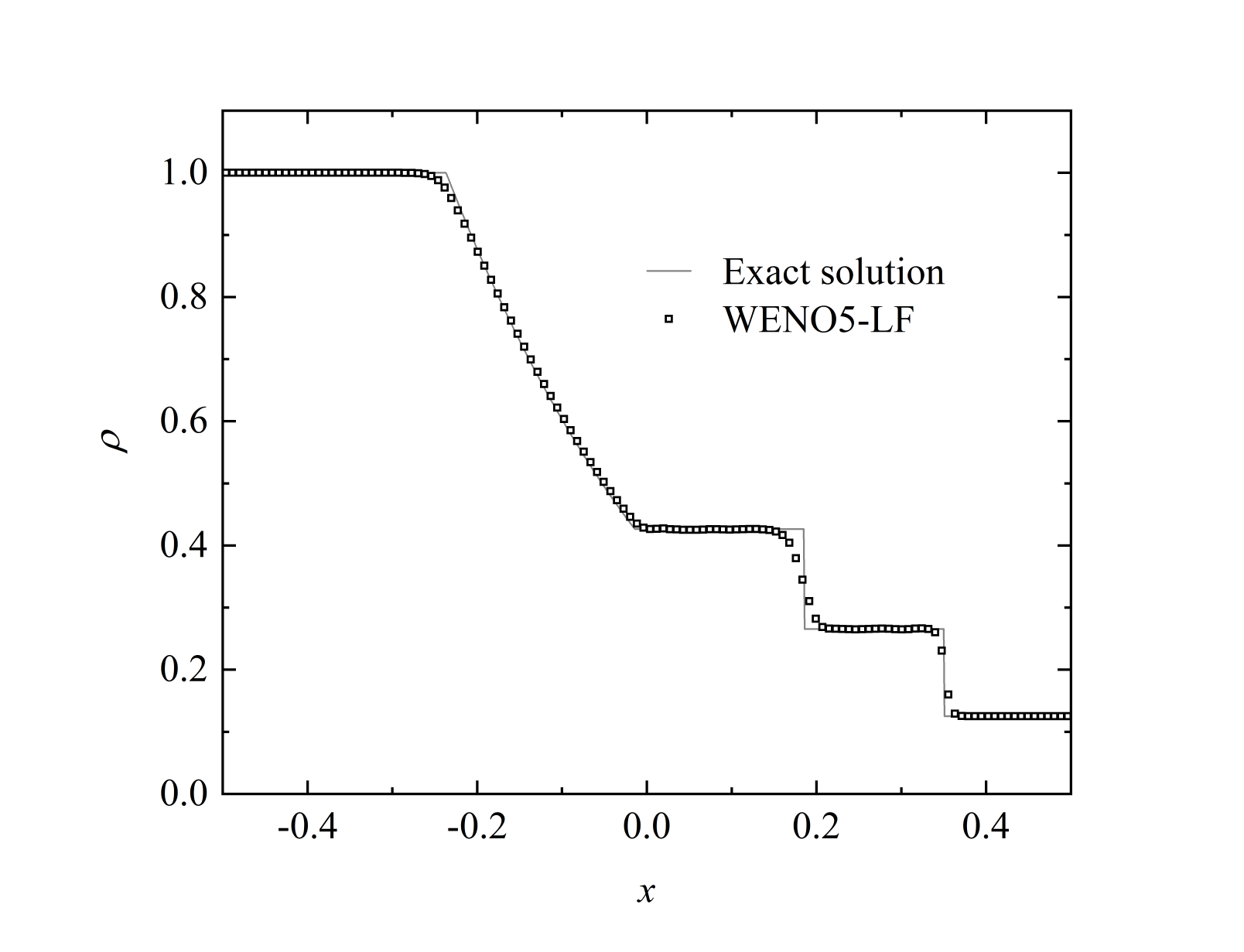}
      \includegraphics[width=7.9cm]{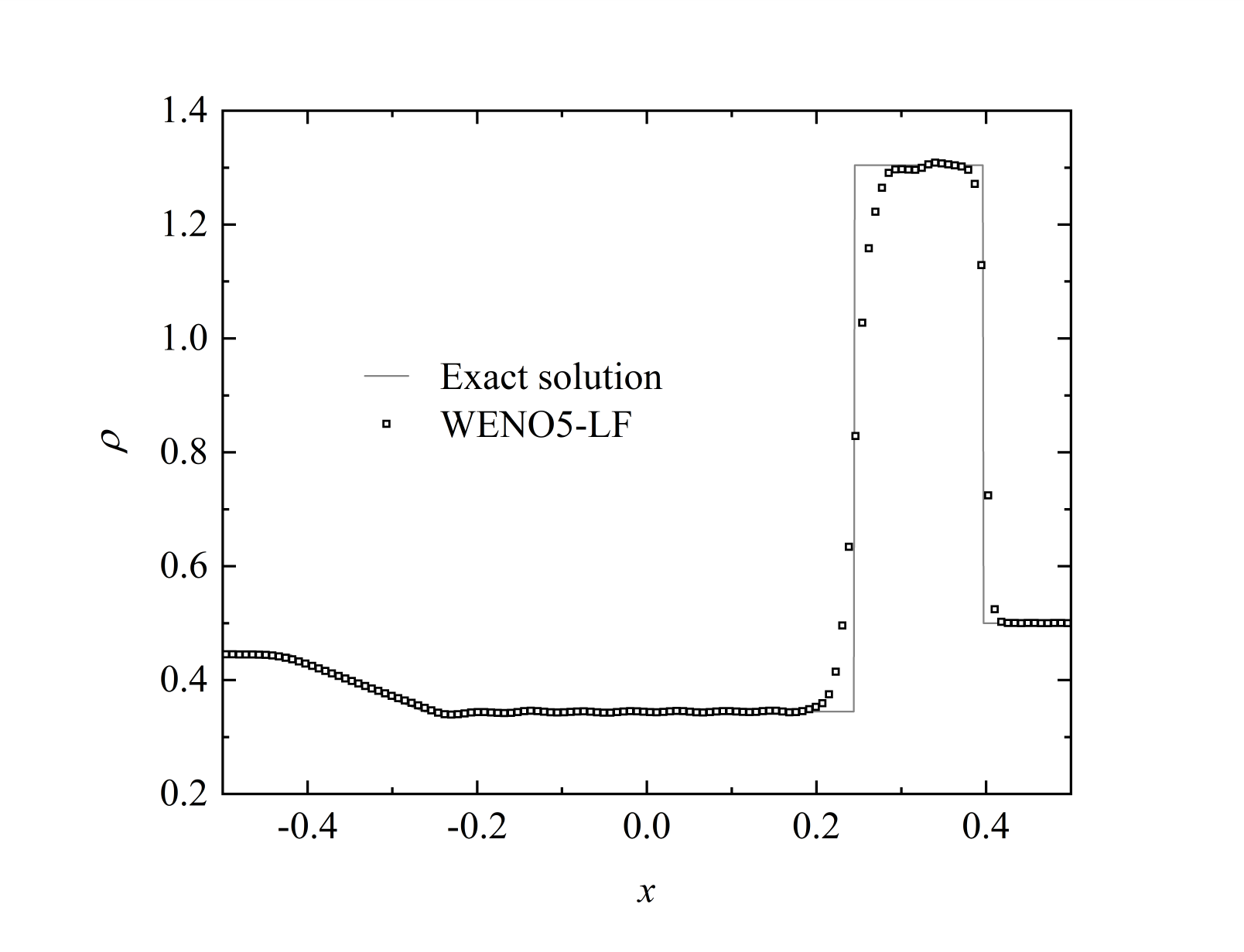}
      \par
      \centering \; (\textit{a}) \hspace{7.4cm} (\textit{b})
    }
  }
	\caption{Density of the one-dimensional Riemann problem. (\textit{a}) Sod problem at 0.20 s, (\textit{b}) Lax problem at 0.16 s.}
	\label{fig:R1D:SL}
\end{figure}

Figure \ref{fig:R1D:SL} illustrates the computational results for two classical one-dimensional Riemann problems: the Sod \citep{Sod:1978} and Lax problems \citep{Lax:1954}. Overall, even with the increased numerical viscosity introduced by the LF scheme, the WENO5-LF Riemann problem solver maintained good accuracy. The captured expansion and shock waves in both the Sod and Lax problems appeared clean and clear in the plot, with only a marginal dissipation observed at the contact discontinuity.

\begin{figure}
	\centering
  \fbox{
    \parbox[b]{16cm}
    {
      \includegraphics[width=7.9cm]{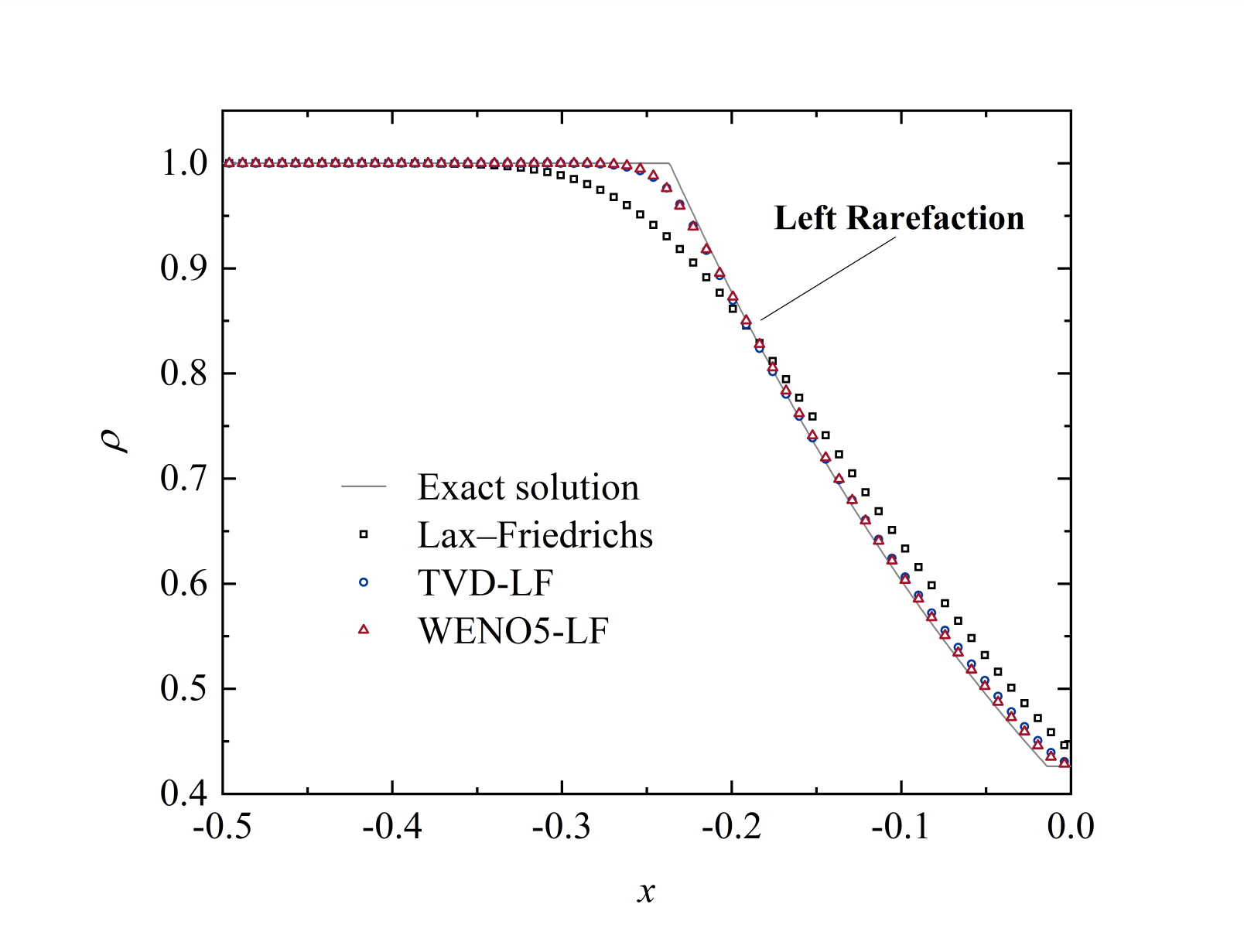}
      \includegraphics[width=7.9cm]{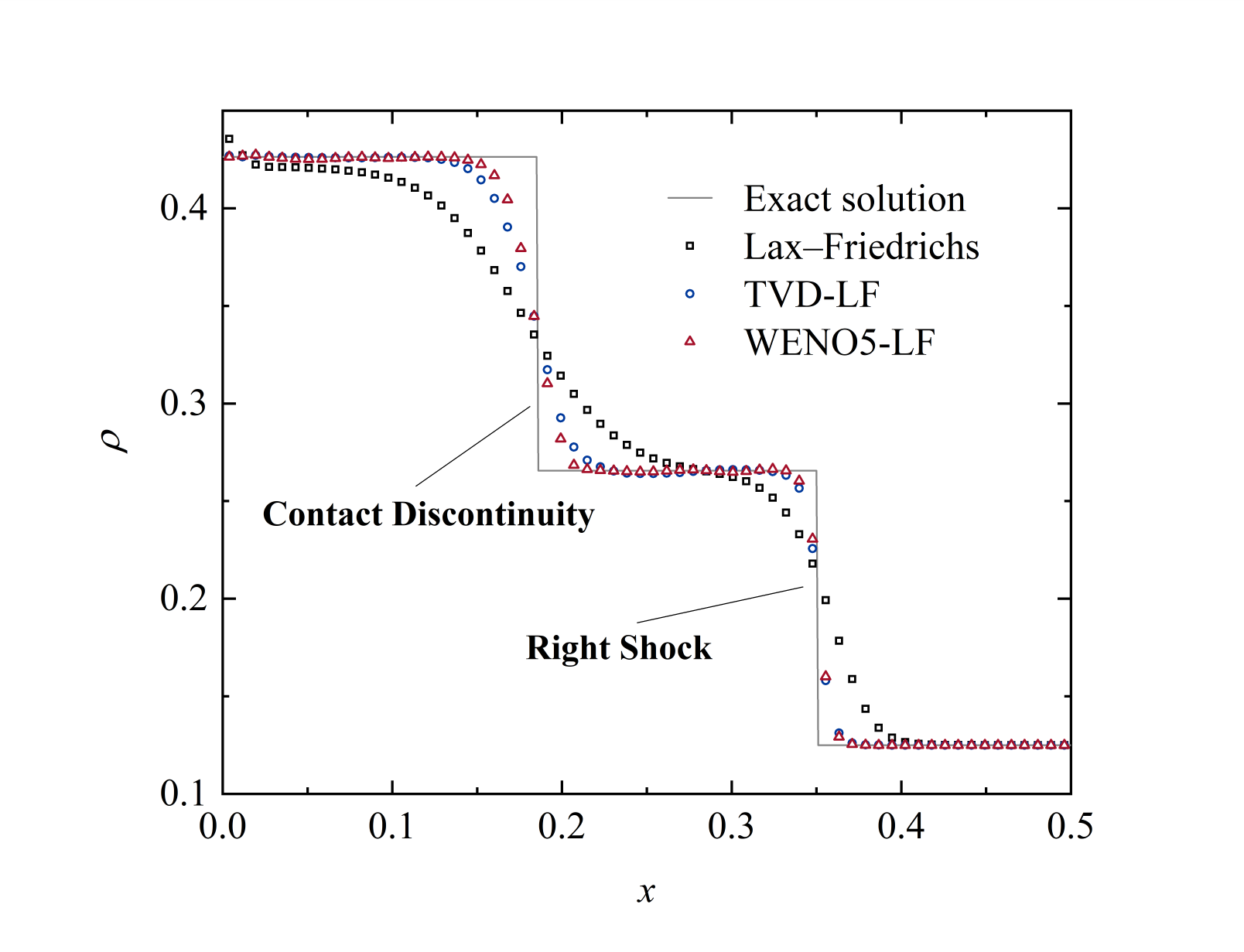}
      \par
      \centering \; (\textit{a}) \hspace{7.4cm} (\textit{b})
    }
  }
	\caption{Results of the Sod problem (one-dimensional shock tube) using different schemes for the Riemann problem. (\textit{a}) Expansion wave on the left, (\textit{b}) shock and contact discontinuity on the right.}
	\label{fig:R1D:SodRL}
\end{figure}

A further comparison of the LF, TVD-LF, and WENO5-LF schemes in terms of computational details is presented in Figure \ref{fig:R1D:SodRL}. In the capture of the left-propagating expansion wave in Figure \ref{fig:R1D:SodRL}(\textit{a}), the combination of the LF scheme with the WENO5 scheme noticeably reduced the numerical dissipation. In Figure \ref{fig:R1D:SodRL}(\textit{b}), which depicts the capture of the right-propagating contact discontinuity and shock wave, the LF scheme with WENO5 reconstruction achieved sharper discontinuity interfaces. The computational accuracy of the TVD scheme was marginally lower than that of the WENO5 scheme; however, overall, the accuracies remained very close, reaffirming that the shortcomings of the TVD scheme are not in capturing discontinuities in Riemann problems but rather in the previously discussed issue of extrema reduction.

\subsubsection{Two-dimensional Riemann Problem}

\begin{figure}
	\centering
  \fbox{
    \includegraphics[width=8.0cm]{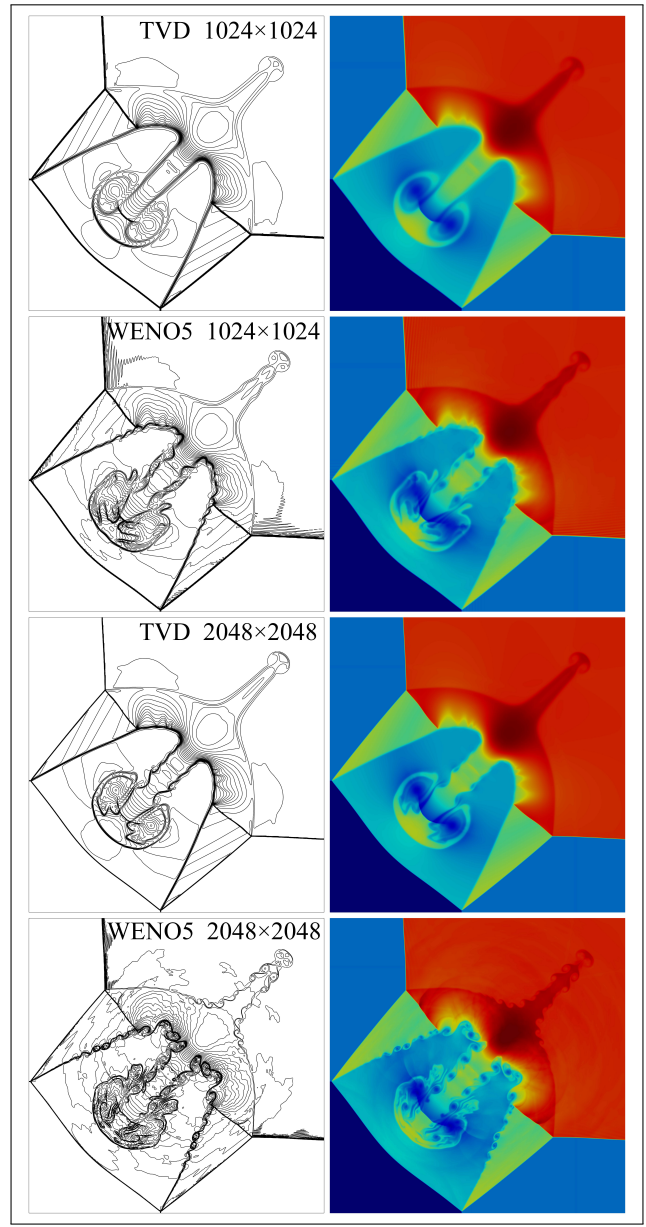}
  }
	\caption{Density of the two-dimensional Riemann problem at 1.1 s (contour lines on the left range from 0.15 to 1.7 with 31 lines; the colormap on the right ranges from 0.14 to 1.75).}
	\label{fig:R2D:compare}
\end{figure}

The two-dimensional Riemann problem proposed by \citet{Schulz:1993} has been widely used to validate the accuracy of numerical methods in studies related to high-order schemes and Riemann solvers \citep{Balsara:2014, Brio:2001, Fleischmann:2019}. However, this problem posed significant challenges for most industrial software packages; thus, achieving high-precision flow substructures using high-order methods is difficult. Although low-order FVM methods can accurately compute shockwaves/expansion waves developed in the flow field, they fail to correctly capture the vortex street generated by the Kelvin--Helmholtz instability owing to their substantial numerical dissipation. The performance difference of the TVD scheme mentioned earlier in capturing the flow-field discontinuities and small-scale flow structures is evident here. In Figure \ref{fig:R2D:compare}, the TVD scheme (based on the VanAlbada limiter) significantly suppressed the Kelvin--Helmholtz instability. In the case with a lower resolution of $1024^2$ grids, the TVD scheme almost eliminated all possible unstable vortex streets associated with the Mach stem and contact discontinuity. The flow characteristics included a mushroom cap and two-dimensional shockwaves. Even when the TVD scheme was applied with a resolution of $2048^2$ grids, the vortex street did not form on either side of the contact discontinuity, and only the Mach stem exhibited marginal fluctuations. However, when the WENO5-DOLINC scheme was used for computation at a resolution of $1024^2$ grids, it successfully captured the Kelvin--Helmholtz instability of both the Mach stem and contact discontinuity. The coherent vortex structures on the coarse grid with the WENO5 scheme were clearly visible and surpassed the results achieved by the TVD scheme on a finer grid of $2048^2$ grids. As the WENO5 scheme was further increased to a resolution of $2048^2$ grids, the flow details induced by the Kelvin--Helmholtz instability became clearer, the unstable position of the contact discontinuity advanced, and vortices emerged within the mushroom cap. At this point, the capture results of the TVD scheme with $2048^2$ grids could not compete with those achieved by the WENO5 scheme with the same resolution.

\begin{figure}
	\centering
  \fbox{
    \includegraphics[width=8.0cm]{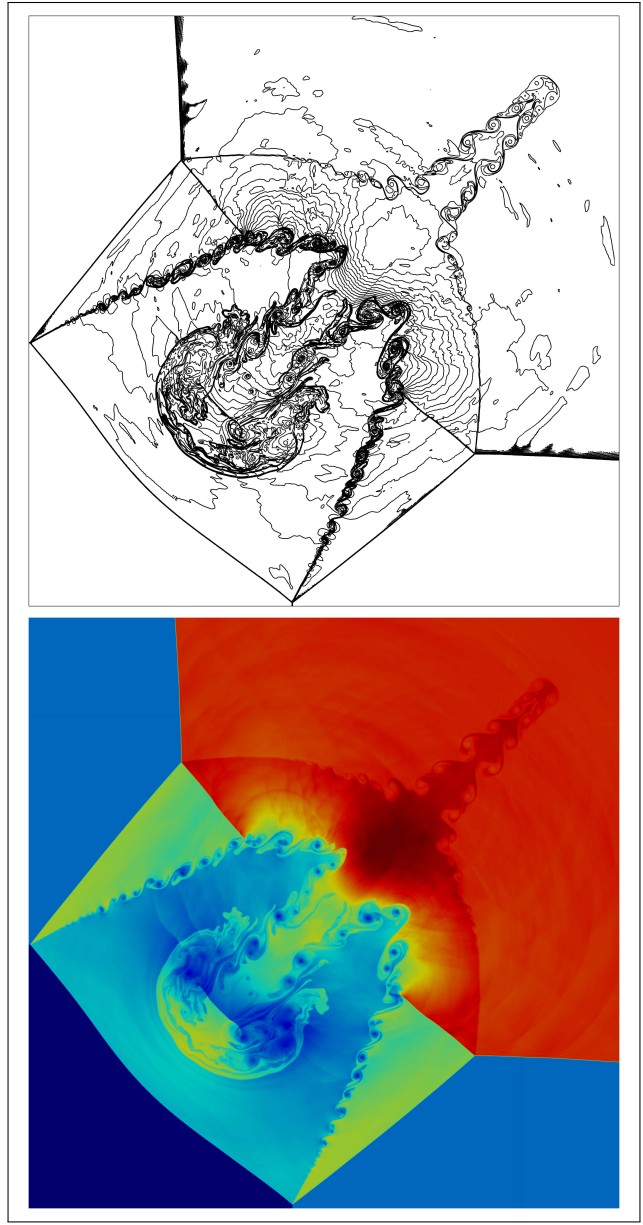}
  }
	\caption{Contour lines and colormap of density obtained through WENO5 scheme at the extreme resolution of $4096\times4096$.}
	\label{fig:R2D:4096}
\end{figure}

High-order schemes implemented based on the DOLINC method also have an advantage over the TVD scheme in terms of efficiency in improving computational accuracy. When 64 processors were used for parallel computation, the CPU time for the TVD scheme with a resolution of $1024^2$ grids was 1290.28 s. Using this as a benchmark for accuracy and speed, only approximate capture results were obtained. When the grid was refined to a resolution of $2048^2$ for more flow details, the computation time increased to 15599.9 s, nearly 11.09 times the baseline cost for additional accuracy. However, when the resolution was maintained at $1024^2$ grids and the high-order WENO5 scheme was chosen for computation, the total time was only 4880.49 s, increasing the accuracy at an additional cost of only 2.78 times the baseline. As discussed in the analysis of Figure \ref{fig:R2D:compare}, the coarse-grid WENO5 not only surpassed fine-grid TVD in terms of accuracy, but also consumed only 31\% of the computation time of the latter, reducing the cost of accuracy improvement by almost 75\%. This again demonstrated that high-order schemes for unstructured grids based on the DOLINC method had better computational efficiency than the existing schemes in conventional low-order FVM architectures. The advantages of DOLINC, in terms of accuracy and speed, along with its relatively simple integration, are expected to accelerate the transition from industrial software to high-order methods. Figure \ref{fig:R2D:4096} shows the computational results of the DOLINC scheme at an extreme grid resolution. The capturing of vortex street instability is comparable to the results obtained using high-order schemes in structured-grid high-precision solvers in previous research \citep{Fleischmann:2019}, reflecting the significant improvement in the accuracy limits achieved by the DOLINC method for low-order FVM methods.

\subsubsection{Rayleigh--Taylor Instability}
\label{sssec:RTI}

\begin{figure}
	\centering
  \fbox{
    \includegraphics[width=8.0cm]{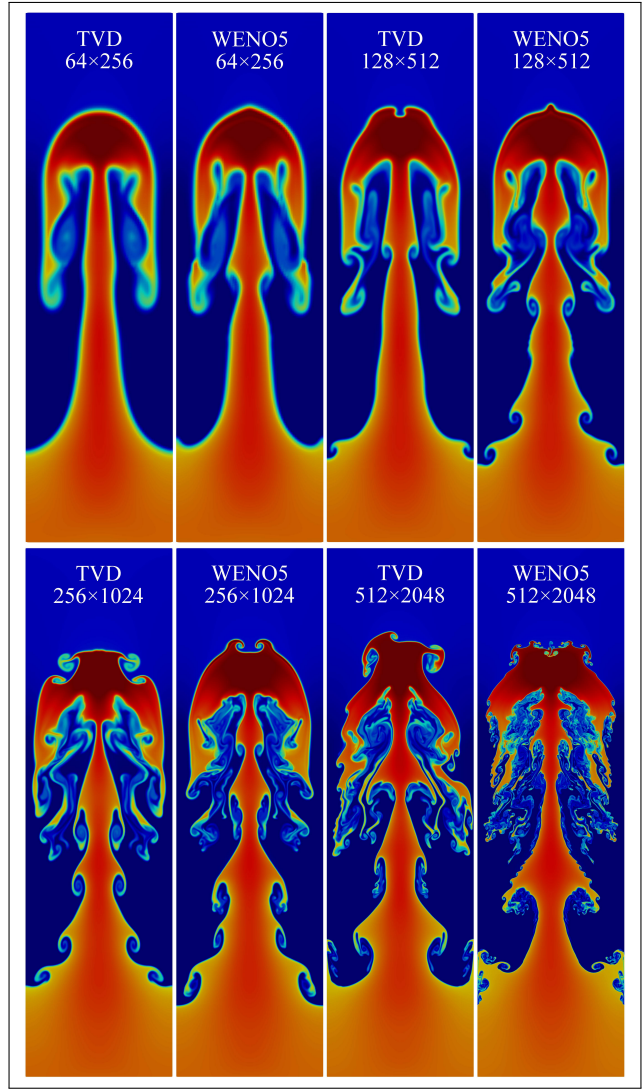}
  }
	\caption{Density colormap at 1.95 s for the Rayleigh--Taylor instability problem (colormap ranges from 0.85 to 2.25).}
	\label{fig:RTI:color}
\end{figure}

\begin{figure}
	\centering
  \fbox{
    \includegraphics[width=8.0cm]{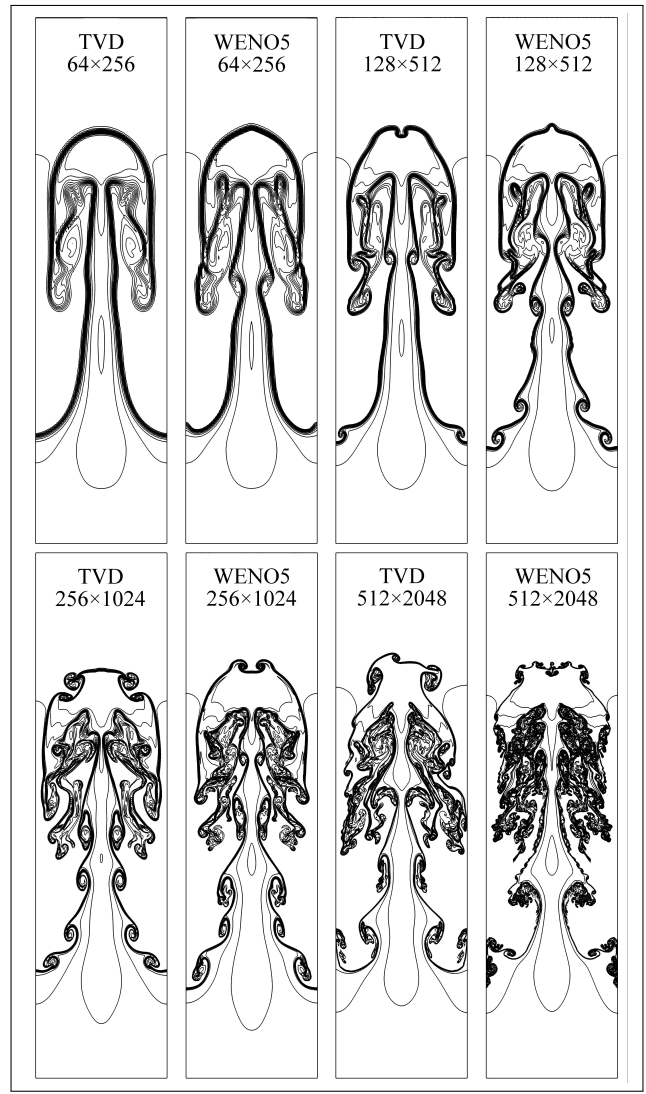}
  }
	\caption{Density contour lines at 1.95 s for the Rayleigh–Taylor instability problem (contour lines range from 0.95 to 2.15 with 12 lines).}
	\label{fig:RTI:line}
\end{figure}

\begin{figure}
	\centering
  \fbox{
    \includegraphics[width=8.0cm]{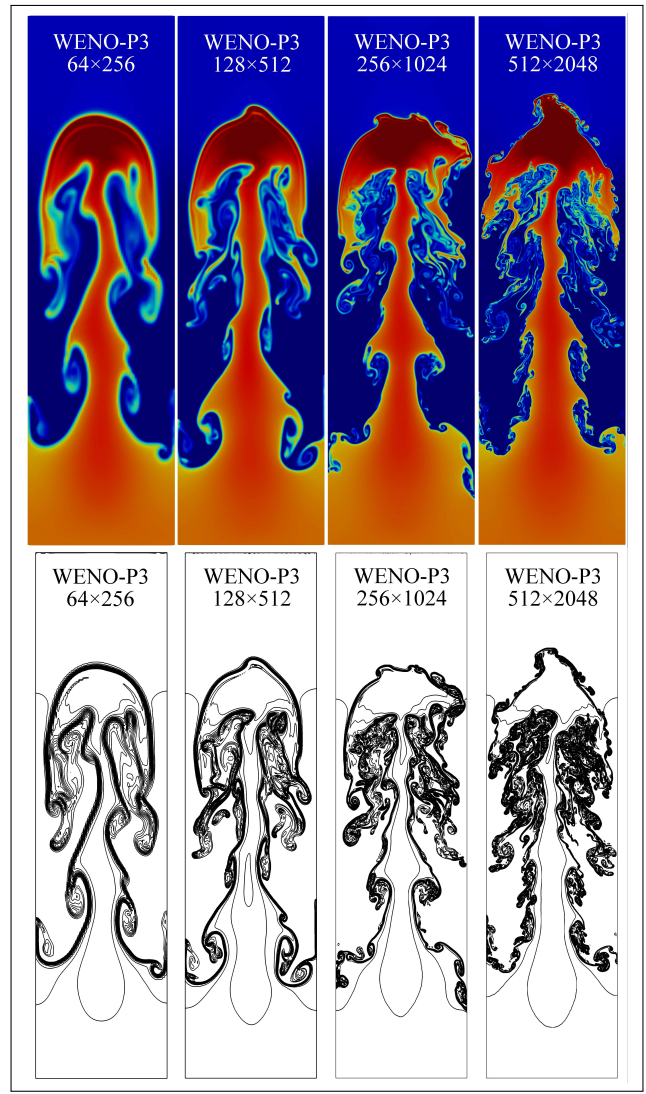}
  }
	\caption{Colormap and contour lines of density obtained using the WENO-P3 scheme.}
	\label{fig:RTI:EXT}
\end{figure}

Another widely studied case in high-order methods research is the Rayleigh--Taylor instability problem \citep{Fleischmann:2019, Remacle:2003, Shi:2003}. In contrast to Riemann problems, the Rayleigh--Taylor instability is induced by gravitational effects on two fluids of different densities. This involves a lower overall Mach number and primarily focuses on the initiation and evolution of fine vortex structures. Because this problem has been extensively described in the literature, we did not go into repetitive details. Specific settings are available in the cited literature and are well-known. The initial perturbation at time zero was set as follows:

\begin{equation}
  v = \varepsilon \cos(8\pi x) \sin^{\tau}(\pi y),
\end{equation}

where $\varepsilon = \mathrm{Ma}_0 \times a_0$. We take $\tau = 6$, $a_0 = \sqrt{0.5\gamma}$, and $\mathrm{Ma}_0 = 0.1$. In contrast to the Riemann solvers used in other studies and LF-based Euler equation solver used in the other cases in this section, this case employs the native rhoPimpleFoam solver based on a pressure-coupling algorithm to showcase the effects of the combinations of different types of solvers with the DOLINC scheme. The calculations were carried out using the TVD scheme, WENO5 scheme implemented based on the DOLINC method, and WENO-P3 scheme based on the k-exact method and Gaussian integration, as shown in Figures \ref{fig:RTI:color}, \ref{fig:RTI:line} and \ref{fig:RTI:EXT}.

In this low-Mach compressible flow problem, the absence of physical viscosity enabled the capture of finer vortex structures, indicating that the solution method possessed higher numerical accuracy and lower numerical viscosity. As shown in Figure \ref{fig:RTI:color}, at the four grid resolutions, the WENO5 scheme captured more flow substructures than the TVD scheme. The difference in capturing the flow details increased with the continuous improvement in grid accuracy, especially at resolutions of $256\times1024$ and $512\times2048$. Similarly, in Figure \ref{fig:RTI:line}, at a resolution of $256\times1024$, the computational accuracy of WENO5 was comparable to that of TVD at $512\times2048$, whereas the analytical accuracy of the WENO5 results at a resolution of $512\times2048$ significantly surpassed the accuracy of TVD at the same grid resolution.

Another noteworthy aspect is the issue of symmetry in the Rayleigh–Taylor instability structure. Although the initial perturbation is generally symmetric, many schemes/solvers could not consistently maintain flow symmetry throughout the computation. Various factors contributed to the appearance of asymmetry. Notably, even for a well-established scheme, such as TVD (which directly uses the original code from the OpenFOAM distribution in this study), asymmetry still occurred at a grid resolution of $512\times2048$, as shown in Figures \ref{fig:RTI:color} and \ref{fig:RTI:line}. This asymmetry was particularly pronounced in the WENO-P3 scheme, as shown in Figure \ref{fig:RTI:EXT}. The computational results of WENO-P3 exhibited significant asymmetrical features at lower grid resolutions due to the use of multi-grid cell templates and triangular meshing steps on grid faces. In practical computations, the data transfer between processes in parallel computing and certain treatments using linear-equation iterative solvers for ill-conditioned matrices can also contribute to the emergence of asymmetry. Although the WENO-P3 scheme in Figure \ref{fig:RTI:EXT} captures more refined flow characteristics, it cannot achieve symmetric results similar to those of high-order structured grid solvers (Fleischmann et al., 2019). However, the WENO5 scheme implemented based on the DOLINC method demonstrated good symmetry, as shown in Figures \ref{fig:RTI:color} and \ref{fig:RTI:line}, even at resolutions for which the TVD schemes exhibited asymmetric vortex structures. This highlights the advantages of the DOLINC method in implementing high-order unstructured grid schemes. For a more in-depth discussion of the various reasons for asymmetry and methods to improve symmetry, further reading of the relevant literature is recommended \citep{Fleischmann:2019, Remacle:2003}.

\begin{figure}
	\centering
  \fbox{
    \includegraphics[width=8.0cm]{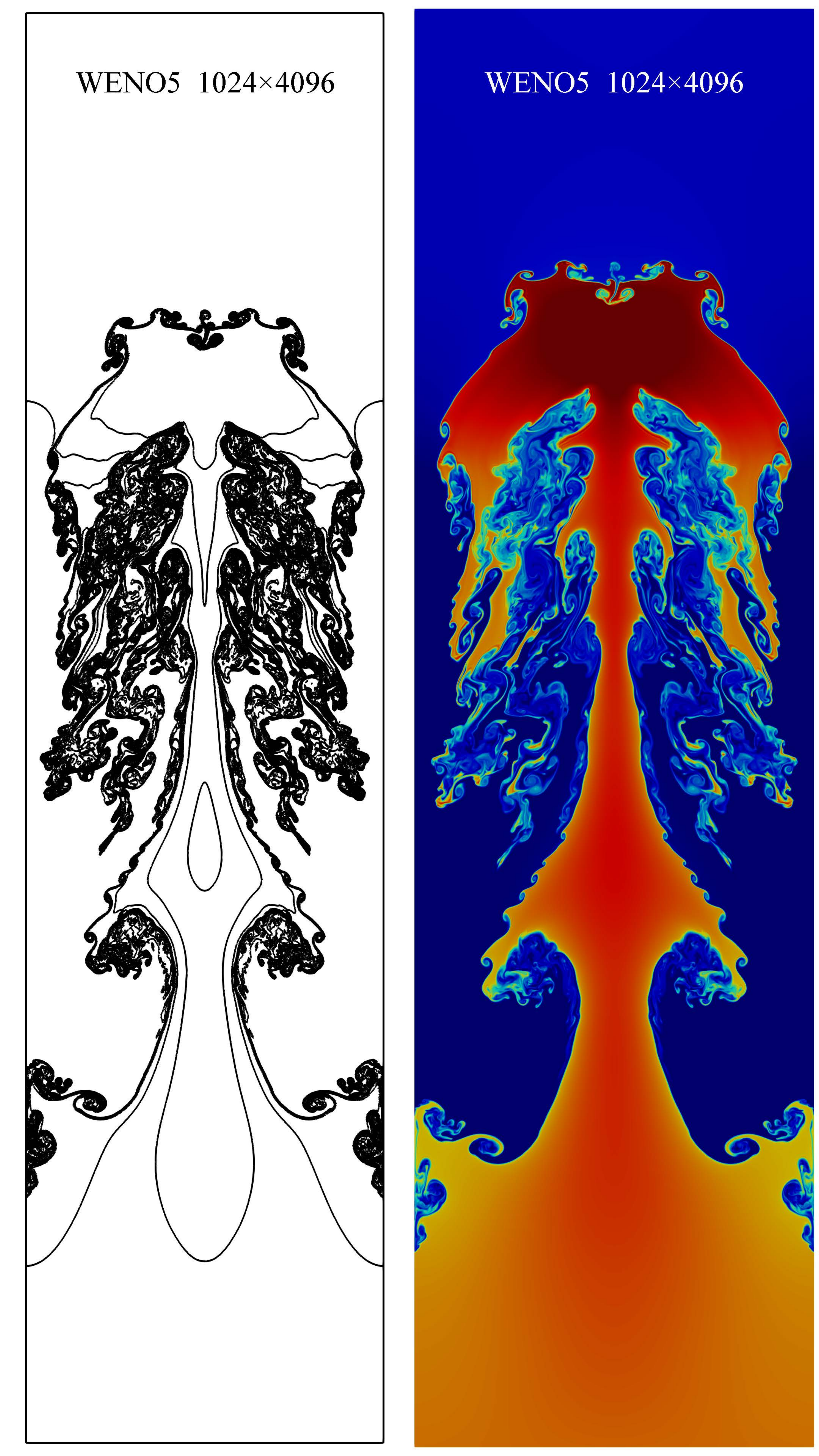}
  }
	\caption{Contour lines and colormap of density obtained using the WENO5 scheme on the $1024\times4096$ grid.}
	\label{fig:RTI:4096}
\end{figure}

In this case, the DOLINC scheme also demonstrated a higher computational efficiency in improving accuracy. The CPU time for the TVD scheme at a grid resolution of $256\times1024$ using 16 cores was 1089.53 s. Considering this as a benchmark for accuracy and speed, when the grid was refined to $512\times2048$ without considering the partial effect of parallel acceleration, the original computation time for the TVD scheme was approximately 12712.68 s, increasing the cost by nearly 10.67 times. Even when considering the impact of parallel acceleration, maintaining the same number of grids per computing thread as in the benchmark case and using 64 cores still requires 3878.17 s. When the WENO5-DOLINC scheme was used for computation based on the same grid resolution of $256\times1024$, then the total time was only 1829.7 s. The cost increased by only 0.68 times compared to the benchmark case and was even faster than the TVD scheme with parallel acceleration at a higher grid resolution. For a high-order unstructured grid scheme implemented using classical k-exact methods and Gauss integration instead of the DOLINC method, the WENO-P3 scheme with similar accuracy at a grid resolution of $256\times1024$ required 11678.9 s. Thus, the cost increased by nearly 9.72 times compared to the benchmark and was almost equivalent to the TVD scheme with a refined grid of $512\times2048$ in the low-order FVM architecture. Considering the complexity of implementing such unstructured grid WENO methods in low-order FVM architectures, we can understand why these methods lack appeal for industrial software. At the same grid resolution, the computation time for WENO5 was only 16\% that of WENO-P3, reducing the additional computational cost by 93 \%. Overall, the DOLINC method had a clear advantage over low-order methods in improving computational-accuracy efficiency and outperformed other high-order unstructured grid methods. Further refining the grid to an ultimate resolution of $1024\times4096$, as shown in Figure \ref{fig:RTI:4096}, the DOLINC scheme exhibited very fine flow-detail capture and better symmetry, achieving high-precision results similar to those of structured-grid high-order solvers based on low-order FVM architectures \citep{Fleischmann:2019, Shi:2003}.

\subsubsection{Scramjet Inlet Simulation}

We test the computational performance of the DOLINC method on more irregular grids and demonstrate its applicability to engineering problems by using the WENO5-DOLINC scheme to compute the flow in the inlet of a more complex scramjet engine. Two different forms of computational domains were utilized to simulate different engine conditions, as shown in Figure 25. Model A was used for the scramjet engine operating under the design conditions with a Mach 8.5 inflow. Three converging compression shocks occurred at the lower edge, point 3. In this case, the polyhedral mesh exhibited a certain degree of skewness and noticeable non-uniformity. Model B was employed for the non-design conditions of the scramjet engine, typically involving a lower cruise Mach number of an aircraft. In this case, the incident position of the converging oblique shock moved forward, striking the lower boundary of the computational domain and requiring the use of non-reflective boundary conditions. The non-uniformity of the mesh in the inlet section of Model B was further enhanced.

\begin{figure}
	\centering
  \fbox{
    \parbox[c][6cm]{12cm}{
      \centering (\textit{a})
      \parbox[c][2.5cm]{10cm}{
        \includegraphics[width=10cm]{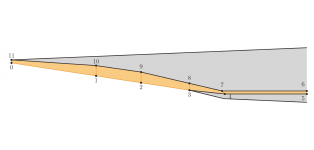}
      }
      \par
      \centering (\textit{b})
      \parbox[c][2.5cm]{10cm}{
        \includegraphics[width=10cm]{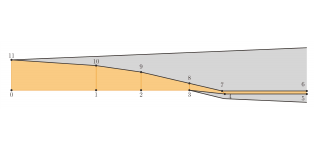}
      }
    }
  }
	\caption{Schematic of the inlet geometry and computational domain. (\textit{a}) Domain model A for the design condition, (\textit{b}) domain model B for the generic condition.}
	\label{fig:SI:geom}
\end{figure}

\begin{figure}
	\centering
  \fbox{
    \includegraphics[width=8.5cm]{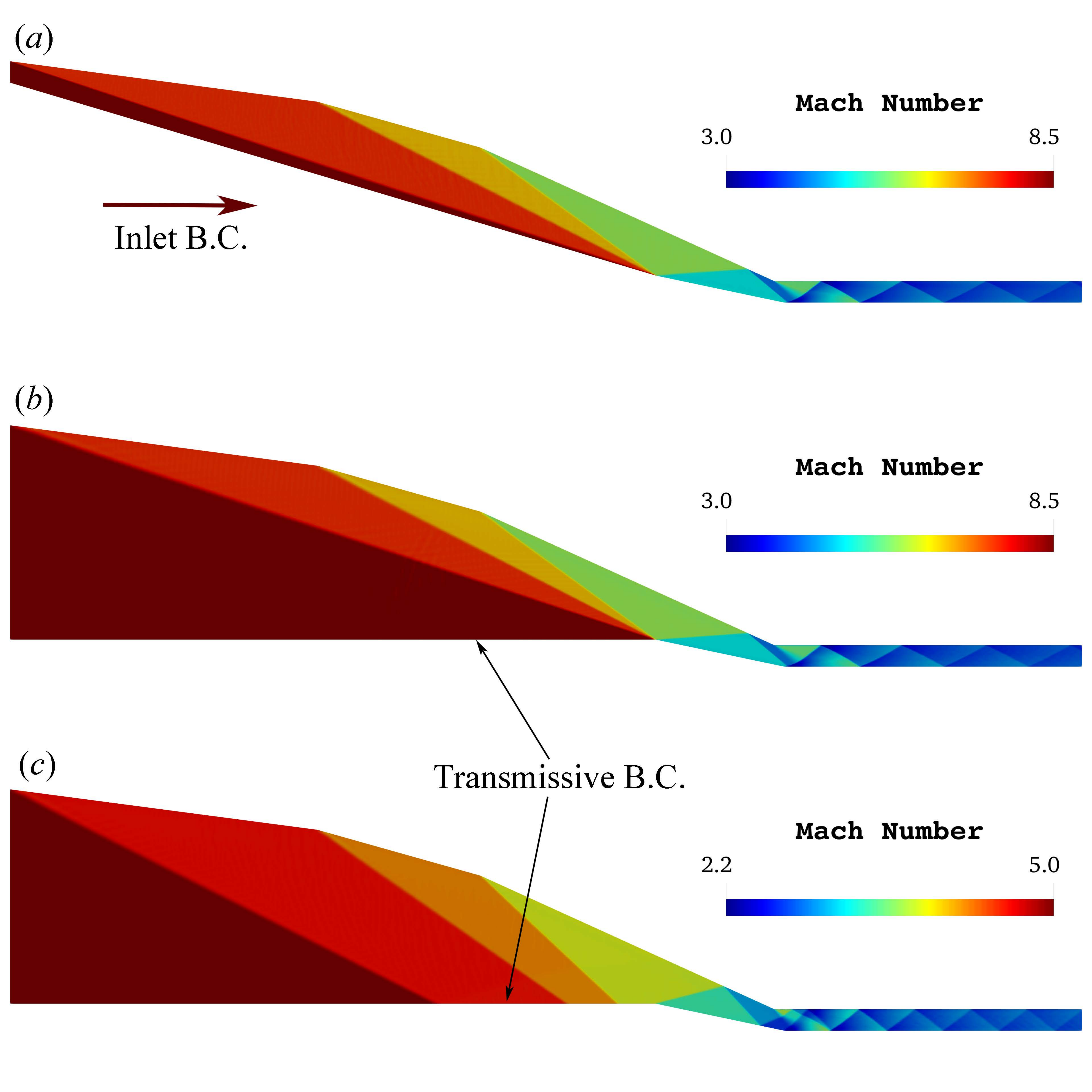}
  }
	\caption{Mach number of the scramjet inlet under different operating conditions. (\textit{a}) Domain model A with inflow Mach number of 8.5, (\textit{b}) domain model B with inflow Mach number of 8.5, and (\textit{c}) domain model B with inflow Mach number of 5.0.}
	\label{fig:SI:Ma}
\end{figure}

The Mach number contours obtained using the WENO5-DOLINC scheme for the design condition with an incoming Mach number of 8.5 and non-design condition with an incoming Mach number of 5.0 are shown in Figure \ref{fig:SI:Ma}. The scales of the horizontal and vertical axes were adjusted to facilitate the display of the morphological structure of the shock/expansion wave systems. For the design condition, the various wave structures in both Models A and B were consistent, confirming that the designed scramjet-engine geometry achieved the intended goal of correctly converging three oblique shocks at the lower-wall front point at the Mach 8.5 inflow. In the results for the Mach 5.0 inflow in Figure \ref{fig:SI:Ma}(\textit{c}), the non-reflective bottom boundary smoothly passed through the oblique compression shock without forming any reflected shock waves to interfere with the downstream computational domain. Overall, even in grids with significant non-uniformity and some skewness, the DOLINC method could still obtain high-precision results. The DOLINC scheme demonstrated high accuracy in capturing multiple complex reflection processes and the interaction between shock/expansion waves in the isolator section without unexpected dissipation or oscillations owing to the large grid-scale differences between the inlet and isolator sections.

\begin{figure}
	\centering
  \fbox{
    \parbox[b]{16cm}
    {
      \includegraphics[width=7.9cm]{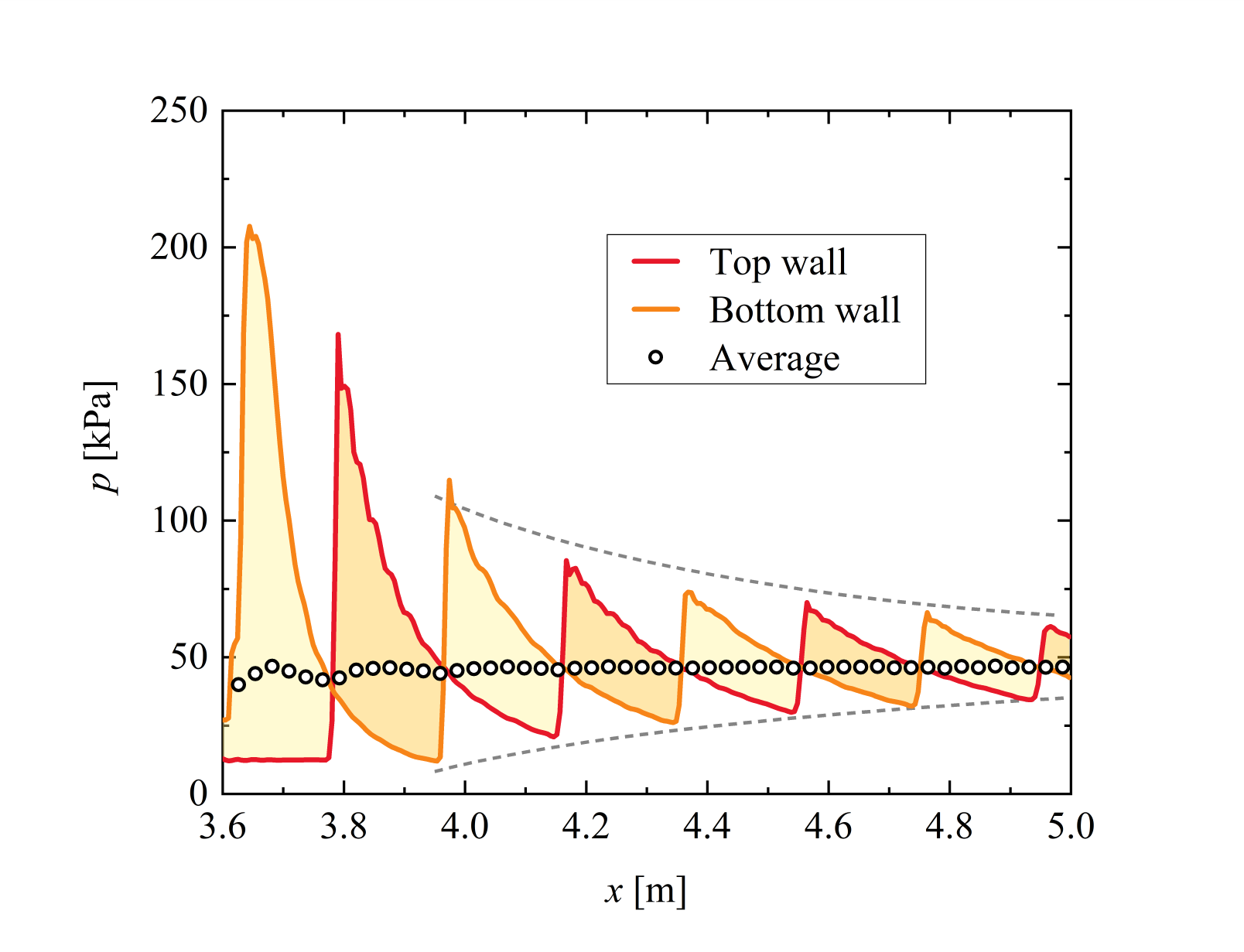}
      \includegraphics[width=7.9cm]{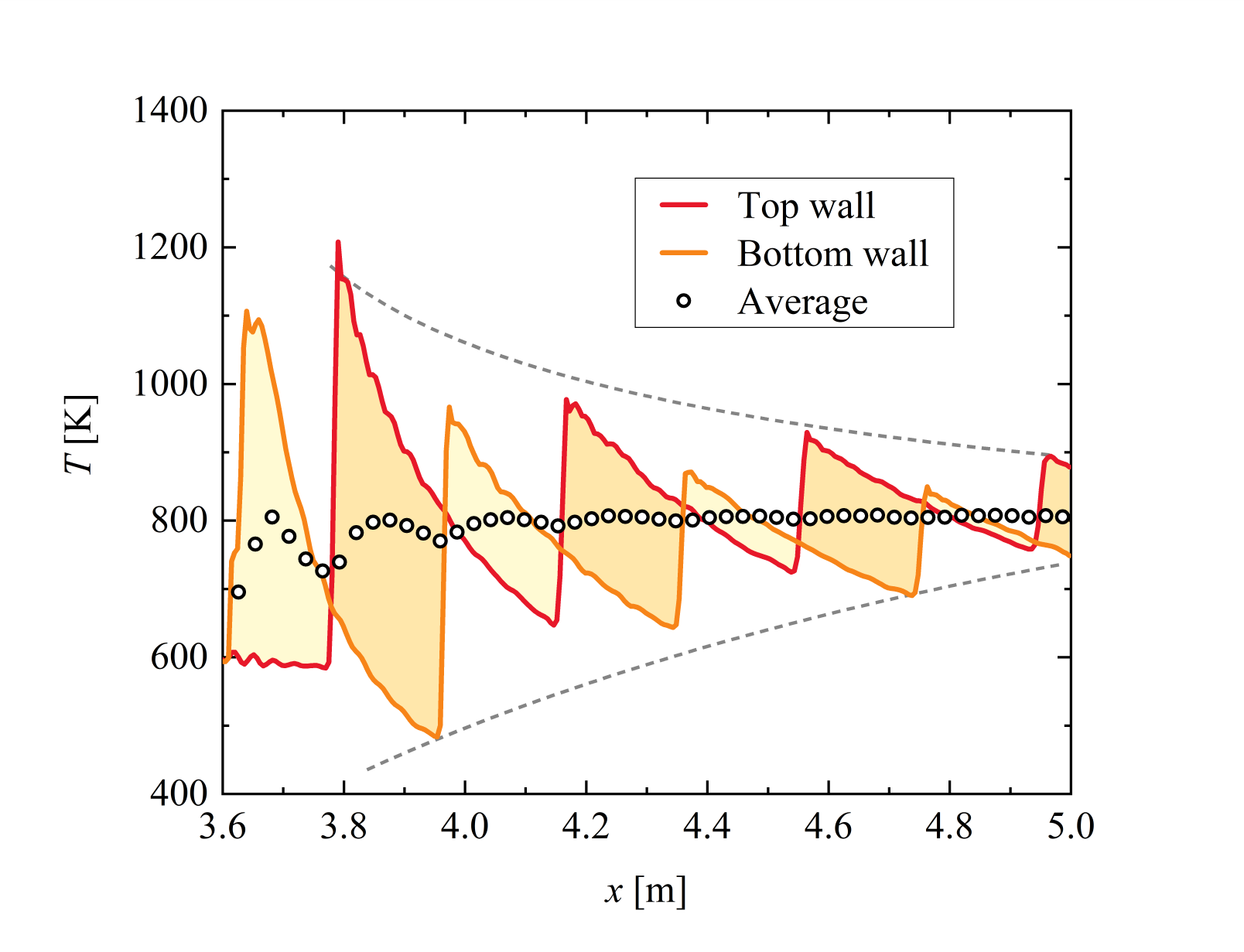}
      \par
      \centering \; (\textit{a}) \hspace{7.4cm} (\textit{b})
    }
  }
	\caption{Parameter distribution along isolator path under Mach 8.5 inflow conditions. (\textit{a}) Static pressure, (\textit{b}) static temperature.}
	\label{fig:SI:lineMa8}
\end{figure}

\begin{figure}
	\centering
  \fbox{
    \parbox[b]{16cm}
    {
      \includegraphics[width=7.9cm]{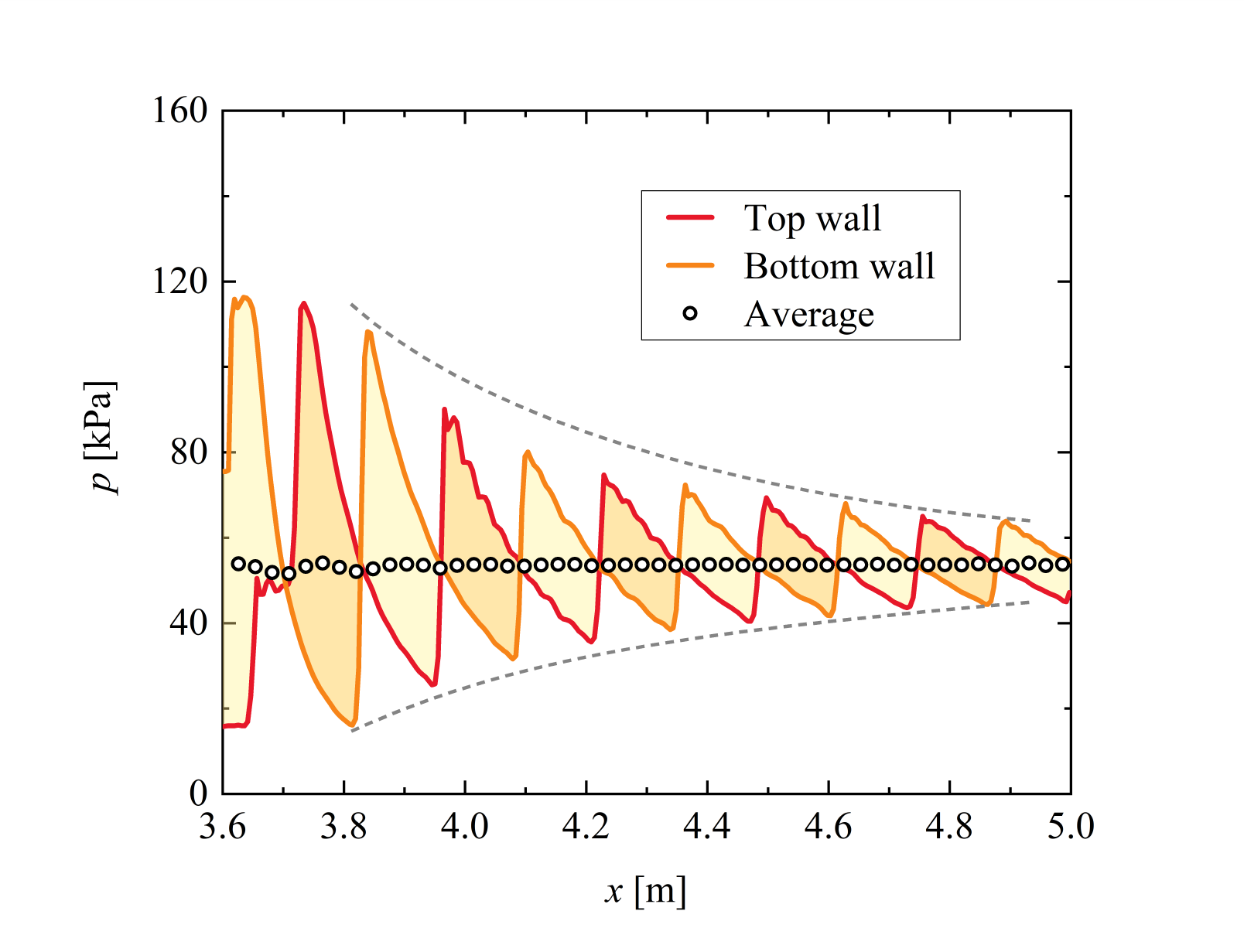}
      \includegraphics[width=7.9cm]{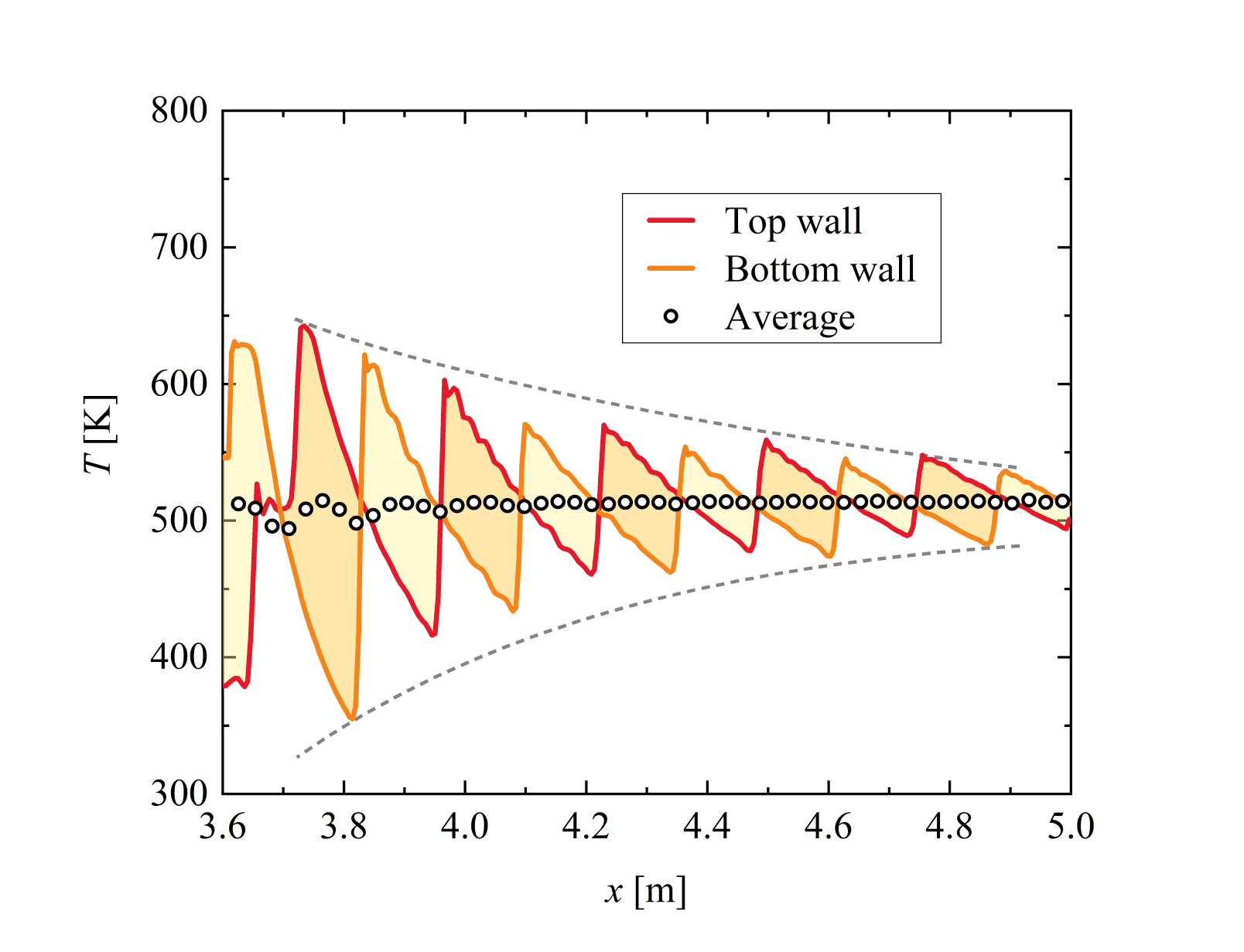}
      \par
      \centering \; (\textit{a}) \hspace{7.4cm} (\textit{b})
    }
  }
	\caption{Parameter distribution along isolator path under Mach 5.0 inflow conditions. (\textit{a}) Static pressure, (\textit{b}) static temperature.}
	\label{fig:SI:lineMa5}
\end{figure}

The outlet parameters of the isolator section are crucial in the system modeling and simulation process of scramjet engines because they are related to the design targets of the downstream combustion chambers. Therefore, a dedicated study was conducted on the variations in the isolator section parameters along the flow path, as shown in Figures \ref{fig:SI:lineMa8} and \ref{fig:SI:lineMa5}. Figure \ref{fig:SI:lineMa8} presents the along-path variations in the static temperature/pressure on the upper wall, lower wall, and cross-sectional average in the isolator section under the design conditions. Although the isolator section experienced continuous shock reflections and interactions between shock and expansion waves, the average pressure and temperature remained relatively stable. In Figure \ref{fig:SI:lineMa5}, under non-design conditions, the front end of the isolator section may exhibit more complex wave structures and parameter fluctuations compared with the design condition; however, the variations in the averages remain stable within a certain range. From the along-path distributions, we can observe that at the Mach 8.5 inflow condition, the pressure-fluctuation range at the isolator section outlet was approximately 15 kPa, and the temperature-fluctuation range was approximately 130 K. When the inflow Mach number was reduced to 5.0, the pressure-fluctuation range decreased to approximately 8 kPa, and the temperature-fluctuation range decreased to approximately 15 K. The design of the downstream combustion chambers should also consider the possible deviations in these corresponding parameters. Therefore, accurate prediction of the flow in the engine inlet using the DOLINC method provides more reliable practical references for simplifying assumptions in system-level simulations (such as which incoming shocks/expansion waves can be ignored or simplified and which must be precisely modeled) and offers more accurate boundary parameters for other component-level design work, thereby improving the overall design effectiveness.

\section{Conclusions}
\label{sec:conclusions}

The present study introduces a face-neighboring cell order-lifted inversion approach, the DOLINC method, to achieve high-order schemes within a low-order FVM framework based on unstructured polyhedral grids. This method successfully implemented various high-order schemes, including the fixed-template reconstruction and ENO/WENO methods, in a multi-dimensional second-order FVM framework. This study evaluated the computational accuracy and efficiency of the DOLINC method using different computational cases, leading to the following main conclusions:

(1) In principle, the DOLINC method is applicable to any polyhedral grid stored in an unstructured data format. By increasing the order, the DOFs required for high-order schemes were converted into DOLINC differentials . The inversion formulas derived in this study enabled the precise computation of the far-field cell data required for high-order schemes on non-uniform hexahedral grids.

(2) The results from one-dimensional and two-dimensional cases indicated that using the DOLINC scheme in unstructured grid solvers could achieve analytical accuracy comparable to that of high-order solvers on structured grids, as reported in existing studies.

(3) In terms of efficiently improving the computational accuracy, the DOLINC method had an advantage over low-order FVM using refined grids. Compared to the baseline schemes, CD2/TVD, on fine grids, CSR4-DOLINC and WENO5-DOLINC achieved faster computation on coarse grids with higher accuracy, reducing the CPU time by approximately 50\% and 70\%, respectively.

(4) When implementing similar high-order schemes with comparable accuracy on unstructured grids, the DOLINC method exhibited a noticeable speed advantage over the classical k-exact method. In addition, it demonstrated better symmetry in regular grid problems.

(5) For meshes characterized by significant variations in local cell size, such as boundary-layer meshes, the accuracy of the DOLINC method was significantly higher than that of the uniform-mesh-based least-squares gradient approximation.

(6) The DOLINC method performed well for a wide range of problems, including pressure-coupling algorithms, Riemann solvers, and passive scalar transport. This effectively balanced accuracy and computational costs. The proposed method was characterized by simple implementation, did not  require special boundary handling, and exhibited high parallel efficiency. It could be easily applied to the existing low-order FVM frameworks adopted by most industrial CFD software programs with minimal modifications.

To summarize, the DOLINC method combined the advantages of easy implementation, high computational speed, and high solution accuracy. This contributed to the current trend of industrial software transitioning to higher-order methods, thereby facilitating efficient and accurate solutions to complex industrial problems. The main limitation of the DOLINC method currently lies in its low-accuracy inversion formula for tetrahedral or hybrid grids, which potentially leads to accuracy degradation. However, this issue can be addressed by developing specific inversion formulas for particular types of elements, indicating a direction for future improvements to the DOLINC method.

\section*{Acknowledgments}

This study was supported by the National Natural Science Foundation of China (U23B20108) and the National Science and Technology Major Project (Grant No. J2019-III-0019-0063).

\section*{Code availability}

The source code used to reproduce the results of this study will be openly available on GitHub at \href{https://github.com/Fracturist/FVM-High-Older-DOLINC-Scheme}{FVM-High-Older-DOLINC-Scheme} upon publication.

\bibliographystyle{unsrtnat}
\bibliography{DOLINC}

\end{document}